\documentclass[12pt,a4paper,twoside,final]{report}
\usepackage{axodraw,latexsym,amssymb,amsmath,bbm,graphicx,graphics,here,subfigure,rotating}
\usepackage[sort&compress,square]{natbib}
\usepackage{showkeys} 
\usepackage{footnote}
\usepackage{floatflt} 
\usepackage{wick}

\allowdisplaybreaks[1] 

\usepackage{fancyhdr}
\newcommand{\Tr}{\mathrm{Tr}}
\newcommand{\STr}{\mathrm{STr}}

\renewcommand{\bar}{\overline}
\newcommand{\pt}{\ensuremath{\partial_{t}}}
\newcommand{\lv}{\ensuremath{\lambda_{V}}}
\newcommand{\ls}{\ensuremath{\lambda_{\sigma}}}

\newcommand{\hs}{\ensuremath{h_{\sigma}}}
\newcommand{\ha}{\ensuremath{h_{A}}}
\newcommand{\hv}{\ensuremath{h_{V}}}

\newcommand{\hsz}{\ensuremath{h^{2}_{\sigma}}}
\newcommand{\haz}{\ensuremath{h^{2}_{A}}}
\newcommand{\hvz}{\ensuremath{h^{2}_{V}}}
\newcommand{\hsv}{\ensuremath{h^{4}_{\sigma}}}
\newcommand{\hav}{\ensuremath{h^{4}_{A}}}
\newcommand{\hvv}{\ensuremath{h^{4}_{V}}}
\newcommand{\afb}[3][4]{\ensuremath{a^{(FB),#1}_{2,2}(#2,#3)}}
\newcommand{\afbb}[4][4]{\ensuremath{a^{(FBB),#1}_{2,1,1}(#2,#3,#4)}}
\newcommand{\vv}{\ensuremath{v_{4}}}
\newcommand{\bfb}[3][4]{\ensuremath{b^{(FB),#1}_{2,2}(#2,#3)}}
\newcommand{\bfbb}[4][4]{\ensuremath{b^{(FBB),#1}_{2,1,1}(#2,#3,#4)}}
\newcommand{\cfb}[3][4]{\ensuremath{c^{(FB),#1}_{2,2}(#2,#3)}}
\newcommand{\cfbb}[4][4]{\ensuremath{c^{(FBB),#1}_{2,1,1}(#2,#3,#4)}}

\newcommand{\dfbb}[4][4]{\ensuremath{d^{(FBB),#1}_{2,1,1}(#2,#3,#4)}}
\newcommand{\efb}[3][4]{\ensuremath{e^{(FB),#1}_{2,2}(#2,#3)}}
\newcommand{\efbb}[4][4]{\ensuremath{e^{(FBB),#1}_{2,1,1}(#2,#3,#4)}}
\newcommand{\betafb}[3][4]{\ensuremath{\beta^{(FB),#1}_{2,2}(#2,#3)}}
\newcommand{\betafbb}[4][4]{\ensuremath{\beta^{(FBB),#1}_{2,1,1}(#2,#3,#4)}}
\newcommand{\gammafb}[3][4]{\ensuremath{\gamma^{(FB),#1}_{2,2}(#2,#3)}}
\newcommand{\gammafbb}[4][4]{\ensuremath{\gamma^{(FBB),#1}_{2,1,1}(#2,#3,#4)}}

\newcommand{\epsilonfb}[3][4]{\ensuremath{\epsilon^{(FB),#1}_{2,2}(#2,#3)}}
\newcommand{\epsilonfbb}[4][4]{\ensuremath{\epsilon^{(FBB),#1}_{2,1,1}(#2,#3,#4)}}

\newcommand{\ban}{\ensuremath{\beta^{(0)}_{\lambda_{A}}}}
\newcommand{\bvn}{\ensuremath{\beta^{(0)}_{\lambda_{V}}}}
\newcommand{\bsn}{\ensuremath{\beta^{(0)}_{\lambda_{\sigma}}}}
\newcommand{\baz}{\ensuremath{\beta^{(2)}_{\lambda_{A}}}}
\newcommand{\bvz}{\ensuremath{\beta^{(2)}_{\lambda_{V}}}}
\newcommand{\bsz}{\ensuremath{\beta^{(2)}_{\lambda_{\sigma}}}}
\newcommand{\aefb}[3][4]{\ensuremath{a^{(FB),#1}_{2,1}(#2,#3)}}

\newcommand{\betaefb}[3][4]{\ensuremath{\beta^{(FB),#1}_{2,1}(#2,#3)}}
\newcommand{\gammaefb}[3][4]{\ensuremath{\gamma^{(FB),#1}_{2,1}(#2,#3)}}

\newcommand{\epsilonefb}[3][4]{\ensuremath{\epsilon^{(FB),#1}_{2,1}(#2,#3)}}
\newcommand{\af}[2][4]{\ensuremath{a^{(F),#1}_{2}(#2)}}

\newcommand{\mev}{\ensuremath{\textrm{meV}}}
\newcommand{\Mev}{\ensuremath{\textrm{MeV}}}
\newcommand{\Nc}{\ensuremath{\textrm{N}_{\textrm{c}}}}
\newcommand{\Gev}{\ensuremath{\textrm{GeV}}}
\newcommand{\fss}[1]{#1\!\!\!/}
\newcommand{\msigma}{\ensuremath{m^{0}_{\sigma}}}
\newcommand{\I}{\text{i}}

\newcommand{\case}[2]{{\scriptstyle \frac{#1}{#2}}}
\newcommand{\Gk}{\Gamma_k}

\newcommand{\yb}{\bar{\psi}}

\newcommand{\pat}{\partial_t}

\newcommand{\SP}{\,(\text{S--P})}
\newcommand{\SPn}{\,(\text{S--P})_{\text{N}}}
\newcommand{\SPN}{\,(\text{S--P})^{\text{N}}}
\newcommand{\VAp}{\,(\text{V+A})}
\newcommand{\VAm}{\,(\text{V--A})}
\newcommand{\VAn}{\,(\text{V--A})_{\text{N}}}

\newcommand{\Nf}{\ensuremath{\textrm{N}_{\text{f}}}}
\newcommand{\lp}{\hat{\lambda}_{+}}
\newcommand{\lm}{\hat{\lambda}_{-}}
\newcommand{\lsc}{\hat{\lambda}_{\sigma}^{\text{c}}}
\newcommand{\lsf}{\hat{\lambda}_{\sigma}^{\text{f}}}
\newcommand{\lva}{\hat{\lambda}_{\text{VA}}}

\newcommand{\bls}{\lambda_{\sigma}}
\newcommand{\blp}{\lambda_{+}}
\newcommand{\blm}{\lambda_{-}}
\newcommand{\blsc}{\lambda_{\sigma}^{\text{c}}}
\newcommand{\blsf}{\lambda_{\sigma}^{\text{f}}}
\newcommand{\blva}{\lambda_{\text{VA}}}

\newcommand{\fsl}[1]{#1\!\!\!\!/}
\newcommand{\lF}{l_1^{\text{(F)},4}}
\newcommand{\lFB}{l^{\textrm{(FB)},4}_{1,2}}

\newcommand{\lsh}{\hat{\lambda}_{\sigma}}

\begin{document}

\pagenumbering{roman}
\begin{titlepage}
\pagestyle{empty}


\pagestyle{empty}
%
%
%
{\vspace*{-2.5cm} \hfill  HD-THEP-03-46}

\vspace{1.0cm}

\begin{center}
  \Huge \bf \sc
  Effective Actions for Strongly\\
  Interacting Fermionic Systems
\end{center}
\vfill
\begin{center}
  {
  \Large Dissertation\\}
  \vspace{0.3cm}
  {\large
  submitted to the\\
  Combined Faculties for the Natural Sciences and for Mathematics\\
  of the Ruperto--Carola University of Heidelberg, Germany\\
  for the degree of\\
  Doctor of Natural Sciences}
\end{center}
\vfill
\begin{center}
\begin{large}
  presented by\\
  \vspace{0.5cm}
  {\Large\sc J\"{o}rg J\"{a}ckel}\\
  \vspace{0.5cm}
\end{large}
\end{center}
\vfill
\begin{center}
\begin{Large}
Supervisor: Prof.~Dr.~Christof Wetterich
\end{Large}
\end{center}

\cleardoublepage


\vspace*{-1.25cm}
\vspace{-1.2cm}
\enlargethispage{0cm}
\begin{center}
  {\bf \large Effektive Wirkungen f\"{u}r stark wechselwirkende fermionische Systeme}\\[.2cm]
  {\bf Zusammenfassung}
\end{center}

\parindent0cm
{\footnotesize
Wir vergleichen verschiedene nicht-st\"{o}rungstheoretische Methoden zur Beschreibung
\mbox{fermionischer} Systeme, die gebundene bosonische Zust\"{a}nde (BBS) und
spontane Symmetriebrechung (SSB) aufweisen. In einer rein fermionischen Sprache
erfordert das Eindringen in die SSB Phase Techniken jenseits
von St\"{o}rungstheorie und Renormierungsgruppengleichungen. Dazu
ist eine Beschreibung, die BBS und elementare Teilchen
gleichberechtigt behandelt, besser geeignet. Die ``Partielle \mbox{Bosonisierung''}
f\"{u}hrt aber zu einer Willk\"{u}r in der Wahl der BBS Felder, da diese
durch die \mbox{klassische} Wirkung nicht eindeutig festgelegt ist.
Die Ergebnisse approximativer Rechnungen, z.B. mean field theory, k\"{o}nnen aber von dieser
Wahl abh\"{a}ngen. Dies beschr\"{a}nkt die quantitative Aussagekraft. Am Beispiel
des Nambu--Jona-Lasinio-Modells zeigen wir, wie diese Abh\"{a}ngigkeit durch geeignet
gew\"{a}hlte Approximationen reduziert und manchmal
sogar zum Verschwinden gebracht werden kann.

Schwinger-Dyson-Gleichungen (SDE) erlauben eine Beschreibung von SSB ohne
Hilfsfelder. Die 2PI-Wirkung erm\"{o}glicht es uns,
verschiedene L\"{o}sungen der SDE zu vergleichen und so die stabile zu finden.
Diese Methode wenden wir auf eine sechs-Fermion Wechselwirkung an, die der
drei-Flavor-Instantonwechselwirkung der QCD \"{a}hnelt. Wir finden einen
Phasen\"{u}bergang erster Ordnung in die chiral gebrochene Phase, aber
keine stabile Phase mit gebrochener color-Symmetrie.

Die Existenz eines elementaren skalaren bosonischen Teilchens im Standardmodell
-- dem Higgs -- f\"{u}hrt zu mehreren Fragen. Die mit fundamentalen Skalen ($\sim M_{\textrm{GUT}}$)
verglichen kleine Masse erfordert ein extremes Ma\ss\ an Finetuning. Au\ss erdem ist
das $\phi^4$-Potential m\"{o}glicherweise nicht renormierbar im strengen Sinne.
Im Hinblick darauf diskutieren wir die M\"{o}glichkeit, da\ss\ das Higgs ein BBS aus
Fermionen ist.

}

\vspace{1.0cm}

\begin{center}
  {\bf \large Effective Actions for Strongly Interacting Fermionic Systems}\\[.2cm]
  {\bf  Abstract}
\end{center}

\parindent0cm
{\footnotesize

We compare different non-perturbative methods for calculating the effective
action for fermionic systems featuring
bosonic bound states (BBS) and spontaneous symmetry breaking (SSB).
In a purely fermionic language proceeding into the SSB phase
requires techniques beyond perturbation
theory and renormalization group equations.
Improvement comes from a description with BBS fields
and elementary fields treated on equal footing.
Yet, ``partial bosonization'' introduces an arbitrariness as the choice for
the composite fields is usually not completely determined by the classical action.
Results of approximate
calculations, e.g. mean field theory, may depend strongly on
this choice, thus limiting their quantitative reliability. Using the Nambu--Jona-Lasinio model
as an example we demonstrate how this dependence can be reduced, sometimes even be eliminated
by suitably chosen approximations.

Schwinger-Dyson equations (SDE) allow for a description of SSB without auxiliary fields.
The 2PI effective action enables us to compare
different solutions of the SDE and find the stable one.
We apply this method to a six-fermion interaction
resembling the three-flavor instanton interaction in QCD. We find a first
order chiral phase transition but no stable phase with broken color symmetry.

The existence of an elementary scalar boson in the Standard Model -- the Higgs --
raises several questions. The smallness of its mass compared to some fundamental
scale ($\sim M_{\textrm{GUT}}$) requires an extreme amount of fine-tuning. Moreover, its
$\phi^4$-potential may not be renormalizable in a strict sense. In view
of this we discuss the possibility of a Higgs as BBS of fermions.

}
 \vfill \cleardoublepage

\end{titlepage}


\tableofcontents


\clearpage{\pagestyle{empty} \cleardoublepage} 
\setcounter{page}{1}
\pagenumbering{arabic}

\chapter{Introduction}
\begin{quotation}\sl

There are no eternal facts, as there are no absolute truths.

\begin{flushright}

Friedrich Nietzsche

\end{flushright}
\end{quotation}
\section{Bosons Made up of Fermions}
\subsection*{Everything is Made of Fermions?}
The search for ``fundamental'' particles is one of the most
ambitious enterprizes in physical research. While it has been very
successful in terms of uncovering new particles it has also led us to
question over and over again what ``fundamental'' really means. The
notion ``fundamental particle'' has changed with time. First it
was atoms then it was electrons and nuclei and later the latter
ones were split into protons and neutrons. Today, we are quite sure
that even neutrons and protons are made up of the more fundamental
quarks. What once had been fundamental particles became bound states.

\begin{figure}[t]
\includegraphics[width=15cm]{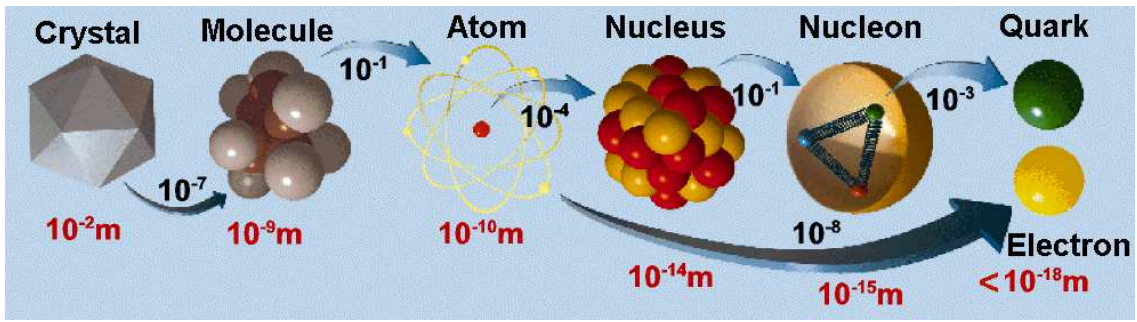}
\caption{With the exploration of smaller and smaller scales it often became apparent
that particles which were thought to be fundamental are instead composed of even smaller particles.
Nevertheless, for a description at a given scale it is often useful to treat
bound states as ``fundamental'' particles, e.g. for a first description
of water vapor it is a very good approximation to treat the water molecules as
fundamental, i.e. we use an ``effective theory'' in which the water molecules are pointlike
and have some kind of effective interaction.
However, if we want to
describe the absorption and emission of electromagnetic waves it becomes necessary
to consider the water molecule as composed of atoms (infrared) and eventually
the latter ones to be composed of electrons and a nucleus (visible, ultraviolet).
Corresponding to the scale we probe, we have to consider different effective theories.
This kind of scale dependent description is one of the main ideas
behind the renormalization group (RG) (cf. Sect. \ref{sec::rg}). The RG provides us with a means
to calculate the couplings of an effective theory at a given scale from the ones
of an effective theory valid at a smaller length scale, and ultimately from the fundamental one.}
\label{fig::microscope}
\end{figure}

Looking at Fig. \ref{fig::microscope} we can see that this evolution
is simply like turning up the resolution of a microscope.
At a low resolution we see nothing but a point (particle) while at
a higher resolution it exhibits structure, i.e. we can see that it is
composed of other particles. In particle physics the ``microscope''
is a scattering experiment and the resolution improves with smaller
wavelength $\lambda\sim\frac{\hbar c}{E}$, and therefore higher
energy $E$ of the scattering particle. Consequently, with progress in accelerator physics,
it might turn out that some (or even all) of the particles of the
Standard Model are not fundamental, but bound states.

What has all that to do with fermions?
At the moment we are in a very peculiar situation. Classifying
particles by spin ($S$) and statistics all particles of the
``Standard Model'' are either vector bosons with $S=1$ or fermions with
$S=\frac{1}{2}$. The only exception is the Higgs as it is supposed to be a
scalar ($S=0$) boson. Although, recent experiments
\cite{Barate:2000ts,Acciarri:2000ke,Abreu:2000fw,Abbiendi:2000ac,:2001xw}
at the final runs of LEP may have detected a Standard Model Higgs
(or a supersymmetric extension of it), the evidence is still
quite shaky, leaving room for speculations.
Being provocative we could state that so far we have not yet observed
any fundamental scalar boson. Hence, one could conjecture that the Higgs might be a bound state of
fermions, more explicitly a top-antitop bound state
\cite{Cvetic:1997eb,Nambu:bk,Miransky:1989ds,Miransky:1988xi,Bardeen:1989ds}.
In addition to this phenomenological aspect, the idea of ``top-quark condensation''
might also be a way to circumvent some technical problems, like the triviality of
$\phi^{4}$-theory (Higgs potential!) \cite{Callaway:ya,Lindner:1985uk,Frohlich:tw,Aizenman:du,Luscher:1987ay},
paving the way to a ``renormalizable'' Standard Model.

\subsection*{Spontaneous Symmetry Breaking}
The phenomenon of spontaneous symmetry breaking (SSB) and the formation
of bosonic bound states in strongly interacting fermionic systems are tightly
connected.

First of all, according to Goldstone's theorem \cite{Goldstone:es,Goldstone:eq} it is
inevitable to have massless Spin-0 particles, i.e. scalar bosons,
if we spontaneously break a continuous global symmetry.
In a purely fermionic theory these must be bound states.
An example for these bosons are the pseudoscalar mesons of QCD, which are the Goldstone bosons of
chiral symmetry breaking.

Secondly, it is impossible for a fermionic (Grassmannian) field to
acquire a non-vanishing vacuum expectation value. Therefore, the simplest
possible symmetry-breaking term is a bosonic operator made up of
two fermions. Thus, even for purely fermionic
theories, the phenomenon of SSB is characterized by bosons.

Physically, we can imagine a Mexican-hat-type potential
for the bosonic composite operator. Depending on whether we have a local or a global
symmetry the Goldstone bosons, corresponding to the angular excitations,
may or may not be eaten up by gauge bosons. But, in addition to those we always have
the radial excitations, corresponding to (usually) massive bosons.
E.g. in the top condensation model this would be the Higgs boson.
So, at this level, there is really no difference to a model with
elementary bosons.

Yet, there is a slight difference in the way the Mexican-hat is generated.
For elementary bosons we quite often simply choose the potential to be a
Mexican-hat, whereas for fermions the generic case is that the
non-trivial minima are generated dynamically by quantum fluctuations. That is
why it is often referred to as dynamical symmetry breaking.

Finally, we note that the case where the described radial boson becomes massless
corresponds to a second order phase transition. With features like universality these
special cases are especially interesting.

Last but not least, of course, there can be bound states not directly linked to SSB,
e.g. the hydrogen atom or positronium.

\subsection*{A lot of Bound States -- Some Models}
Of course, a speculative model of the Higgs and chiral symmetry breaking are not the only
situations where we encounter the mechanisms described above.

\textbf{Color superconductivity:} At very high density it is expected that the QCD ground state
is a color superconducting phase
\cite{Bailin:bm,Alford:1997zt,Berges:1998rc,Alford:1998mk,Schafer:1998ef,Son:1998uk,Alford:1999pa,Schafer:1999jg,Oertel:2002pj}.
Depending on the specifics of temperature, number of flavors etc.
there are several different phases. Let us just mention
the color flavor locking phase characterized by a non-vanishing expectation value of\footnote{$\textrm{L}$
refers to left handed, $\textrm{R}$ to right handed.}
$\langle \psi^{a}_{\textrm{L}i}\psi^{b}_{\textrm{L}j}\rangle
\sim(\delta_{ia}\delta_{bj}-\frac{1}{\Nc}\delta_{ij}\delta_{ab})$ as an example. In this phase the
$SU(3)_{\textrm{color}}\times SU(3)_{\textrm{L}}\times SU(3)_{\textrm{R}}$ is broken down
to a vectorlike $SU(3)_{V}$. Breaking global (chiral) as well as local (color) symmetries,
we have Goldstone bosons as well as massive gauge bosons (gluons).

\textbf{Color symmetry breaking in the vacuum:} It was conjectured \cite{Wetterich:2000pp,Wetterich:1999vd}
that the QCD
ground state at zero density might also have a broken color symmetry. In this case
the order parameter is $\langle\bar{\psi}^{a}_{\textrm{L}i}\psi^{b}_{\textrm{R}j}\rangle$
with the same type of expectation value and symmetry breaking pattern as for the
color flavor locking phase in color superconductivity. It is worth noting that the massive
gauge bosons can then be associated to the vector mesons of QCD. We will investigate
this possibility briefly in Chap. \ref{chap::bea}.

\textbf{Chiral symmetry breaking:} This is probably one of the most
studied cases of SSB \cite{Nambu:1961tp}.
A vacuum expectation value
$\langle \bar{\psi}^{a}_{\textrm{L}i}\psi^{b}_{\textrm{R}i}\rangle\sim\delta_{ab}$ gives
a mass to the (nearly) massless quarks. The pseudoscalar mesons (pions etc.) bear physical
witness of this process.

\textbf{Superconductivity:} Yeah, just plain old ordinary superconductivity \cite{Bardeen:1957mv}
in ordinary matter like
metals at temperatures of some Kelvin. It comes about due to the formation
of Cooper pairs and the condensation a non-vanishing $\langle\psi\psi\rangle$ where
$\psi$ is now an ordinary electron field. This gives us an example where
we are at a scale of some $\mev$ instead of several hundred $\Mev$ (chiral SSB) or
even $\Gev$ (Higgs model).

\section{Describing Bound States}
As there are plenty of systems featuring bosonic bound states we better start looking for ways
to calculate something useful. Interesting quantities are, of course,
masses and couplings of the bosons, and a potential vacuum expectation value.

A very useful tool to study such quantities is the effective action \cite{Jona-Lasinio:1964cw,Goldstone:es,dewitt},
replacing the action of classical field theory, it allows for a simple description of SSB, yet it includes
quantum effects. Hence, this is the object we would like to calculate.

\subsection*{The NJL model}
Many of the problems associated with the description of bound states can be studied
already in a very simple
NJL\footnote{NJL stands for Nambu and Jona-Lasinio who used this model to study
chiral symmetry breaking \cite{Nambu:1961tp}. Due to its simplicity this and similar
models are still very popular, e.g. a sizable part of the studies on
color superconductivity is based on this model
\cite{Alford:1997zt,Bailin:bm,Berges:1998rc,Alford:1998mk,Schafer:1998ef,Oertel:2002pj,Alford:1999pa}.}-type
model (for only one fermion species) with a chirally invariant pointlike four-fermion interaction:
\begin{eqnarray}
\label{equ::faction}
\textrm{S}_{\textrm{F}}=\int d^{4}x
&\bigg\{&\bar{\psi}i\fss{\partial}\psi
\\\nonumber
&+&\frac{1}{2}\lambda_{\sigma}[(\bar{\psi}\psi)^{2}-(\bar{\psi}\gamma^{5}\psi)^2]
-\frac{1}{2}\lambda_{V}[(\bar{\psi}\gamma^{\mu}\psi)^2]
-\frac{1}{2}\lambda_{A}[(\bar{\psi}\gamma^{\mu}\gamma^{5}\psi)^{2}]\bigg\}.
\end{eqnarray}

Depending on the value of $\overrightarrow{\lambda}=(\lambda_{\sigma},\lambda_{V},\lambda_{A})$
we are in a symmetric phase or in a phase with broken chiral symmetry. As it turns
out, the critical $\overrightarrow{\lambda}_{\textrm{crit}}$ separating these two phases is an interesting
but relatively easy to calculate quantity.

\subsection*{Arbitrary Parameters -- Fierz Ambiguity}
The simplest calculation which comes into ones mind is probably a mean field calculation. We will apply this
method (well-known from statistical physics) to the model Eq. \eqref{equ::faction}
at the beginning of Chap. \ref{chap::njl}.
We find that there is a basic ambiguity connected to the possibility to perform Fierz transformations (FT)
on the initial Lagrangian -- we will refer to it as Fierz ambiguity. This Fierz ambiguity can influence the
value of the critical coupling quite dramatically, severly limiting the applicability
of MFT \cite{Baier:2000yc,tobidoktor}.

The origin of the Fierz ambiguity can be understood quite well when looking at the model of Eq. \eqref{equ::faction}.
Due to the Fierz identity (s. App. \ref{app::fermion} for our conventions on $\gamma$-matrices)
\begin{equation}
\label{equ::fierz}
\left [(\bar{\psi}\gamma^{\mu}\psi)^{2}-(\bar{\psi}
\gamma^{\mu}\gamma^{5}\psi)^{2}\right]
+2\left[(\bar{\psi}\psi)^{2}-(\bar{\psi}\gamma^{5}\psi)^{2}\right ]=0
\end{equation}
only two of the quartic couplings are independent and we write
\begin{equation}
\label{equ::invariant}
\lambda_{\sigma}=\bar{\lambda}_{\sigma}+2\gamma\bar{\lambda}_{V},
\quad \lambda_{V}=(1-\gamma)\bar{\lambda}_{V},\quad \lambda_{A}=\gamma\bar{\lambda}_{V}.
\end{equation}
Where $\gamma$ parametrizes the ``symmetry'', and the $\bar{\lambda}$
are ``invariant'' couplings.

In a naive way one would like
to combine the fermions in the four-fermion interaction of Eq. \eqref{equ::faction} into pairs and interpret those as bosons,
e.g. one would like to take the term multiplying $\lambda_{A}$
pair it into two $\bar{\psi}\gamma^{\mu}\gamma^{5}\psi\sim A^{\mu}$ and hence interpret this
term as an interaction (mass term) for ``axial vector bosons''. However, using Eq. \eqref{equ::fierz}
we can now transform this term to zero, eliminating the ``axial vector bosons''.
As this pairing is the basic idea of MFT, an ambiguity seems inevitable.

\subsection*{Auxiliary Fields}
There are several standard methods (perturbation
theory, RG equations, Schwinger-Dyson equations (SDE)) which allow us to calculate
the critical coupling without any reference to a pairing into bosons. The
results are then naturally unambiguous. Yet, these
descriptions have their problems, too. Perturbation theory is unable to describe SSB at all, the RG calculation
cannot be extended into the broken phase without considerable calculational difficulty and SDE's beyond the
simplest approximation become quite
difficult to solve, too. At least part of the problem is that we lack an intuitive understanding of the
momentum dependence of complicated fermionic operators.

Looking at the example of mesons again it is known that an effective theory with bosonic meson fields
interacting with themselves and the quarks works quite well without having very complicated terms in the effective
Lagrangian. In addition, bosonic fields allow for a good understanding of SSB. Thus, it seems a reasonable
step to introduce auxiliary fields to describe the bound states.
Formalizing the naive pairing procedure,
partial bosonization
\cite{stratonovich,Hubbard:1959ub,Klevansky:1992qe,Alkofer:1996ph,Berges:1999eu}
(cf. Chap. \ref{chap::boso1}) leads to a model with massive bosonic fields
and Yukawa-type interactions, but no four-fermion interactions,
\begin{eqnarray}
\label{equ::baction}
\textrm{S}_{\textrm{B}}=\int d^{4}x&\bigg\{&i\bar{\psi}\fss{\partial}\psi
+\mu^{2}_{\sigma}\phi^{\star}\phi
+\frac{\mu^{2}_{V}}{2}V_{\mu}V^{\mu}+\frac{\mu^{2}_{A}}{2}A_{\mu}A^{\mu}
\\\nonumber
+h_{\sigma}\!\!\!\!\!&&\!\!\!\!\!\!\!\!\!\left[ \bar{\psi}\left(\frac{1+\gamma^{5}}{2}\right)\phi\psi
-\bar{\psi}\left(\frac{1-\gamma^{5}}{2}\right)\phi^{\star}\psi\right]
-h_{V}\bar{\psi}\gamma_{\mu}V^{\mu}\psi
-h_{A}\bar{\psi}\gamma_{\mu}\gamma^{5}A^{\mu}\psi\bigg\}.
\end{eqnarray}
The identification
\begin{equation}
\label{equ::bosocouplings}
\mu^{2}_{\sigma}=\frac{h^{2}_{\sigma}}{2\lambda_{\sigma}},
\quad\mu^{2}_{V}=\frac{h^{2}_{V}}{\lambda_{V}},
\quad\mu^{2}_{A}=\frac{h^{2}_{A}}{\lambda_{A}}
\end{equation}
makes this model equivalent to the NJL-type model \eqref{equ::faction}.

However, due to the Fierz identity \eqref{equ::fierz} the couplings in \eqref{equ::baction} are not unique, bringing
back the ambiguity of MFT. Indeed, MFT appears as a simple approximation to this model, neglecting all
bosonic fluctuations.

This is not the only situation in physics where an arbitrary parameter (in our case $\gamma$) appears in
calculations.
Prominent examples are
the renormalization point $\mu$ or the gauge fixing $\alpha$ in gauge theories.
Ultimately, in an exact calculation physical quantities should not depend on such a parameter.
Nevertheless, approximate calculations usually do. Improvement in the approximation
often tends to reduce the dependence on such parameters. Sometimes, there are
even special approximation schemes where we can achieve
independence of such parameters, e.g. every order of perturbation theory provides such a scheme for
the gauge fixing parameter. Beside the practical advantages, finding such an approximation scheme
also shows that there is nothing fundamentally wrong with the method in question.
From another point of view, considering an approximation which is not
independent of the arbitrary parameter, we can say that the spread of the results under a variation of the
arbitrary parameter gives us an estimate of the minimal uncertainty of a given approximation.

We will find a very simple example of an approximation which is
independent of the Fierz parameter \cite{Jaeckel:2002yy,Jaeckel:2002rm}.
Interestingly, it turns out that this introduces the concept of scale dependent degrees
of freedom.
Technically we use the scale dependence of our auxiliary fields to
keep the form of the (effective action) simple \cite{Gies:2002nw}. However,
on a deeper level we would like to interpret this simplicity as a first step to
a description with the ``right'' degrees of freedom at every scale.

Unfortunately, there is a huge
number of possible field redefinitions and in simple approximations, e.g. so
called local potential approximation (LPA) \cite{Morris:1994ki,Tetradis:1993ts,Hasenfratz:1986dm},
it is a priori not clear which is the ``correct'' one. A
criterium can be obtained only by the consideration of terms with derivatives of the fields.
In the end this leaves us to choose between high algebraic (and/or numerical) difficulty
and a ``physical guess''.

\section*{2PI Effective Action}
In view of the many problems connected with auxiliary fields it seems prudent to look for
alternatives. One possibility is the 2PI effective action \cite{Cornwall:vz,Baym:1961b,Baym:1961a}.
After the introduction of sources for the composite operators bilinear in the fields we can omit the introduction
of auxiliary fields and directly perform an additional Legendre transformation with respect to
the sources of the composite operators.
For purely fermionic theories it turns out that this description is redundant and the
dependence on the propagators is sufficient. Hence, we can omit the dependence on the fields,
leaving us with a description
completely in terms of bosonic variables (the propagators).
Nevertheless, this method is naturally not bothered by the Fierz ambiguity as we do not introduce
auxiliary fields.
In general the sources in question do not need to be local.
However, for an interpretation
as a potential this is useful, but even with this restrictions
the 2PI effective action has its own problems. Simple approximations are often unbounded from below raising
serious questions about the interpretation of the 2PI effective action.

Nevertheless, we do not want to leave this model without at least one not completely
trivial application. We will investigate an interaction resembling the three-flavor three-color instanton
interaction of QCD. This six-fermion interaction provides for a mechanism to
have a first order chiral phase transition. Furthermore, it was conjectured to
have a color-symmetry breaking vacuum \cite{Wetterich:1999vd,Wetterich:2000pp,Wetterich:2000ky}.

\section*{Outline}
As the effective action is the central object of our interest, we
will briefly review its basic definitions and some simple ways to
calculate it in Chap. \ref{chap::effectiveaction}. In Chap.
\ref{chap::njl} we will present some very simple calculations of
the critical coupling of the model Eq. \eqref{equ::faction}. In
particular, we will perform a MFT calculation and encounter the Fierz ambiguity.
Chap. \ref{chap::boso1} introduces the concept of
partial bosonization, and clarifies the origin of the
Fierz ambiguity.
A description with scale-dependent bosonic degrees of freedom allow us to cure
the Fierz ambiguity for the RG calculation in Chap. \ref{chap::boso2}.
The following Chap.
\ref{chap::bea} turns to the concept of the 2PI effective action
to circumvent the problems of partial bosonization.
As an application of the 2PI effective action we investigate an
instanton-like interaction. In particular,
we focus on chiral symmetry breaking and a possible breaking
of color symmetry by a non-vanishing color-octet condensate. Finally, in
Chap. \ref{chap::quest} we will return to the question of a
possible composite Higgs. In particular, we are interested in
the possibility of a non-perturbative renormalizable ``Standard Model''.
We will only sketch some of the ideas
and possible problems, accordingly this will be more like an
outlook. Chap. \ref{chap::final} summarizes and concludes this work.

\chapter{1PI Effective action} \label{chap::effectiveaction}

The effective action $\Gamma$ \cite{Jona-Lasinio:1964cw,Goldstone:es,dewitt} is a
very useful tool in quantum field theory (QFT). It allows us to
calculate interesting quantities like vacuum expectation values,
propagators and correlation functions more or less by simply
taking (functional) derivatives. Indeed, we can promote a
classical equation to full quantum status by replacing the action
by the effective action $S\rightarrow \Gamma$ and the fields by
their expectation values $\phi\rightarrow \langle\phi\rangle$.
E.g. the equation of motion becomes
\begin{equation}
\frac{\delta\Gamma[\langle\phi\rangle]}{\delta\langle\phi\rangle}=0.
\end{equation}

Knowledge of the effective action is equivalent to knowledge of
the full quantum theory. From this one can already deduce that
calculating the effective action is a quite difficult task and can
usually be done only approximately.

\begin{figure}[t]
\begin{center}
\subfigure[]{\scalebox{1.0}[1.0]{
\begin{picture}(80,50)
\SetOffset(35,30)
\CArc(0,0)(35,0,360)
\Vertex(-35,0){2}
\Vertex(35,0){2}
\Line(-35,0)(35,0)
\end{picture}}
\label{subfig::1PIa}}
\subfigure[]{\scalebox{1.0}[1.0]{
\begin{picture}(220,50)
\SetOffset(105,30)
\CArc(-60,0)(35,0,360)
\CArc(60,0)(35,0,360)
\Vertex(-25,0){2}
\Vertex(25,0){2}
\Line(-25,0)(25,0)
\end{picture}}
\label{subfig::1PIb}}
\end{center}
\vspace{-0.5cm}
\caption{An example of a diagram which is 1PI \ref{subfig::1PIa} and one which is not
\ref{subfig::1PIb}. The latter one can be split into two by cutting the line
between the two \mbox{bubbles.}}
\label{fig::1PI}
\end{figure}
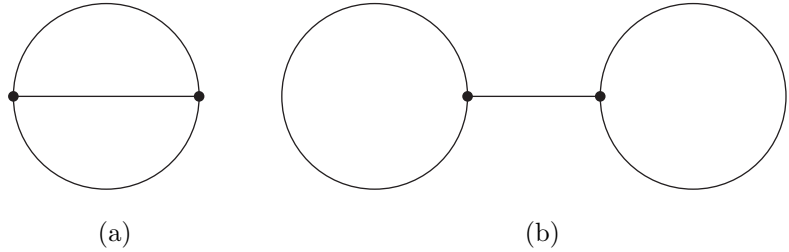

Before going into more detail let us briefly review the definition
of the \mbox{(1PI) effective} action\footnote{1PI abbreviates one
particle irreducible (cf. Fig. \ref{fig::1PI}).}, and some of its
basic properties. In the following we will write $\tilde{\phi}$
for the fluctuating quantum field and
$\phi=\langle\tilde{\phi}\rangle$. We suppress all indices.
Indeed, $\phi$ might also contain fermionic degrees of
freedom. A typical $\phi$ might therefore look like
$(\sigma,\sigma^{\star},V^{\mu},A^{\mu},\ldots,\psi_{i},\bar{\psi}_{j}\ldots)$
with several bosonic and fermionic species. If we keep track of
the order of fields and differential operators, no problems arise
from this notation.

Moreover, we work in Euclidean space. That is why we have a minus sign in the path integrals
instead of an
$i$ in front of the action. The transition to Euclidean time is
usually done via a Wick rotation. We do not want to go into detail,
here, however, for fermions there are some slight difficulties
because the action is no longer necessarily Hermitian \cite{Osterwalder:dx,zinn-justin1995}.

The generating functional (or partition function if one prefers the statistical mechanics
language) of a quantum field theory is defined by
the following functional integral
\begin{equation}
\label{equ::gen}
Z[j]=\int{\mathcal{D}}\tilde{\phi}\exp(-S[\tilde{\phi}]+j\tilde{\phi}).
\end{equation}
Here, $S[\tilde{\phi}]$ is the classical action and $j$ is an
external source. We recall that in our matrix notation $j\tilde{\phi}=\int d^{d}x j(x)\tilde{\phi}(x)$.

Using the generating functional, expectation values of fields (and
products of fields) can be calculated by taking derivatives with
respect to $j$
\begin{equation}
\label{equ::expectation}
\phi[j](q)=\langle\tilde{\phi}(q)\rangle=
\frac{\int{\mathcal{D}}\tilde{\phi}\,\tilde{\phi}(q)\exp(-S[\tilde{\phi}]+j\tilde{\phi})}
{\int{\mathcal{D}}\tilde{\phi}\exp(-S[\tilde{\phi}]+j\tilde{\phi})}
=\frac{1}{Z[j]}\frac{\delta Z[j]}{\delta j(q)}=\frac{\delta W[j]}{\delta j(q)},
\end{equation}
with
\begin{equation}
\label{equ::w}
W[j]=\ln(Z[j]).
\end{equation}
Physical values are obtained at vanishing external sources e.g. $\phi[0]$.

The (1PI) effective action is now the Legendre transform of $W$,
\begin{equation}
\label{equ::gamma}
\Gamma[\phi]=-W[j[\phi]]+j[\phi]\phi
\end{equation}
and depends on the expectation value of the field.
Combining \eqref{equ::gen}, \eqref{equ::w}, \eqref{equ::gamma} and
shifting the integration variable to
$\hat{\phi}=\tilde{\phi}-\phi$ we obtain the
following very useful formula
\begin{equation}
\label{equ::int}
\Gamma[\phi]=
-\ln\int{\mathcal{D}}\hat{\phi}
\exp\left(-S[\phi+\hat{\phi}]+\frac{\delta\Gamma[\phi]}{\delta\phi}\hat{\phi}\right),
\end{equation}
since
\begin{equation}
\frac{\delta\Gamma[\phi]}{\delta\phi}=j.
\end{equation}

We note that due to the shift of the integration variable
$\langle\hat{\phi}\rangle=0$. Finally, we would like to comment on
the notion of one particle irreducibility (1PI). This is explained
most easily in terms of Feynman diagrams. A diagram is one
particle irreducible if it is impossible to split it into two
parts by cutting an internal line (cf. Fig \ref{fig::1PI}). The
effective action is now the generating functional of the 1PI
diagrams. We stress this point because later in Chap. \ref{sec::bea}
we will encounter a 2PI effective action. For
a proof of this statement and further details about the 1PI
effective action we refer to textbooks as e.g.
\cite{zinn-justin1995,steven,peskin,hatfield}.

\section{Calculating the Effective Action}
In the following we will shortly present some of the standard methods to calculate
the effective action. Before we start let us mention that
in this section we will not explain mean field theory
as it will later (Chap. \ref{chap::boso1}) be shown to be a one-loop approximation
of a modified theory with auxiliary fields. Instead, we will explain by
example a common method to obtain it in the next chapter and postpone a somewhat
more thorough discussion until we have introduced partial bosonization
in Chap. \ref{chap::boso1}.
\subsection{Loop Expansion}
The loop expansion is a perturbative technique to calculate the effective action.
It can be shown that the effective action is nothing
but the sum of all 1PI vacuum diagrams in a background field $\phi$ as depicted in
Fig. \ref{fig::pert}
\begin{figure}[t]
\begin{center}
\scalebox{1.1}[1.1]{
    \begin{picture}(250,25)(0,0)
    \SetOffset(-30,-17)
    \Text(-10,20)[]{$\Gamma[\phi]=$}
    \Text(20,20)[]{$S[\phi]$}
    \SetOffset(-20,-17)
    \Text(35,20)[]{$+$}
    \CArc(70,20)(20,0,360)
    \Text(105,20)[]{$+$}
    \CArc(140,20)(20,0,360)
    \Vertex(120,20){2}
    \Vertex(160,20){2}
    \Line(120,20)(160,20)
    \SetOffset(50,-17)
    \Text(105,20)[]{$+$}
    \CArc(140,20)(20,0,360)
    \Vertex(160,20){2}
    \CArc(180,20)(20,0,360)
    \Text(215,20)[]{$+$}
    \Text(235,20)[]{$\cdots$}
    \end{picture}}
\end{center}
\vspace{0.5cm}
\caption{Perturbative expansion of the effective action. To be explicit we choose
a theory which has a quartic interaction. The propagators are propagators
in a background field $\phi$. For the example of a theory with a
$\frac{\lambda}{12}\tilde{\phi}^{4}$ interaction the propagator in the background field would be
$(p^2+m^2+\lambda\phi^2)^{-1}$.}
\label{fig::pert}
\end{figure}
An easy way to obtain this expansion is to make a saddle-point approximation
of Eq. \eqref{equ::int}.
In lowest non-trivial order one obtains,
\begin{eqnarray}
\label{equ::perti}
\Gamma[\phi]=S[\phi]+\frac{1}{2}\STr\ln(S^{(2)}[\phi])+\cdots,
\end{eqnarray}
where
\begin{equation}
S^{(2)}[\phi]=
\frac{\overrightarrow\delta}{\delta\phi^{\textrm{T}}}S[\phi]\frac{\overleftarrow\delta}{\delta\phi}.
\end{equation}
In the diagrammatic language (Fig. \ref{fig::pert}) the second term of Eq. \eqref{equ::perti}
corresponds to the one-loop diagram, and we have omitted the higher loop
diagrams in \mbox{Eq. \eqref{equ::perti}}.
The supertrace ($\STr$) comes around due to our notation where bosonic as well as fermionic
degrees of freedom are contained in $\phi$. However, its effect is very simple
as it only provides a minus sign in the fermionic sector of the matrix.

An advantage of perturbation theory is that it can usually be constructed to
preserve symmetries order by order in the expansion.
However, as we will see in the next chapter
it has severe shortcomings if we want to go into the non-perturbative domain
(hence the name) where the coupling is not small. Indeed, we will see that it is
unable to describe the interesting phenomenon of SSB.

\subsection{Renormalization Group Equations} \label{sec::rg}
Originally devised as a tool to hide the infinities of quantum field theory the renormalization group
has shown to give us a much deeper insight into the scale dependence of physics.
Especially, the understanding of critical phenomena has profited immensely. In addition,
in the last fifteen years or so the renormalization group has been established as a powerful
tool for doing actual calculations in the non-perturbative regime. It is the latter
aspect on which we want to focus.

The basic idea of renormalization group equations is to introduce scale dependent effective
couplings. Roughly speaking we look at the physical system of interest with a microscope
with varying resolution. E.g. consider a lattice where a particle
with spin sits at each lattice site. With a very high resolution we can see every particle and can measure each spin
independently. However, shifting to a lower resolution things become a little bit blurry
and we can only resolve regions which already contain several particles. What we then measure
is something like an effective spin (more or less the sum of the individual spins) of
several particles combined. Indeed, the first papers
\cite{Kadanoff:1966wm,Wilson:1973jj,Wilson:1971bg,Wegner} on renormalization group equations
considered spin models like the Ising model.

There is a wide variety of means for putting this intuitive picture into a mathematical form.
A very convenient picture to do this is the path integral formulation.
For a theory with an UV cutoff $\Lambda$ one can write the
partition function as
\begin{equation}
Z=\prod_{p\leq\Lambda}\int d\phi(p)\exp(-S[\phi]),
\end{equation}
where the symbolic notation shall indicate that we integrate only over momentum modes
with $p\leq\Lambda$.

We can now implement the idea of decreasing the resolution of our microscope
by integrating out modes in a small momentum shell
$[\Lambda^{\prime}=\Lambda-\Delta\Lambda,\Lambda]$.
This procedure averages over the small length scales we do not want to see,
\begin{equation}
\label{equ::ergebasic}
Z=\prod_{p\leq\Lambda^{\prime}}\int d\phi(p)\exp(-S^{\prime}[\phi])
\end{equation}
with
\begin{equation}
\label{equ::ergebasic2}
\exp(-S^{\prime}[\phi])=\prod_{\Lambda^{\prime}\leq p\leq\Lambda}
\int d\phi(p) \exp(-S[\phi]).
\end{equation}
Where the so-called Wilson effective action $S^{\prime}$ is now integrated only over a smaller
range of momenta. Considering a $\Lambda^{\prime}$ which is infinitesimally close to $\Lambda$
one can derive an evolution equation for $S^{\prime}$ in the form of a differential equation.

Following this general procedure (often followed by a
re-scaling $p\rightarrow \frac{\Lambda}{\Lambda^{\prime}}p$ to recover the initial
momentum range) allows to derive a
variety of equations for different physical quantities like the
Hamilton operator, correlation functions, coupling constants etc..
In addition, we do not need to restrict ourselves to the sharp momentum cutoff
indicated above. Indeed, a smooth momentum cutoff is often much more convenient, as the
sharp one has a tendency to introduce non-localities in position space.

All equations derived in this way are exact, i.e. if they could be solved exactly
they would give the same results for physical quantities. Hence, they are
called exact renormalization group equations (ERGE). In principle they
are all equivalent. Nevertheless, in practical computations where we have
to use approximations they usually differ. For a review and
a comparison s. \cite{Bagnuls:2000ae}.

Now, let us go directly to a formulation
for the effective action or, more precisely, the effective average action
as introduced in
\cite{Bonini:1992vh,Ellwanger:mw,Wetterich:1989xg,Wetterich:yh,Wetterich:1993be,Wetterich:an}.
Let us begin by noting that for the effective action the running cutoff $\Lambda^{\prime}$
becomes an infrared cutoff. This is due to the fact that the effective action contains
the modes which are already integrated out, thus those in the
range $[\Lambda^{\prime},\Lambda]$. $\Lambda^{\prime}$ becomes a lower bound
for the modes included in $\Gamma_{\Lambda^{\prime}}$ and therefore an infrared cutoff.
We remark that $\Gamma_{\Lambda^{\prime}}$ depends both on the
IR cutoff $\Lambda^{\prime}$ as well as on the physical UV cutoff $\Lambda$.
The first is quite clear since $\Lambda^{\prime}$ simply measures how far we have
progressed in integrating out modes. The dependence on the UV cutoff is actually more
physical. E.g. if we want to describe interacting particles on a lattice
a continuum description might be quite good for length scales larger than the
lattice spacing. However, roughly speaking this model does not contain any momentum modes
higher than the inverse lattice spacing. Placing the same particles on lattices
with different spacing will obviously produce different results. Thus, we have a
dependence on the UV cutoff (for some more details s. also App. \ref{app::infanduv}).
Indeed, only in very special theories is it possible
to send the UV cutoff to infinity and still obtain finite results. These are, of
course, the renormalizable theories. In this case we can have $\Lambda=\infty$.

A big advantage of the formulation in terms of the effective action is that
$\Gamma_{\Lambda^{\prime}}$ has a direct physical interpretation.
Having included all quantum fluctuations above $\Lambda^{\prime}$ we can now view
$\Gamma_{\Lambda^{\prime}}$ as the ``microscopic action'' on the scale $\Lambda^{\prime}$ where
we have averaged over volumes of a size $\sim\Lambda^{\prime(-d)}$.
Hence, the name effective average action or coarse-grained effective action.
In an ideal description we would be able to observe the change of the relevant degrees of freedom
from one scale to another, e.g. at very small scales we would have
quark and gluon degrees of freedom while at a larger scale we observe mesons and nucleons,
and at even larger scales we would have atoms or molecules, putting
the basic idea of the renormalization group into full effect.

Let us now get started and derive an explicit equation. To establish more clearly
its function as an IR cutoff
we write $k$ instead of $\Lambda^{\prime}$ or use the convenient $t=\ln(k)$.

In order to achieve a suppression of the low-momentum modes in the functional integral
we add an effective momentum-dependent mass term
\begin{equation}
\label{equ::masscutoff}
\Delta S_{k}[\phi]=\frac{1}{2}\int_{p} \phi^{\textrm{T}}(-p)R_{k}(p)\phi(p)
\end{equation}
to the initial action. The idea is to add a high mass to the momentum modes $p\leq k$
and a small or zero mass to those with $p\gg k$, therefore effectively removing
(or at least suppressing) the low-momentum modes in the functional integral.
To render this more precise we demand the following constraints for the
function $R_{k}(p)$,
\begin{equation}
\label{equ::conditions}
\mathbf{1.}\,\,\lim_{p^2/k^2\rightarrow 0} R_{k}(p)>0,\quad
\mathbf{2.}\,\,\lim_{p^2/k^2 \rightarrow \infty}R_{k}(p)=0,
\quad \mathbf{3.}\,\,\lim_{k\rightarrow\infty}R_{k}(p)\rightarrow \infty.
\end{equation}
The first condition is the statement that we want to suppress the low momentum
modes by an additional mass term. Having a mass term for the zero-momentum modes
has another very nice effect as it removes all IR divergences produced by
massless particles.
The second condition ensures that the high-momentum modes (high compared to $k$)
are not suppressed and that the cutoff is removed in the limit $k\rightarrow 0$.
As we will later see when we have the explicit expression
for the flow equation it is useful to choose a cutoff
that vanishes sufficiently fast (e.g. exponentially) in the UV to
avoid UV divergences.
The third condition ensures that at $k\rightarrow\infty$ no modes are integrated out. However, a comment
is in order as
we want to send $k$ to infinity. Originally, we wanted to suppress all modes
with $p<k$. But, if we have a finite physical cutoff $\Lambda$ there are
no modes with $p>\Lambda$. Thus, one might want to rewrite
condition 3 as
\begin{equation*}
\mathbf{3^{\prime}.}\,\,\lim_{k\rightarrow\Lambda}R_{k}(p)\rightarrow \infty,
\end{equation*}
and indeed this is a quite common form given and used e.g. in
\cite{Ellwanger:wy,Gies:2002af,Berges:2000ew}. However, both conditions are equivalent
as we can use every bijective function $k^{\prime}(k)$ which maps $[0,\Lambda]$ into
$[0,\infty]$ to re-scale $k$ such that either $3$ or $3^{\prime}$ is fulfilled.
Using a smooth cutoff the notion of modes being included or not included is anyway
somewhat blurry. With a little bit of discretion in
the choice of $k^{\prime}(k)$ we do not distort the picture of integrating out
modes down to the scale $k$ very much for $k<\Lambda$,
making the choice between $3$ and $3^{\prime}$ a matter of convenience\footnote{Some additional
details concerning different possible UV regularizations are provided in \mbox{App. \ref{app::infanduv}.}}. In this
work we will choose the former since it makes some of the analytical expressions easier.
More important for the interpretation of $k$ as the coarse graining
with modes $p\lesssim k$ not yet being integrated
out is that the shift between small values of $R_{k}$ and large values of $R_{k}$
occurs roughly at $p\approx k$. This is not a necessary condition to define the flow
equation but useful for the interpretation. We leave it at the somewhat rough statement
\begin{equation}
\label{equ::contribcond}
\mathbf{4.}\,\,R_{k}(p)\,\, \textrm{large for}\,\,p<k,\quad R_{k}(p)\,\,
\textrm{small for}\,\, p>k.
\end{equation}
Having talked so much about the cutoff let us finally give an explicit example of a very
convenient one introduced in \cite{Litim:2001up},
\begin{equation}
\label{equ::optcutoff}
R_{k}(p)=Z_{k}(1-p^2)\Theta\left(1-\frac{p^2}{k^2}\right).
\end{equation}
Here we have included a factor of the wave function renormalization $Z_{k}$. This choice
allows us to write
\begin{equation}
Z_{k}p^2+R_{k}(p)=Z_{k}P(p),
\end{equation}
and guarantees that the reparametrization invariance of physical quantities $\phi\rightarrow\alpha\phi$
is respected.

As discussed in the last few paragraphs the term \eqref{equ::masscutoff} acts more or less like
an additional mass term. In consequence we might get concerned about violating symmetries
like chiral or gauge symmetry. In principle there are two possible ways to tackle this problem.
The first is to construct a cutoff function $R_{k}(p)$ which fulfills Eq. \eqref{equ::conditions} but
does not violate
the symmetry. This is relatively simple for chiral fermions
\cite{Wetterich:an,Bornholdt:1992up,Bornholdt:za,Comellas:1995ea}. An example is
\begin{equation}
\label{equ::fermioncutoff}
R_{k}(p)=Z_{k}\fss{p}\left(\sqrt{\frac{k^2}{p^2}}-1\right)\Theta(1-\frac{p^2}{k^2}),
\end{equation}
which is more or less \eqref{equ::optcutoff} adapted to chiral fermions.
The second strategy (usually used for gauge theories) is
to accept the fact that during the flow the symmetry might be violated
by the cutoff and is completely symmetric only at the endpoint when the cutoff vanishes. Nevertheless,
the symmetry is only hidden during the flow and reveals itself in modified Ward-Takashi identities
\cite{Attanasio:1996jd,Bonini:1994dz,Bonini:1994kp,Ellwanger:1996wy,Ellwanger:1995qf}.
Taking these into account it is possible to construct an invariant flow.

Adding the cutoff to the initial action we get a set of \emph{different}
actions parametrized by the scale $k$,
\begin{equation}
S_{k}[\phi]=S[\phi]+\Delta S_{k}[\phi].
\end{equation}
Correspondingly, we obtain
\begin{equation}
W_{k}[j]=\ln(Z_{k}[j])=\int {\mathcal{D}}\tilde{\phi}\exp(-S_{k}[\tilde{\phi}]+j\tilde{\phi})
\end{equation}
and
\begin{equation}
\phi=\langle\tilde{\phi}\rangle=\frac{\overrightarrow{\delta}}{\delta j}W_{k}[j].
\end{equation}
Now, we want to introduce the effective average action by a modified Legendre transform
\begin{equation}
\label{equ::average}
\Gamma_{k}[\phi]=-W_{k}[j[\phi]]+j[\phi]\phi-\Delta S_{k}[\phi],
\end{equation}
where we have substracted $\Delta S_{k}[\phi]$ in order to remove the
cutoff effects from $\Gamma_{k}$.
This is particularly clear in the case $k\rightarrow\infty$.
Consider the formula analogous to Eq. \eqref{equ::int},
\begin{equation}
\Gamma_{k}=-\ln\int{\mathcal{D}}\hat{\phi}\exp\left(-S[\phi+\hat{\phi}]
+\frac{\delta\Gamma_{k}[\phi]}{\delta\phi}\hat{\phi}-\Delta S_{k}[\hat{\phi}]\right).
\end{equation}
Due to the third condition in \eqref{equ::conditions} the additional term $\exp(-\Delta S_{k}[\hat{\phi}])$
in the functional integral acts like a functional $\delta(\hat{\phi})$-function for $k\rightarrow\infty$
and we obtain
\begin{equation}
\label{equ::infty}
\Gamma_{\infty}[\phi]=S[\phi],
\end{equation}
the microscopic action without the cutoff.

\begin{figure}[t]
\begin{center}
\begin{picture}(160,150)
\includegraphics[width=6cm]{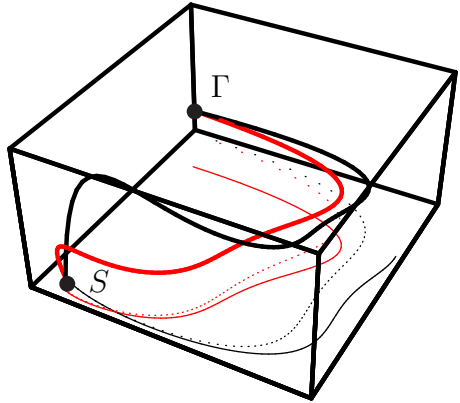}
\SetOffset(-170,-1)
\Vertex(22,45){3}
\Text(34,46)[]{$S$}
\Vertex(70,110){3}
\Text(80,120)[]{$\Gamma$}
\end{picture}
\end{center}
\caption{RG flow in the space of all action functionals. The thick line is the exact flow in the
(here 3-dimensional) space, the thin line is the approximate flow in the 2-dimensional truncation.
We point out that the approximate flow does, of course, not coincide with the projection of the
exact flow on the subspace of the truncation (dotted line), as the latter is driven also by the operator
in the third direction. The red set of lines, shows the same but for an ``optimized'' IR regulator.
The exact flows coincide only at the start- and endpoint. The optimized flow is generally closer to the plane
of the truncation, and the approximate flow is improved.}
\label{fig::approxflow}
\end{figure}

Since the cutoff vanishes for $k\rightarrow 0$ (second condition) we have in addition
\begin{equation}
\label{equ::zerok}
\Gamma_{0}[\phi]=\Gamma[\phi].
\end{equation}
Thus, the effective average action interpolates between the classical or bare action and
the full effective action (cf. Fig. \ref{fig::approxflow}).

Now, let us come to the final piece, the ERGE which governs the evolution from $k=\infty$
to $k=0$. Taking a derivative with respect to $k$,
\begin{eqnarray}
\frac{\partial}{\partial k}(\Gamma_{k}[\phi])&=&-(\partial_{k}W_{k})[j]
-(\partial_{k}j)\frac{\overrightarrow\delta}{\delta j}W_{k}[j]+(\partial_{k}j)\phi-\partial_{k}\Delta S_{k}[\phi]
\\\nonumber
&=&\langle \partial_{k}\Delta S_{k}[\tilde{\phi}]\rangle-\partial_{k}\Delta S_{k}[\phi]
\\\nonumber
&=&\frac{1}{2}\left(\langle\tilde{\phi}^{\textrm{T}}(\partial_{k}R_{k})\tilde{\phi}\rangle
-\langle\tilde{\phi}^{\textrm{T}}\rangle(\partial_{k}R_{k})\langle\tilde{\phi}\rangle\right)
\\\nonumber
&=&\frac{1}{2}\left(\STr\left[(\langle\tilde{\phi}\tilde{\phi}^{\textrm{T}}\rangle
-\langle\tilde{\phi}\rangle\langle\tilde{\phi}^{\textrm{T}}\rangle)\partial_{k}R_{k}\right]\right).
\end{eqnarray}
Expressing this with
\begin{equation}
W^{(2)}_{k}=\frac{\overrightarrow{\delta}}{\delta j}\frac{\overrightarrow{\delta}}{\delta j^{\textrm{T}}}W_{k}
=\langle\tilde{\phi}\tilde{\phi}^{\textrm{T}}\rangle-\langle\tilde{\phi}\rangle\langle\tilde{\phi}^{\textrm{T}}\rangle.
\end{equation}
gives
\begin{equation}
\label{equ::vorstufel}
\partial_{k}\Gamma_{k}=\frac{1}{2}\STr(W^{(2)}_{k}\partial_{k}R_{k}).
\end{equation}
Since
\begin{equation}
j_{k}=(\Gamma_{k}+\Delta S_{k})\frac{\overleftarrow{\delta}}{\delta \phi}
\end{equation}
we have
\begin{equation}
\mathbf{1}=\left(\frac{\overrightarrow{\delta}}{\delta j}\phi^{\textrm{T}}\right)
\left(\frac{\overrightarrow{\delta }}{\delta \phi^{\textrm{T}}}j\right)
=W^{(2)}_{k}(\Gamma^{(2)}_{k}+R_{k})
\end{equation}
with
\begin{equation}
\Gamma^{(2)}_{k}=\frac{\overrightarrow{\delta}}{\delta\phi^{\textrm{T}}}
\Gamma_{k}\frac{\overleftarrow{\delta}}{\delta\phi}.
\end{equation}
Inserting this into Eq. \eqref{equ::vorstufel} yields, finally, the flow equation
\begin{equation}
\label{equ::flow1}
\partial_{k}\Gamma_{k}=\frac{1}{2}\STr\left\{(\Gamma^{(2)}_{k}+R_{k})^{-1}\partial_{k}R_{k}\right\}.
\end{equation}
Defining the operator
\begin{equation}
\tilde{\partial}_{t}=(\partial_{t}R_{k})\frac{\partial}{\partial R_{k}}
\end{equation}
we can obtain an even more compact form
\begin{equation}
\label{equ::flow2}
\partial_{t}\Gamma_{k}=\frac{1}{2}\STr[\tilde{\partial}_{t}\ln(\Gamma^{(2)}_{k}+R_{k})].
\end{equation}
This looks quite similar to Eq. \eqref{equ::perti}, and indeed if we neglect the change of $\Gamma_{k}$
on the right hand side, integrate and use \eqref{equ::infty} at $k=\infty,\,t=-\infty$, we recover \eqref{equ::perti}.

In short, substituting $\Gamma^{(2)}_{k}+R_{k}$ for $S^{(2)}$ and writing it in a differential form
turns the one-loop expression \eqref{equ::perti} into an exact equation. The one-loop
form of Eqs. \eqref{equ::flow1}, \eqref{equ::flow2} is depicted in Fig. \ref{fig::flowequation}.

\begin{figure}[t]
\begin{center}
\scalebox{0.65}[0.65]{\fbox{
\begin{picture}(600,80)(40,-30)
\SetOffset(12,10)
\Text(140,-1)[]{\scalebox{1.5}[1.5]{$\partial_{k}\Gamma_{k}=$}}
\SetOffset(172,10)
\CArc(60,0)(-30,0,360)
\GOval(60,-30)(8,8)(0){0.5}
\Vertex(60,30){3}
\Text(200,0)[]{\scalebox{1.5}[1.5]{$\partial_{k}$}}
\SetOffset(392,-20)
\Line(0,30)(60,30)
\GOval(30,30)(8,8)(0){0.5}
\Text(90,29)[]{\scalebox{1.5}[1.5]{$=$}}
\SetOffset(482,-20)
\Line(20,0)(60,0)
\Vertex(60,60){3}
\Line(60,0)(100,0)
\CArc(60,30)(-30,-90,90)
\CArc(60,30)(-30,90,270)
\GOval(60,0)(8,8)(0){0.5}
\GOval(30,30)(8,8)(0){0.5}
\GOval(90,30)(8,8)(0){0.5}
\end{picture}}}
\end{center}
\caption{Depiction of the flow equation \eqref{equ::flow1}. The line
with the shaded circle is the full field dependent and IR regularized propagator $\Gamma^{(2)}_{k}+R_{k}$. The dot
denotes the insertion of $\partial_{k}R_{k}$. Taking functional derivatives with respect to the
field $\phi$ adds external legs, i.e. we obtain flow equations for the propagators and vertices.
An example for the flow of the propagator is shown on the right side. The shaded circle
denotes the full $k$-dependent vertex.}
\label{fig::flowequation}
\end{figure}

In addition to being exact, Eq. \eqref{equ::flow1} has two more nice features. First, due
to the presence of $R_{k}$ the expression is IR finite. Second, for $R_{k}$ decreasing
sufficiently fast in the UV, $\partial_{k}R_{k}$ provides
an UV regularization. The flow equation is therefore completely finite. Of course,
those divergences must still be included in some way. While the IR divergences
might reappear in the integration of the flow equation for $k\rightarrow 0$, the
UV divergences have been absorbed in the initial condition for some finite $k$.
Specifying the initial conditions of the flow at some finite $k$ gives a special
regularization scheme called the ERGE scheme (cf. App. \ref{app::regularization}).

Having a flow equation is only part of the game. Since it is a functional differential equation,
it is in most cases impossible to solve it analytically, and we remember once again that this
would amount to solving the quantum field theory in question.
So, it will be impossible or at least difficult to avoid using approximations in
most of the physically relevant cases.

A consistent and systematic approach is the use of truncations. In a truncation we restrict the space of
all possible actions ($\Gamma_{k}$), spanned by all possible combinations of field operators compatible with
the symmetries to a (very often finite dimensional) subspace given by a subset of operators.
The approximate flow equation now is the projection of the flow onto
this subspace (cf. Fig. \ref{fig::approxflow}). From this we can calculate flow equations for
the coefficients (generalized couplings) in front of the operators.
We stress that the approximate flow is only driven by the operators in the subset.
An easy, nevertheless usually quite tedious, way to improve the
approximation and to check for errors is to enlarge the subspace.
Doing this successively we may find a ``convergence'' of the results, and we may be tempted to
interpret this as the approach to the right result.
Still, we should be careful with this as it may well be, that we have
indeed convergence, but convergence to the wrong result. This is usually
the case when we have missed a relevant operator. As the number of all possible
linearly independent operators is infinite we can always add operators but still miss the
relevant one.

From a physics point of view it is clear that a good approximation should include all relevant
degrees of freedom, i.e. the corresponding operators.
However, as we discussed in the introduction the relevant degrees of freedom can change with scale,
making it necessary to adapt the description during the flow. For our case of interest
a step in this direction was taken in \cite{Gies:2002nw} and we will
discuss it at length in Chap. \ref{chap::boso2}.

Finally, let us come to the role the cutoff plays in
approximations of the flow equation. By construction, an exact
solution has a $\Gamma_{0}=\Gamma$ independent of the cutoff. But,
the trajectory $\Gamma_{k}$ is not independent of the cutoff. As
depicted in Fig. \ref{fig::approxflow} it may well be that a
certain choice of IR cutoff may bring the real trajectory closer
to the subspace defining our truncation, therefore usually
improving the approximation. A systematic study to exploit this
possibility has been put forward in
\cite{Gies:2002af,Canet:2002gs,Litim:2002cf,Litim:2001dt,Litim:2001up,Litim:2000ci}.
A more traditional approach would be to use the IR cutoff
dependence as a measure of uncertainty, s. e.g. \cite{Aoki:1997fh}.
\subsection{Schwinger-Dyson Equations}
Schwinger-Dyson equations (SDE) \cite{Schwinger:1951ex,Dyson:1949ha}
(for a review and some applications s. \cite{Roberts:dr,Roberts:2000aa,Alkofer:2000wg})
were one of the first really non-perturbative tools
in quantum field theory. For a theory with polynomial interactions up to $\phi^{m}$
they provide an (infinite) hierarchy of equations which connect a
1PI Greens function of order $n$  ($n$th derivative of the effective action)
on the one hand with a set of 1PI Greens functions up to order
$(n+m)$ on the other hand.

In principle they are a consequence of the fact, that
the functional integral over a total derivative vanishes, if
the functional vanishes at the boundary (like in normal calculus),
\begin{eqnarray}
\label{equ::ze}
0&=&\int{\mathcal{D}}\phi\frac{\overrightarrow{\delta}}{\delta\phi}\exp(-S[\phi]+j\phi)
\\\nonumber
&=&\int{\mathcal{D}}\phi\left(-\frac{\overrightarrow{\delta} S[\phi]}{\delta\phi}\pm j\right)\exp(-S[\phi]+j\phi)
\\\nonumber
&=&\left(\frac{\overrightarrow{\delta} S}{\delta\phi}[\frac{\delta}{\delta j}]\pm j\right)Z[j],
\end{eqnarray}
where $+$ is for bosons and $-$ is for fermions, respectively.
Examining the last line in Eq. \eqref{equ::ze} it becomes clear that the SDE's are the Euler Lagrange
equations of quantum field theory.

\begin{figure}[t]
\begin{center}
\scalebox{0.65}[0.65]{\fbox{
\begin{picture}(640,80)(0,-30)
\SetOffset(12,10)
\Line(0,0)(52,0)
\Line(68,0)(120,0)
\GOval(60,0)(8,8)(0){0.5}
\Text(140,-1)[]{\scalebox{1.5}[1.5]{$=$}}
\SetOffset(172,10)
\Line(0,0)(60,0)
\Line(60,0)(120,0)
\Vertex(60,0){2}
\Text(140,0)[]{\scalebox{1.5}[1.5]{$+$}}
\SetOffset(332,-20)
\Line(0,0)(60,0)
\Vertex(60,0){2}
\Line(60,0)(120,0)
\CArc(60,30)(-30,-90,90)
\CArc(60,30)(-30,90,270)
\GOval(60,60)(8,8)(0){0.5}
\Text(140,30)[]{\scalebox{1.5}[1.5]{$+$}}
\SetOffset(492,10)
\Line(0,0)(60,0)
\Vertex(30,0){2}
\ArrowLine(60,0)(120,0)
\CArc(60,0)(-30,0,360)
\GOval(90,0)(8,8)(0){0.5}
\GOval(60,30)(8,8)(0){0.5}
\GOval(60,0)(8,8)(0){0.5}
\GOval(60,-30)(8,8)(0){0.5}
\end{picture}}}
\end{center}
\caption{SDE for the propagator (in $\phi^4$-theory). Full propagators and vertices are depicted with
a shaded circle while bare quantities are represented by a dot. This equation is exact. However, it involves
the full 4-point function which is given by another SDE. A simple approximation would be to neglect
the last diagram on the right hand side giving us a closed equation.}
\label{fig::SDE}
\end{figure}
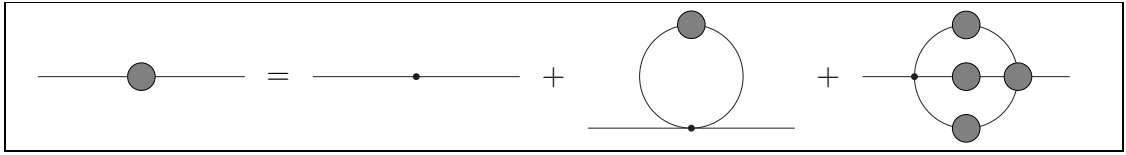

Due to the appearance of $S$ on the right hand side of Eq. \eqref{equ::ze} we have always exactly one
bare vertex in every expression contributing to the right hand side. A typical SDE therefore looks like
in Fig. \ref{fig::SDE}.

In its basic form the SDE allow us to calculate only the derivatives of the effective action. At first
sight this may not seem like a major weakness, but in situations where we have multiple solutions
for the SDE we have no way to compare them without knowledge of the value of the effective action.
However, the case of multiple solutions is one of the most interesting ones, as it usually signals
the possibility for spontaneous symmetry breaking. In Chap. \ref{chap::bea} we will
discuss the 2PI effective action \cite{Cornwall:vz,Baym:1961b,Baym:1961a} as a remedy to this problem.

As in the case of the ERGE most of the time we are unable to solve the complete
set of SDE. A popular approximation scheme is to neglect all 1PI Greens functions
starting from a certain order. This gives a closed set of integral equations.
More generally, similar to a truncation for an ERGE, one can restrict the space of
all possible $\Gamma$ and hence its derivatives to a subspace.

On the exact level the RG and SD approaches are equivalent in the sense that the propagator and higher N-point functions
calculated using the flow \mbox{equation \eqref{equ::flow1}} are also solutions of the
SDE
\cite{Ellwanger:1996wy,Terao:2000ae}. Nevertheless, once approximations are used the results will, in general,
differ.

\chapter{A Simple Example: The NJL Model} \label{chap::njl}
In this chapter we want to get a grasp of the problems associated with the introduction
of bosonic composite fields by studying the
model, \mbox{Eq. \eqref{equ::faction}} in some very simple approximations. In particular,
we use this as an opportunity to introduce MFT.
\section{Critical Couplings from Mean Field Theory} \label{sec::mean}
A mean-field calculation treats the fermionic fluctuations in a homogenous background of
fermion bilinears $\tilde{\phi}=\langle\bar{\psi}\left(\frac{1-\gamma^{5}}{2}\right)\psi\rangle$,
$\tilde{\phi}^{\star}=-\langle\bar{\psi}\left(\frac{1+\gamma^{5}}{2}\right)\psi\rangle$,
$\tilde{V}_{\mu}=\langle\bar{\psi}\gamma_{\mu}\psi\rangle$ and
\mbox{$\tilde{A}_{\mu}=\langle\bar{\psi}\gamma_{\mu}\gamma^{5}\psi\rangle$.} It seems straightforward
to replace in the four-fermion interaction in Eq. \eqref{equ::faction} one factor by the bosonic mean field,
i.e.
\begin{eqnarray}
\label{equ::meanferm}
\nonumber
(\bar{\psi}\psi)^{2}-(\bar{\psi}\gamma^{5}\psi)^{2}
&\rightarrow&2\tilde{\phi}\bar{\psi}(1+\gamma^{5})\psi-2\tilde{\phi}^{\star}\bar{\psi}(1-\gamma^{5})\psi,
\\\nonumber
(\bar{\psi}\gamma_{\mu}\psi)^{2}&\rightarrow&2\tilde{V}_{\mu}\bar{\psi}\gamma^{\mu}\psi,
\\
(\bar{\psi}\gamma_{\mu}\gamma^{5}\psi)^{2}&\rightarrow&2\tilde{A}_{\mu}\bar{\psi}\gamma^{\mu}\gamma^{5}\psi.
\end{eqnarray}
The partition function becomes then a functional of $\tilde{\phi}$, $\tilde{V}_{\mu}$,
$\tilde{A}_{\mu}$,
\begin{equation}
\label{equ::part}
Z[\tilde{\phi},\tilde{V},\tilde{A}]=\int {\mathcal{D}}\bar{\psi}{\mathcal{D}}\psi
\exp\left(-\textrm{S}[\bar{\psi},\psi,\tilde{\phi},\tilde{V},\tilde{A}]\right),
\end{equation}
where S is given by \eqref{equ::faction}, with the
replacements \eqref{equ::meanferm}. Self consistency for the expectation values of
the fermion bilinears requires
\begin{eqnarray}
\label{equ::self}
\tilde{\phi}&=&\frac{1}{2}\langle\bar{\psi}(1-\gamma^{5})\psi\rangle
=\frac{1}{2}\lambda^{-1}_{\sigma}\frac{\partial}{\partial\tilde{\phi}^{\star}}\ln Z,
\\\nonumber
\tilde{V}_{\mu}&=&\langle\bar{\psi}\gamma_{\mu}\psi\rangle
=\lambda^{-1}_{V}\frac{\partial}{\partial\tilde{V}^{\mu}}\ln Z,
\end{eqnarray}
and similar for the other bilinear $\tilde{A}_{\mu}$.
Chiral symmetry breaking by a nonzero $\tilde{\phi}$ requires that the ``field equation''
\eqref{equ::self} has a nontrivial solution. We note that $Z[\tilde{\phi},\tilde{V},\tilde{A}]$
corresponds to a one-loop expression for the fermionic fluctuations in a bosonic background.
With $\Gamma^{(\textrm{F})}_{1}=-\ln Z$ the field equation is equivalent to an extremum
of
\begin{equation}
\label{equ::extremo}
\Gamma^{(\textrm{F})}=
\int d^{4}x\left\{2\lambda_{\sigma}\tilde{\phi}^{\star}\tilde{\phi}
+\frac{1}{2}\lambda_{V}\tilde{V}_{\mu}\tilde{V}^{\mu}
+\frac{1}{2}\lambda_{A}\tilde{A}_{\mu}\tilde{A}^{\mu}\right\}
+\Gamma^{(\textrm{F})}_{1}.
\end{equation}
A discussion of spontaneous symmetry breaking in MFT amounts therefore to a calculation
of the minima of $\Gamma^{(\textrm{F})}$.

As we already noted in the introduction this calculation can be done equivalently in the Yukawa
theory \eqref{equ::baction}, \eqref{equ::bosocouplings}.
The mapping of the bosonic fields reads $\phi=\left(h_{\sigma}/\mu^{2}_{\sigma}\right)\tilde{\phi}$,
$V_{\mu}=\left(h_{V}/\mu^{2}_{V}\right)\tilde{V}_{\mu}$, \mbox{$A_{\mu}=\left(h_{A}/\mu^{2}_{A}\right)\tilde{A}_{\mu}$.}
Keeping the bosonic fields fixed and performing the remaining Gaussian fermionic functional
integral yields precisely Eq. \eqref{equ::extremo}. Mean field theory therefore corresponds
precisely to an evaluation of the effective action in the partially bosonized Yukawa model
in a limit where the bosonic fluctuations are neglected.

We want to compute here the critical couplings (more precisely, the critical line in the plane of
couplings $\bar{\lambda}_{\sigma}$, $\bar{\lambda}_{V}$) for which a nonzero expectation value
$\phi\neq 0$ indicates the onset of spontaneous symmetry breaking. For this purpose
we calculate the mass term $\sim \phi^{\star}\phi$ in $\Gamma^{(\textrm{F})}$
and look when it turns negative. This defines the critical couplings. We assume here a situation
where the expectation values of other bosonic fields like $V_{\mu}$ or $A_{\mu}$
vanish in the relevant range of couplings. It is then sufficient to evaluate
$\Gamma^{(\textrm{F})}$ for $V_{\mu}=A_{\mu}=0$.

In a diagrammatic language Gaussian integration over the fermionic variables corresponds to evaluating
the diagram of Fig. \ref{fig::mass}. We define our model with a fixed ultraviolet momentum cutoff
$q^{2}<\Lambda^{2}$, such that the MFT result becomes ($v_{4}=1/(32\pi^{2})$, $x=q^2$):
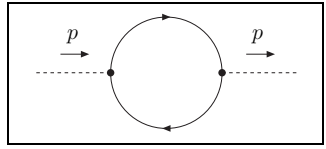
\begin{figure}[t]
\begin{center}
\scalebox{0.7}[0.7]{
\fbox{
\begin{picture}(160,70)
\SetOffset(-82,-25)
\DashLine(90,60)(130,60){2}
\Vertex(130,60){2}
\ArrowArc(160,60)(-30,0,180)
\ArrowArc(160,60)(-30,180,0)
\Vertex(190,60){2}
\DashLine(190,60)(230,60){2}
\LongArrow(103,70)(117,70)
\LongArrow(203,70)(217,70)
\Text(110,80)[]{$p$}
\Text(210,80)[]{$p$}
\end{picture}}}
\end{center}
\caption{Bosonic mass correction due to fermion fluctuations. Fermionic lines are solid with an arrow, bosonic
or ``mean field lines'' $\sim\langle\bar{\psi}\left(\frac{1-\gamma^{5}}{2}\right)\psi\rangle$ are dashed.
For use in Chap. \ref{chap::boso2} we have indicated an external momentum $p$.
A MFT calculation corresponds to an evaluation for $p=0$.}
\label{fig::mass}
\end{figure}
\begin{equation}
\label{equ::potential}
\Gamma^{(\textrm{F})}_{1}=-4v_{4}\int^{\Lambda^{2}}_{0}dx\,x\ln(x+h^{2}_{\sigma}\phi^{\star}\phi).
\end{equation}
From this one finds the mean field effective action
\begin{eqnarray}
\Gamma^{(\textrm{F})}&=&\Gamma^{(\textrm{F})}_{0}+\Gamma^{(\textrm{F})}_{1}
\\\nonumber
&=&\left(\mu^{2}_{\sigma}-4v_{4}h^{2}_{\sigma}\Lambda^{2}\right)\phi^{\star}\phi
+\textrm{const}+{\mathcal{O}}\left((\phi^{\star}\phi)^{2}\right),
\end{eqnarray}
where we have expanded in powers of $\phi$.
The mass term turns negative if
\begin{equation}
\label{equ::crit}
\frac{2\mu^{2}_{\sigma}}{h^{2}_{\sigma}\Lambda^{2}}<8v_{4},
\end{equation}
As it should be this result
only depends on the ratio $h^{2}_{\sigma}/\mu^{2}_{\sigma}=2\lambda_{\sigma}$.

We now want to determine the critical line in the plane of invariant couplings $\bar{\lambda}_{\sigma}$,
$\bar{\lambda}_{V}$ from the condition  \eqref{equ::crit}, i.e.
\begin{equation}
\lambda^{\textrm{crit}}_{\sigma}=\frac{1}{8v_{4}\Lambda^{2}}.
\end{equation}
Using the relation \eqref{equ::invariant} we infer a linear dependence
of $\bar{\lambda}^{\textrm{crit}}_{\sigma}$ on the arbitrary unphysical parameter
$\gamma$ whenever $\bar{\lambda}_{V}\neq 0$
\begin{equation}
\label{equ::MFT}
\bar{\lambda}^{\textrm{crit}}_{\sigma}=\frac{1}{8v_{4}\Lambda^{2}}
-2\gamma\bar{\lambda}_{V}.
\end{equation}
(For numerical values see Tabs. \ref{tab::crit} and \ref{tab::crit2}).
This dependence is a major shortcoming of MFT. We will refer to it as ``Fierz ambiguity''. The
Fierz ambiguity does not only affect the critical couplings but also influences the values of masses,
effective couplings etc..

The origin of the Fierz ambiguity can be traced back to the treatment of fluctuations.
A FT of the type \eqref{equ::fierz} changes the effective mean field.
In a symbolic language a FT maps
$(\bar{\psi}_{a}\psi_{a})(\bar{\psi}_{b}\psi_{b})\rightarrow(\bar{\psi}_{a}\psi_{b})(\bar{\psi}_{b}\psi_{a})$
where the brackets denote contraction over spinor indices and matrices $\sim\gamma_{\mu}$
or $\sim\gamma^{5}$ are omitted.
A mean field $\bar{\psi}_{a}\psi_{a}$, appears
after the FT as $\bar{\psi}_{a}\psi_{b}$. From the viewpoint
of the fluctuations one integrates out different fluctuating fields
before and after the FT.
It is therefore no surprise that all results depend on $\gamma$.
\begin{table}[!t]
\begin{center}
\begin{tabular}{|c|c|c|c|c|c|c|c}
  \hline
  Approximation & Chap.  & $\gamma=0$ & 0.25 & 0.5 & 0.75 & 1 \\
  \hline
  MFT &\ref{sec::mean} & 39.48 & 38.48 & 37.48 & 36.48 & 35.48 \\
  Ferm. RG & \ref{sec::fermion}  & 41.54 & 41.54 & 41.54 & 41.54 & 41.54 \\
  Bos. RG & \ref{sec::bosoflow}& 36.83 & 36.88 & 36.95 & 37.02 & 37.12 \\
  Adapted Bos. RG &\ref{sec::redef} & 41.54 & 41.54 & 41.54 & 41.54 & 41.54 \\
  \hline
  SD &\ref{sec::schwinger} &   37.48 & 37.48 & 37.48 & 37.48 & 37.48 \\
  \hline
\end{tabular}
\end{center}
\caption{Critical values $\bar{\lambda}^{\textrm{crit}}_{\sigma}$ for $\bar{\lambda}_{V}=2$ and for
various values of the unphysical
\mbox{parameter $\gamma$} (with $\Lambda=1$). In anticipation of Chaps. \ref{chap::boso1}, \ref{chap::boso2}
we give also results
for the (adapted) bosonic RG.
Progressing from MFT to the bosonic RG and adapted bosonic RG the dependence on $\gamma$ decreases as more
and more diagrams are included.
The Schwinger-Dyson result is independent of $\gamma$ but contains no vertex corrections in contrast to the
RG calculations.}
\label{tab::crit}
\end{table}
\begin{table}[!t]
\begin{center}
\begin{tabular}{|c|c|c|c|c|c|c|}
  \hline
  Approximation & Chap. &  $\gamma=0$ & 0.25 & 0.5 & 0.75 & 1 \\
  \hline
  MFT &\ref{sec::mean}& 39.48 & 29.48 & 19.48 & 9.48 & -0.52 \\
  Ferm. RG &\ref{sec::fermion} & 14.62 & 14.62 & 14.62 & 14.62 & 14.62 \\
  Bos. RG &\ref{sec::bosoflow} & 15.44 & 13.39 & 13.45 & 15.55 & 19.46 \\
  Adapted Bos. RG &\ref{sec::redef}& 14.62 & 14.62 & 14.62 & 14.62 & 14.62 \\
  \hline
  SD &\ref{sec::schwinger}&   19.48 & 19.48 & 19.48 & 19.48 & 19.48 \\
  \hline
\end{tabular}
\end{center}
\caption{The same\protect\footnotemark \hspace{0.0ex} as in Tab. \ref{tab::crit} but with $\bar{\lambda}_{V}=20$.}
\label{tab::crit2}
\end{table}
\section{Perturbation Theory} \label{sec::perturbation}
\footnotetext{The negative sign for the critical coupling at $\gamma=1$ in the MFT calculation means that
the system is in the broken phase for any positive value of $\bar{\lambda}_{\sigma}$ in this calculation.}

\begin{figure}[t]
\subfigure{\scalebox{0.65}[0.65]{\fbox{
\begin{picture}(150,120)
\SetOffset(-87,0)
\Text(90,90)[]{a}
\Text(90,30)[]{a}
\Text(160,100)[]{c}
\Text(160,20)[]{c}
\Text(230,90)[]{b}
\Text(230,30)[]{b}
\ArrowLine(130,60)(100,90)
\ArrowLine(100,30)(130,60)
\Vertex(130,60){2}
\ArrowArc(160,60)(30,180,0)
\ArrowArc(160,60)(30,360,180)
\Vertex(190,60){2}
\ArrowLine(220,90)(190,60)
\ArrowLine(190,60)(220,30)
\end{picture}}}}
\subfigure{\scalebox{0.65}[0.65]{\fbox{
\begin{picture}(150,120)
\SetOffset(-87,0)
\Text(90,90)[]{a}
\Text(90,30)[]{a}
\Text(160,100)[]{b}
\Text(160,20)[]{b}
\Text(230,90)[]{b}
\Text(230,30)[]{b}
\ArrowLine(100,90)(130,60)
\ArrowLine(130,60)(100,30)
\Vertex(130,60){2}
\ArrowArc(160,60)(30,180,0)
\ArrowArc(160,60)(30,360,180)
\Vertex(190,60){2}
\ArrowLine(220,90)(190,60)
\ArrowLine(190,60)(220,30)
\end{picture}}}}
\subfigure{\scalebox{0.65}[0.65]{\fbox{
\begin{picture}(150,120)
\SetOffset(-87,0)
\Text(90,90)[]{a}
\Text(90,30)[]{b}
\Text(160,100)[]{a}
\Text(160,20)[]{b}
\Text(230,90)[]{a}
\Text(230,30)[]{b}
\ArrowLine(130,60)(100,90)
\ArrowLine(100,30)(130,60)
\Vertex(130,60){2}
\ArrowArc(160,60)(30,180,0)
\ArrowArc(160,60)(30,360,180)
\Vertex(190,60){2}
\ArrowLine(220,90)(190,60)
\ArrowLine(190,60)(220,30)
\end{picture}}}}
\subfigure{\scalebox{0.65}[0.65]{\fbox{
\begin{picture}(150,120)
\SetOffset(-87,0)
\Text(90,90)[]{a}
\Text(90,30)[]{b}
\ArrowLine(100,90)(130,60)
\ArrowLine(100,30)(130,60)
\Vertex(130,60){2}
\ArrowArc(160,60)(30,180,0)
\Text(160,20)[]{b}
\Text(160,100)[]{a}
\ArrowArc(160,60)(-30,180,360)
\Vertex(190,60){2}
\Text(230,90)[]{a}
\Text(230,30)[]{b}
\ArrowLine(190,60)(220,90)
\ArrowLine(190,60)(220,30)
\end{picture}}}}
\caption{Perturbative correction to the four-fermion interaction.
Solid lines with an
arrow denote fermionic lines. The letters in the diagrams are given to visualize the
ways in which the fermionic operators are contracted, e.g. the first diagram in the second row results from a
term
$[(\bar{\psi}_{a}\psi_{a})
(\protect\Wwick{1}{<*{\bar{\psi}_{c}}\psi_{c})][({\bar{\psi}_{c}}>*\psi_{c}}
{1}{{\bar{\psi}_{c}}<+\psi_{c})][(>+{\bar{\psi}_{c}}\psi_{c}})(\bar{\psi}_{b}\psi_{b})]$. Evaluating the
diagrams for $k$-dependent ``full'' vertices and IR regularized propagators the above set of
diagrams gives us the flow equation of Sect. \ref{sec::fermion}.}
\label{fig::fermion1}
\end{figure}
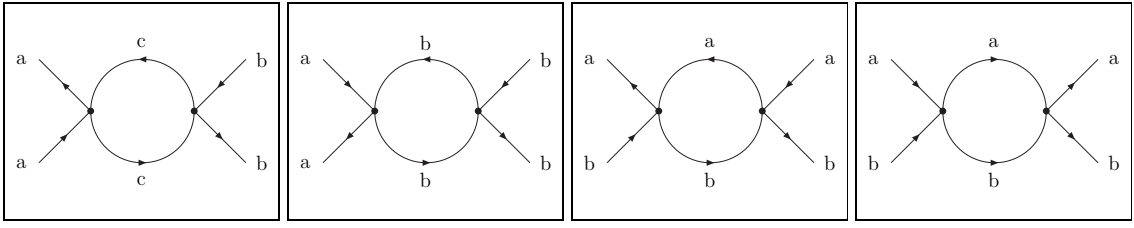

In order to cure the unpleasant dependence of the MFT result on $\gamma$ we will include
part of the bosonic fluctuations in Chaps. \ref{chap::boso1} and \ref{chap::boso2}. Some
guidance for the level of approximations needed can be gained from
perturbation theory in the fermionic language.
Since the four-fermion vertex is uniquely characterized by $\bar{\lambda}_{\sigma}$ and
$\bar{\lambda}_{V}$ the perturbative result must be independent of $\gamma$ at
any given loop order. The lowest-order
corrections to the four-fermion couplings are obtained by expanding the one-loop
expression for the effective action Eq. \eqref{equ::perti}\footnote{
We remember that in the full $\textrm{S}^{(2)}$ matrix we have a term from the $\delta^{2}/\delta\bar{\psi}\delta\psi$
derivative ($\textrm{S}^{(2)}_{\bar{F}F}$) and a term from $\delta^{2}/\delta\psi\delta\bar{\psi}$,
accounting for a factor of 2 in the language with the normal trace. Moreover, the trace includes momentum integration
and summation over internal indices.}
\begin{equation}
\label{equ::oneloop}
\Delta\Gamma^{\textrm{(1-loop)}}=\frac{1}{2}\STr\left[\ln\left(\textrm{S}^{(2)}\right)\right]
=-\Tr\left[\ln\left(\textrm{S}^{(2)}_{\bar{F}F}\right)\right]
\end{equation}
up to order $(\bar{\psi}\psi)^{2}$. For this it is useful to
decompose $S^{(2)}$ according to
\begin{equation}
S^{(2)}={\mathcal{P}}+{\mathcal{F}},
\end{equation}
into a field-independent part ${\mathcal{P}}$ (inverse popagator) and a field-dependent
part ${\mathcal{F}}$. The RHS can then be expanded as follows,
\begin{eqnarray}
\label{equ::pertexpansion}
\Delta\Gamma
=\frac{1}{2}\STr\bigg\{\left(\frac{1}{\mathcal{P}}\mathcal{F}\right)\bigg\}
-\frac{1}{4}\STr\bigg\{\left(\frac{1}{\mathcal{P}}\mathcal{F}\right)^{2}\bigg\}
&+&\frac{1}{6}\STr\bigg\{\left(\frac{1}{\mathcal{P}}\mathcal{F}\right)^{3}\bigg\}
\\\nonumber
&-&\frac{1}{8}\STr\bigg\{\left(\frac{1}{\mathcal{P}}\mathcal{F}\right)^{4}\bigg\}+\cdots.
\end{eqnarray}
This amounts to an expansion in powers of fields and we can compare the coefficients of the four-fermion
terms with the couplings specified by Eq. \eqref{equ::faction}.
Only the second term on the RHS contributes to order $(\bar{\psi}\psi)^{2}$.
The corresponding graphs with two vertices are shown in
Fig. \ref{fig::fermion1}.
From
\begin{eqnarray}
\Delta\Gamma^{\textrm{(1-loop)}}
=v_{4}\Lambda^{2}&\bigg\{&
[4\lambda^{2}_{\sigma}-4\lambda_{\sigma}(\lambda_{A}-2\lambda_{V})]
\left[\left(\bar{\psi}\psi\right)^{2}-\left(\bar{\psi}\gamma^{5}\psi\right)^{2}\right]
\\\nonumber
&+&[-2\lambda_{\sigma}\lambda_{V}+4(\lambda_{A}-\lambda_{V})\lambda_{V}]
\left[\left(\bar{\psi}\gamma^{\mu}\psi\right)^{2}\right]
\\\nonumber
&+&\left[-\lambda^{2}_{\sigma}+2\lambda_{\sigma}\lambda_{A}
+3\lambda^2_{V}-2\lambda_{A}\lambda_{V}-\lambda^{2}_{A}\right]
\left[\left(\bar{\psi}\gamma^{\mu}\gamma^{5}\psi\right)^{2}\right]\bigg\}.
\end{eqnarray}
we can read off the corrections $\Delta\lambda_{\sigma}$,
$\Delta\lambda_{V}$ and $\Delta\lambda_{A}$ to the coupling constants.
In order to establish that our result is independent of $\gamma$ we use the freedom of FT
to bring our results into a standard form, such that
$\frac{\Delta\lambda_{A}}{\Delta\lambda_{V}}=\frac{\gamma}{1-\gamma}$. Inserting next the invariant variables
\eqref{equ::invariant} leads us to:
\begin{eqnarray}
\label{equ::pert}
\Delta\bar{\lambda}_{\sigma}&=&4v_{4}\Lambda^{2}
(\bar{\lambda}^{2}_{\sigma}+4\bar{\lambda}_{\sigma}\bar{\lambda}_{V}
+3\bar{\lambda}^{2}_{V}),
\\\nonumber
\Delta\bar{\lambda}_{V}&=&2v_{4}\Lambda^{2}(\bar{\lambda}_{\sigma}+\bar{\lambda}_{V})^{2}.
\end{eqnarray}
In contrast to MFT the result does not depend on $\gamma$.

The perturbative result, Eq. \eqref{equ::pert}, always leads to finite corrections to the coupling constants. Remembering
that in the fermionic language the onset of SSB is marked by a divergence of the coupling constants,
it becomes  clear that we will never get SSB in perturbation theory. No critical couplings can be calculated.
This is a severe shortcoming of perturbation theory which cannot be overcome by calculating higher
loop orders. Only an infinite number of loops can give SSB.
In the next section we establish how a renormalization group treatment can overcome this difficulty
without encountering the Fierz ambiguity of MFT. A calculation of the critical coupling becomes
feasible. Nevertheless, even this RG treatment has its limitations once the couplings diverge.
In particular, it does
not allow us to penetrate the phase with SSB. In Chaps. \ref{chap::boso1} and \ref{chap::boso2} this
shortcoming will be cured by a RG treatment in the partially bosonized language. In particular,
we will see in Chap. \ref{chap::boso2} which diagrams are needed in order to maintain the independence
of results on $\gamma$ in analogy to perturbation theory.
\section{Renormalization Group for Fermionic Interactions} \label{sec::fermion}
Let us now turn to an RG equation. More explicitly the ERGE for the effective average
action discussed in Chap. \ref{chap::effectiveaction}. Neglecting all change on the RHS
leads to the perturbative result of the previous section. Consequently,
in this approximation we cannot observe SSB.
For a better approximation
we restrict $\Gamma_{k}$ to the terms specified in Eq. \eqref{equ::faction} but take
all couplings to be explicitly $k$-dependent. In the action \eqref{equ::faction} we have only local
interactions. Expressed in momentum space the four-fermion interactions have no
momentum dependence. This is often referred to as the local potential approximation (LPA)
\cite{Tetradis:1993ts,Hasenfratz:1986dm, Morris:1994ki}.

Performing the decomposition into a field-dependent part and a field-independent part as in the previous
section,
\begin{equation}
\Gamma^{(2)}_{k}+R_{k}={\mathcal{P}_{k}}+{\mathcal{F}_{k}}
\end{equation}
we obtain an expansion of the flow equation \eqref{equ::flow2}
\begin{eqnarray}
\label{equ::expansion}
\partial_{t}\Gamma_{k}
\!=\!\frac{1}{2}\STr\bigg\{\tilde{\partial}_{t}\left(\frac{1}{\mathcal{P}}_{k}\mathcal{F}_{k}\right)\bigg\}
\!-\!\frac{1}{4}\STr\bigg\{\tilde{\partial}_{t}\left(\frac{1}{\mathcal{P}}_{k}\mathcal{F}_{k}\right)^{2}\bigg\}
\!\!\!\!&+&\!\!\!\!\frac{1}{6}\STr\bigg\{\tilde{\partial}_{t}\left(\frac{1}{\mathcal{P}}_{k}\mathcal{F}_{k}\right)^{3}\bigg\}
\\\nonumber
&-&\!\!\!\!\frac{1}{8}\STr\bigg\{\tilde{\partial}_{t}\left(\frac{1}{\mathcal{P}}_{k}\mathcal{F}_{k}\right)^{4}\bigg\}+\cdots.
\end{eqnarray}
This is in complete analogy to the previous section, only written in a differential form and with
$k$-dependent vertices.
We obtain
a set of ordinary differential equations for the couplings:
\begin{eqnarray}
\label{equ::fermionflow}
\nonumber
\partial_{t}\bar{\lambda}_{\sigma,k}&=&-8v_{4}l^{(F),4}_{1}(s)k^{2}
(\bar{\lambda}^{2}_{\sigma,k}+4\bar{\lambda}_{\sigma,k}\bar{\lambda}_{V,k}
+3\bar{\lambda}^{2}_{V,k}),
\\
\partial_{t}\bar{\lambda}_{V,k}
&=&-4v_{4}l^{(F),4}_{1}(s)k^{2}(\bar{\lambda}_{\sigma,k}+\bar{\lambda}_{V,k})^{2},
\end{eqnarray}
in agreement with \cite{Aoki:1997fh} where the same model has been
studied. The threshold functions $l^{(F),4}_{1}$
(for our conventions cf. App. \ref{app::cutoff} or \cite{Berges:2000ew})
originate from the momentum space integration over
the IR regulated propagators
and replace the factor $\frac{\Lambda^{2}}{2}$ in Eq.
\eqref{equ::pert}.
For our actual calculation we use a linear
cutoff\footnote{The threshold functions depend on the precise
choice of the cutoff. For the very simple truncation used in this
section this dependence can actually be absorbed by a suitable
rescaling of $k$, cf. App. \ref{app::cutoff}.} \cite{Litim:2000ci}
and adapt the threshold functions to our setting with fixed
momentum cutoff $q^{2}<\Lambda^{2}$ in App. \ref{app::cutoff}. The
dependence on $s=k^{2}/\Lambda^{2}$ becomes relevant only for
$k>\Lambda$ whereas for $k<\Lambda$ one has constants
$l^{(F),4}_{1}=1/2$. Although useful for the comparison
of the RG with MFT or perturbation theory, the explicit $k$-dependence
of the threshold functions makes the fixed momentum cutoff somewhat cumbersome.
An alternative is to use the so called ERGE scheme for
the UV regularization. The basic idea (for details s. App. \ref{app::regularization})
is to use standard threshold functions without
a UV cutoff in the momentum integral and implement the UV regularization
by specifying the initial conditions for $\Gamma_{k}$ at some \emph{finite} $k=\Lambda$.
This has the advantage that threshold functions without a finite UV cutoff are not
explicitly $k$-dependent, greatly simplifying numerical calculations. The price
to pay is that it is in general impossible to compare non-universal quantities like
critical couplings for different choices of the IR cutoff function $R_{k}(p)$.

The fermionic flow equations\footnote{As discussed above, the perturbative result Eq. \eqref{equ::pert} can be recovered
from Eq. \eqref{equ::fermionflow} if we neglect
the $k$-dependence of the couplings on the RHS and perform the
$t$-integration.} \eqref{equ::fermionflow} do not depend on $\gamma$.
In a diagrammatic language we again have evaluated the diagrams of Fig. \ref{fig::summ}
but now with \mbox{$k$-dependent} vertices. In the RG formulation we only go a tiny step $\Delta k$, and reinsert
the resulting couplings (one-loop diagrams) before we do the next step.
This leads to a resummation of loops.
Since Eq. \eqref{equ::fermionflow} is now nonlinear (quadratic terms on the RHS) the couplings
can and do diverge for a finite $k$ if the initial couplings are large enough. Therefore
we observe the onset of SSB and find a critical coupling.
Since Eq. \eqref{equ::fermionflow} is invariant this critical coupling does not
depend on $\gamma$! Values for the critical coupling obtained by numerically solving
\mbox{Eq. \eqref{equ::fermionflow}} can be found in Tabs. \ref{tab::crit} and \ref{tab::crit2}.

The next step in improving this calculation in the fermionic language would
be to take the momentum dependence of the couplings
into account (e.g. \cite{Meggiolaro:2001kp}) or to include higher orders of the fermionic
fields into the truncation. This seems quite complicated and at first sight we have no physical
guess of what is relevant. The renormalization group treatment of the bosonic formulation in
Chap. \ref{chap::boso1} seems much more promising in this respect.

\section{Gap Equation}\label{sec::schwinger}
Let us finally turn to the SDE as the last method discussed in Chap. \ref{chap::effectiveaction}.
For the model Eq. \eqref{equ::faction} the SDE, approximated to lowest
order, is depicted in Fig. \ref{fig::schwingerferm}. It is a
closed equation since only the bare four-fermion vertex appears.
(Only higher order terms involve the full four-fermion vertex.)
We write the full fermionic propagator $G_{\textrm{F}}$ as
\begin{equation}
G^{-1}_{\textrm{F}}(p)=G^{-1}_{\textrm{F}0}(p)+\Sigma_{\textrm{F}}(p)
\end{equation}
with the free propagator $G_{\textrm{F}0}$ and self energy
$\Sigma_{\textrm{F}}$. Using this one obtains a gap equation for
the self energy which can be solved self consistently. To simplify
the discussion we make an ansatz for the self energy:
\begin{equation}
\label{equ::ansatzsde} \Sigma_{\textrm{F}}=M_{\textrm{F}}\gamma^{5},
\end{equation}
where the effective fermion mass $M_{\textrm{F}}$ obeys the gap
equation
\begin{equation}
\label{equ::complete}
M_{\textrm{F}}=8v_{4}\left[\bar{\lambda}_{\sigma}+\bar{\lambda}_{V}\right]
\int^{\Lambda^2}_{0}
dx\,x\frac{M_{\textrm{F}}}{x+M^{2}_{\textrm{F}}}.
\end{equation}
The onset for nontrivial solutions determines the critical
couplings:
\begin{equation}
\left[\bar{\lambda}_{\sigma}+\bar{\lambda}_{V}\right]_{\textrm{crit}}=\frac{1}{8v_{4}\Lambda^{2}}.
\end{equation}
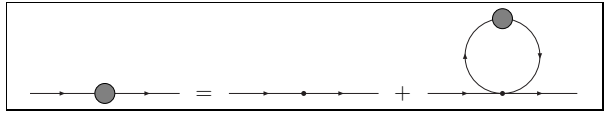
\begin{figure}[!t]
\begin{center}
\scalebox{0.47}[0.47]{\fbox{
\begin{picture}(470,80)
\SetOffset(12,10) \ArrowLine(0,0)(52,0) \ArrowLine(68,0)(120,0)
\GOval(60,0)(8,8)(0){0.5}
\Text(140,-1)[]{\scalebox{1.5}[1.5]{$=$}} \SetOffset(172,10)
\ArrowLine(0,0)(60,0) \ArrowLine(60,0)(120,0) \Vertex(60,0){2}
\Text(140,0)[]{\scalebox{1.5}[1.5]{$+$}} \SetOffset(332,10)
\ArrowLine(0,0)(60,0) \Vertex(60,0){2} \ArrowLine(60,0)(120,0)
\ArrowArc(60,30)(-30,-90,90) \ArrowArc(60,30)(-30,90,270)
\GOval(60,60)(8,8)(0){0.5}
\end{picture}}}
\end{center}
\caption{Diagrammatic representation of the lowest order
Schwinger-Dyson equations for the fermionic model Eq.
\eqref{equ::faction}. The shaded circles depict the full
propagator.}
\label{fig::schwingerferm}
\end{figure}
This result is shown in Tabs. \ref{tab::crit}, \ref{tab::crit2}
and does not depend on $\gamma$, as expected for a  fermionic
calculation. We observe that the MFT result  for the
$\bar{\lambda}^{\textrm{crit}}_{\sigma}$ coincides with the
SD approach for a particular choice $\gamma=1/2$. However, in
general MFT is not equivalent to the lowest-order SDE.
This can be seen by computing also the critical coupling for the
onset of SSB in the vector channel. The MFT and SD results do not
coincide for the choice $\gamma=1/2$.
This becomes evident if we use a vectorlike ansatz instead of Eq. \eqref{equ::ansatzsde}
for the self-energy:
\begin{equation}
\label{equ::vansatz}
\Sigma_{\textrm{F}}=V\!\!\!\!/.
\end{equation}
Using this ansatz for the self energy we find for the onset of non trivial solutions
\begin{equation}
[\bar{\lambda}_{\sigma}+3\bar{\lambda}_{V}]^{\textrm{crit}}_{V}=\frac{1}{2v_{4}\Lambda^{2}}.
\end{equation}
This is different from the MFT result
\begin{equation}
[\bar{\lambda}_{V}]^{\textrm{crit,MFT}}_{V}=\frac{1}{4v_{4}\Lambda^{2}(1-\gamma)}
\end{equation}
obtained from Eq. \eqref{equ::self} by setting $\tilde{\phi}=\tilde{A}^{\mu}=0$.
Again it would be possible to find a choice for $\gamma$ which
makes both results equal. But, in general, this will not be $\gamma=1/2$.
We note that in this channel the dependence of the MFT result is even worse than in the
scalar channel.

At last, let us note that MFT and the SDE gap equation share a nice feature, both allow us
to proceed into the region of broken symmetry. The self consistency conditions
\eqref{equ::self} and \eqref{equ::complete} are valid for non-trivial solutions and
hence finite values of the bosonic condensate,
whereas the simple flow equation \eqref{equ::fermionflow} breaks
down (the couplings become infinite) when we approach the phase with broken symmetry.

\chapter{Partial Bosonization I: Basic Idea}\label{chap::boso1}
The MFT calculation introduces ''mean fields'' composed of fermion
- antifermion (or fermion - fermion) bilinears.
This is motivated by the fact that in many physical systems the
fermions are not the only relevant degrees of freedom at low
energies. Bosonic bound states become important and may condense.
Examples are Cooper pairs in superconductivity or mesons in QCD.
For a detailed description of the interplay between fermionic and
composite bosonic fluctuations it seems appropriate to treat both
on equal footing by introducing explicit fields for the relevant
composite bosons. This will also shed more light on the status of
MFT.

\section{Calculating the Bosonized Action}
A method for introducing the desired composite fields is partial bosonization
\cite{stratonovich,Hubbard:1959ub,Klevansky:1992qe,Alkofer:1996ph,Berges:1999eu}, sometimes also referred
to as a Hubbard Stratonovich transformation. Regardless of the name, in principle
it is nothing else but the insertion of a nicely written factor of unity in the functional integral
for the partition function.

\subsection{A Toy Model Bosonization $(d=0)$}\label{sec::toy}
Before going into the details, let us
demonstrate the idea on a $0$-dimensional toy model. In $0$ dimensions
the field variable $\phi(p)$ is replaced by a simple $c$- or Grassmann number $x$ depending
on whether we deal with fermions or bosons\footnote{Of course, if $x$ is a Grassmann number
it is necessary to have several different components $x_{i}$ in order to have non-trivial
interactions.}. To be explicit let us consider an ``action''
\begin{equation}
\label{equ::oneaction}
S(x)=\frac{1}{2}x^{\textrm{T}}Mx+ \frac{\lambda}{2} (x^{\textrm{T}}Sx)^2
\end{equation}
where $M$ and $S$ are matrices simulating the mass and kinetic terms ($M$), and
the spin structure of the interactions ($S$).
Since we live in $0$ dimensions the momentum or position space integral is trivial, i.e. absent.
This action models a massive field with an inverse propagator $M$ and a quartic interaction with
coupling strength $\lambda$. In the following we want to introduce an auxiliary ``composite'' field
$y$ for the operator,
\begin{equation}
\label{equ::op}
O_{S}(x)=-\frac{h}{m^2}x^{\textrm{T}}Sx,
\end{equation}
where we have introduced the for the moment arbitrary constants $h$, $m$ for later convenience.
To study a composite operator $O_{S}(x)$ it is useful to introduce a
source term $kO_{S}(x)$
for this composite operator in addition to the ordinary source term $jx$.
The functional integral over the field variables becomes an ordinary integral over $x$,
and the partition function reads,
\begin{equation}
\label{equ::onedimensional}
Z(j,k)=\int dx \exp(-S(x)+jx+kO_{S}(x)+\frac{a}{2}k^{2}),
\end{equation}
where we have used the freedom to add a field independent term quadratic in $k$ to the action.

Using the translation invariance of the integral we can obtain the following rather trivial identity,
\begin{equation}
\label{equ::oneidentity}
1=N\int dy \exp(-\frac{m^2}{2}y^2)=N\int dy\exp(-\frac{m^2}{2}(y-O_{S}(x)+d)^2),
\end{equation}
where $N$ is nothing but a normalization constant $N=\left(\int dy \exp(-\frac{m^2}{2} y^2)\right)^{-1}$
and $d$ is for the moment arbitrary, but will be determined later.
Inserting this under the integral in \eqref{equ::onedimensional} yields,
\begin{eqnarray}
Z(j,\hat{k})&=&N\int dx dy\exp(-\hat{S}(x,y)+jx+\hat{k}y),
\\\nonumber
\hat{S}(x,y)&=&\frac{1}{2}x^{\textrm{T}}Mx+\frac{\lambda}{2}(x^{\textrm{T}}Sx)^{2}+k\frac{h}{m^2}x^{\textrm{T}}Sx+\frac{a}{2}k^{2}
+\frac{m^2}{2}y^2
\\\nonumber
&&+hyx^{\textrm{T}}Sx+\frac{1}{2}\frac{h^2}{m^2}(x^{\textrm{T}}Sx)^{2}
+m^{2}dy+hdx^{\textrm{T}}Sx+\frac{m^2d^2}{2}+\hat{k}y.
\end{eqnarray}
The first line looks promising, as it is the partition function for a theory with two ``fields'' $x$, $y$ and
an action $\hat{S}(x,y)$. The second line is still a mess which over and above depends explicitly on the sources
$k$ and $\hat{k}$. However, remembering that we have introduced several arbitrary parameters we can choose
those to our convenience,
\begin{equation}
a=-\frac{1}{m^2},\quad d=-\frac{k}{m^2}, \quad \hat{k}=k,
\end{equation}
simplifying
\begin{equation}
\hat{S}(x,y)=\frac{1}{2}x^{\textrm{T}}Mx+\frac{1}{2}\left(\lambda+\frac{h^2}{m^2}\right)(x^{\textrm{T}}Sx)^{2}
+\frac{m^2}{2}y^2+hyx^{\textrm{T}}Sx.
\end{equation}
Finally, employing the choice
\begin{equation}
\label{equ::cancel}
\frac{h^2}{m^2}=-\lambda,
\end{equation}
cancels all quartic interactions of $x$, leaving us with a mass term for the ``composite'' field $y$,
a Yukawa-type interaction between $y$ and $x$ in addition to the propagator term $x^{\textrm{T}}Mx$ for
the ``elementary'' field $x$,
\begin{equation}
\label{equ::bosonized}
\hat{S}(x,y)=\frac{1}{2}x^{\textrm{T}}Mx
+\frac{m^2}{2}y^2+hyx^{\textrm{T}}Sx.
\end{equation}
This also explains why we have introduced the constants $h$ and $m$ in the normalization of
$O_{S}[x]$, Eq. \eqref{equ::op}.

Having accomplished the ``partial bosonization'' of our $0$-dimensional model we would like to
comment on some rather technical points:
\begin{enumerate}
\item{Physically it is clear that with respect to the symmetries $y$ should
have the same transformation properties as the
composite operator $O_{S}(x)$. From a more technical point of view this is necessary as
we would otherwise be unable to perform the shift in the integration variable in Eq. \eqref{equ::oneidentity}.}
\item{In the derivation given above we did not specify if $x$ is bosonic or fermionic.
We can use the same procedure to introduce composites made up of fermions or bosons. However,
we should be careful. If the integral in \eqref{equ::onedimensional} is fermionic,
it is convergent for all possible choices of $\lambda$, because of the rules
of Grassmannian integration. The integral over the
auxiliary field $y$, Eq. \eqref{equ::oneidentity}, is only convergent for $m^2>0$. This gives us the
condition that $\lambda<0$, in order to render everything finite.
For bosons, however, a $\lambda<0$ leads to a divergence
in Eq. \eqref{equ::onedimensional}. So, naively our bosonization procedure works
only for fermions and a certain region of the coupling constant. Although it is possible to circumvent
these naive arguments by an integration along the complex axis, any interpretation of $y$ as a bound
state is still awkward. Therefore, we will restrict ourselves to stable potentials of the
composite field (s. below), i.e. the integration over the composite is convergent.}
\item{It is not necessary that the integration over the auxiliary field is Gaussian as in Eq. \eqref{equ::oneidentity}.
Indeed, we can replace the term $-\frac{m^2}{2}y^2$ by any potential $-V(y)$ as long
as,
\begin{equation}
\label{equ::bounded}
V(y)>c,\quad \lim_{|y|\to\infty}\frac{V(y)}{(\ln(y))^{2}}\rightarrow\infty,
\end{equation}
i.e. $V(y)$ is bounded from below and grows sufficiently fast for $y\rightarrow\infty$. This allows us to
absorb also higher order interaction as e.g. a term $-\kappa(x^{\textrm{T}}Sx)^{4}$
by a term $\kappa\frac{m^8}{h^4}y^{4}$ in $V(y)$. In general,
\begin{equation}
\label{equ::higherorder}
-F(O_{S}(x))\rightarrow -F(O_{S}(x))+F(O_{S}(x)-y),
\end{equation}
such that the purely fermionic terms are cancelled. Of course, this leaves us with non-linear
Yukawa couplings, as e.g. $\sim y^3 x^2$ and other complicated interactions $\sim y^2x^4$ or $\sim yx^6$.}
\item{As can be seen from \eqref{equ::higherorder} we can also treat terms linear in the composite
operator $O_{S}(x)$, removing e.g. all parts $\sim x^{\textrm{T}}Sx$ from $\frac{1}{2}x^{\textrm{T}}Mx$.
A typical example for this would be the translation of a fermionic mass term $m\bar{\psi}\psi$ into
a source term $j\phi$ for a boson corresponding to $\bar{\psi}\psi$.}
\item{Although it is the most common case, it is not necessary that the composite
operators that we want to bosonize are made up of exactly two field operators. In principle, they
can contain an arbitrary number of fields. The composites can even be fermionic.}
\item{Using the translation invariance of the integral as in Eq. \eqref{equ::oneidentity} is the
simplest but not the only possible way to obtain an identity useful for the introduction
of composite fields. In general, any identity
\begin{equation}
\label{equ::moregeneral}
1=N\exp(F(O_{S}(x)))\int dy \exp(-V(x,y)),
\end{equation}
can be used to cancel a part $F(O_{S}(x))$ in the initial action.
However, the direct interpretation $y\sim O_{S}(x)$ will usually be lost. Of course the $V(x,y)$
in Eq. \eqref{equ::moregeneral} is far from unique. One possibility is always
$V(x,y)=F(O_{S}(x)-y)$ as obtained in \eqref{equ::higherorder}.
In practice it quite difficult to find a $V(x,y)$ with a suitably
simple form like $V(x,y)\sim xy+V^{\prime}(y)$.}\label{item::general}
\item{Sometimes, it might seem useful to add some form of interaction, e.g. $\sim y^{4}$, between the
composite fields to the bosonized action \eqref{equ::bosonized}. We can then use the argument
of \ref{item::general} in a backward way to determine what kind of (higher order) interactions
this would introduce into the initial unbosonized action.}
\item{We can recover the initial action by performing the integration over the auxiliary field $y$,
\begin{equation}
\label{equ::recover}
\exp(-S(x))=\int dy \exp(-\hat{S}(x,y)).
\end{equation}}
\end{enumerate}

\subsection{The Fierz Ambiguity}\label{subsec::fierz}
Already in this simple model we can get a grasp how the Fierz ambiguity arises. Let us take a look
at the ``four-fermion interaction'' in Eq. \eqref{equ::oneaction},
\begin{eqnarray}
\label{equ::introfierz}
\frac{\lambda}{2}(x^{\textrm{T}}Sx)^{2}
=\frac{\lambda}{2}S_{i_{1}i_{2}}S_{i_{3}i_{4}}x_{i_{1}}x_{i_{2}}x_{i_{3}}x_{i_{4}}
&=&\frac{\lambda}{2}(S\otimes S)_{i_{1}i_{2}i_{3}i_{4}}x_{i_{1}}x_{i_{2}}x_{i_{3}}x_{i_{4}}
\\\nonumber
&=&\frac{1}{2}\Lambda_{i_{1}i_{2}i_{3}i_{4}}x_{i_{1}}x_{i_{2}}x_{i_{3}}x_{i_{4}},
\end{eqnarray}
where the LHS defines $\Lambda$. We can now permute the $x_{i}$, e.g. let us exchange $x_{i_2{}}$ and $x_{i_{4}}$,
\begin{equation}
\label{equ::permuted}
\Lambda_{i_{1}i_{2}i_{3}i_{4}}x_{i_{1}}x_{i_{2}}x_{i_{3}}x_{i_{4}}
=\pm\Lambda_{i_{1}i_{2}i_{3}i_{4}}x_{i_{1}}x_{i_{4}}x_{i_{3}}x_{i_{2}}
=\Lambda^{\prime}_{i_{1}i_{2}i_{3}i_{4}}x_{i_{1}}x_{i_{2}}x_{i_{3}}x_{i_{4}},
\end{equation}
where
\begin{equation}
\Lambda^{\prime}_{i_{1}i_{2}i_{3}i_{4}}=\pm\Lambda_{i_{1}i_{4}i_{3}i_{2}},
\end{equation}
and the sign is $+$ for bosons and $-$ for fermions, respectively.
Now, let us assume that $\Lambda^{\prime}$ has a decomposition
(this assumption is nearly always fulfilled),
\begin{equation}
\label{equ::simplefierz}
\Lambda^{\prime}=\lambda_{TT^{\prime}}(T\otimes T^{\prime})\neq \lambda (S\otimes S).
\end{equation}
Accordingly, we would bosonize the RHS of Eq. \eqref{equ::permuted}
with fields corresponding to the operators $O_{T}(x)$ and
the coupling strengths $\lambda_{TT^{\prime}}$. Hence, we obtain a different set of composite fields and
coupling strengths for the \emph{identical} action. In general, we can perform an arbitrary
permutation of the $x_{i}$, and we can obtain not only two but several different bosonized
actions. This is even worse for higher order (e.g. $x^{6}$) interactions.

Comparing Eq. \eqref{equ::simplefierz} with the Fierz identity Eq. \eqref{equ::fierz1} it becomes clear that
an exchange of fields like in Eq. \eqref{equ::permuted} is a Fierz transformation.
Since the bosonized action may look quite different, it is no big surprise that
simple approximations might yield different results. This is what we call ``Fierz Ambiguity''.

One might wonder about the fact that different $\Lambda$, $\Lambda^{\prime}$ describe
the same (unbosonized) action. For fermions this is quite easy to understand.
Due to the Grassmann identity $x_{i}x_{j}=-x_{j}x_{i}$ only the completely
antisymmetric parts of $\Lambda$ give non-vanishing contributions to the action.
Hence, all
\begin{equation}
\Lambda=\hat{\Lambda}+\Sigma
\end{equation}
yield identical action, as long as $\Sigma$ is a sum of terms which are
symmetric in at least two indices. If we want, we can choose $\hat{\Lambda}$ to
be completely antisymmetric. Any Fierz transformation described above can
be obtained by adding a suitable chosen $\Sigma$. The problem is that a non-vanishing $\Sigma$
usually does \emph{not} give a vanishing contribution in the partially bosonized
action. As there is great freedom in choosing $\Sigma$ we can get a nearly arbitrary bosonized action.
For bosons the story is essentially the same, only
one has to replace symmetric by antisymmetric and vice versa.
\subsection{The Case $d>0$, MFT Revisited} \label{sec::largerzero}
It is straightforward to generalize the procedure described in the previous section to the case of $d>0$.
Indeed, the change is more or less only a matter of semantics, as we replace functions by functionals
and integration by functional integration,
\begin{equation}
F(\,\,\,\,)\rightarrow F[\,\,\,\,], \quad d\rightarrow {\mathcal{D}}.
\end{equation}
To demonstrate this let us repeat the procedure for the action
\eqref{equ::faction}\footnote{For simplicity we skip
the introduction of bosonic sources.}.
Introducing bosonic fields for the composite operators
corresponding to scalar, vector and axial vector bosons,
\begin{equation}
O_{\phi}[\psi]=\frac{h_{\sigma}}{2\mu^{2}_{\sigma}}\bar{\psi}(1-\gamma^{5})\psi
=O^{\dagger}_{\phi^{\star}}[\psi],\quad
O_{V}[\psi]=\frac{h_{V}}{\mu^{2}_{V}}\bar{\psi}\gamma^{\mu}\psi,\quad
O_{A}[\psi]=\frac{h_{A}}{\mu^{2}_{A}}\bar{\psi}\gamma^{\mu}\gamma^{5}\psi
\end{equation}
we obtain,
\begin{eqnarray}
\label{equ::partition}
\!\!\!\!Z\!\!\!&=&\!\!\!\!\!\int{\mathcal{D}}\bar{\psi}{\mathcal{D}}\psi\exp\left(-\textrm{S}[\psi]\right)
=\!\int{\mathcal{D}}\bar{\psi}{\mathcal{D}}\psi{\mathcal{D}}\phi{\mathcal{D}}V^{\mu}{\mathcal{D}}A^{\mu}
{\mathcal{N}}_{\phi}{\mathcal{N}}_{V}{\mathcal{N}}_{A}
\exp\left(-\textrm{S}[\psi]\right)
\end{eqnarray}
with
\begin{eqnarray}
\label{equ::factors}
\nonumber
{\mathcal{N}}_{\phi}&=&\exp\bigg[-\int_{x}\mu^{2}_{\sigma}
\left(\phi^{\star}
+\frac{h_{\sigma}}{2\mu^{2}_{\sigma}}\bar{\psi}(1+\gamma^{5})\psi\right)
\left(\phi-\frac{h_{\sigma}}{2\mu^{2}_{\sigma}}\bar{\psi}(1-\gamma^{5})\psi\right)\bigg],
\\\nonumber
{\mathcal{N}}_{V}&=&\,\exp\bigg[-\int_{x}\frac{\mu^{2}_{V}}{2}
\left(V^{\mu}-\frac{h_{V}}{\mu^{2}_{V}}\bar{\psi}\gamma^{\mu}\psi\right)
\left(V_{\mu}-\frac{h_{V}}{\mu^{2}_{V}}\bar{\psi}\gamma_{\mu}\psi\right)\bigg],
\\
{\mathcal{N}}_{A}&=&\,\exp\bigg[-\int_{x}\frac{\mu^{2}_{A}}{2}
\left(A^{\mu}-\frac{h_{A}}{\mu^{2}_{A}}\bar{\psi}\gamma^{\mu}\gamma^{5}\psi\right)
\left(A_{\mu}-\frac{h_{A}}{\mu^{2}_{A}}\bar{\psi}\gamma_{\mu}\gamma^{5}\psi\right)\bigg].
\end{eqnarray}
Collecting the terms in the exponentials and using Eq. \eqref{equ::bosocouplings}
as the equivalent to Eq. \eqref{equ::cancel}, the four-fermion interaction is cancelled.
As expected, it is now replaced by mass terms for the bosons and Yukawa couplings
between bosons and fermions as given by the expression
\eqref{equ::baction}. We note, that the bosons do not yet have a non-trivial kinetic term
and the propagator is simply $\frac{1}{\mu^2}$.

Having arrived at the partially bosonized action \eqref{equ::baction} for our model
\eqref{equ::faction} we can use it to gain new insight into MFT.
The action Eq. \eqref{equ::baction} is quadratic in the
fermionic fields, hence the functional integral over the fermionic
degrees of freedom is Gaussian and can be done in one step. As we
have seen in the previous chapter this leads exactly to the MFT
results. More precisely, we understand now that for different
choices of $\gamma$ the MFT treatment leaves out different bosonic
fluctuations.

In this context we note that the condition \eqref{equ::bounded}
restricts the possible couplings to $\lambda_{\sigma},\lambda_{V},
\lambda_{A}>0$. In the invariant variables this restriction
translates to $\bar{\lambda}_{\sigma}, \bar{\lambda}_{V}>0$ and
for $\gamma$ it implies $0<\gamma<1$.

Incidentally, we note that in the case of a four-fermion interaction we have
an alternative to Eq. \eqref{equ::recover} to recover the
initial action from the partially bosonized one. Instead of integrating
over the bosonic auxiliary fields we can simply solve the classical
field equations for the bosonic fields in terms of the fermionic fields
and reinsert them in the partially bosonized action. Starting from Eq. \eqref{equ::baction}
this returns us to Eq. \eqref{equ::faction}.

The crucial advantage of the bosonic formulation is that it can easily be generalized. For
example, the bosonic bound states become dynamical fields if we allow for appropriate
kinetic terms in the truncation, i.e.
\begin{equation}
\label{equ::kinetic}
\Delta\Gamma_{\textrm{kin}}=\int d^{4}x\{Z_{\phi}\partial_{\mu}\phi^{\star}\partial^{\mu}\phi
+\frac{Z_{V}}{4}V_{\mu\nu}V^{\mu\nu}
+\frac{Z_{A}}{4}A_{\mu\nu}A^{\mu\nu}
+\frac{Z_{V}}{2\alpha_{V}}(\partial_{\mu}V^{\mu})^{2}
+\frac{Z_{A}}{2\alpha_{A}}(\partial_{\mu}A^{\mu})^{2}\bigg\}
\end{equation}
with
\begin{equation}
V_{\mu\nu}=\partial_{\mu}V_{\nu}-\partial_{\nu}V_{\mu},\quad
A_{\mu\nu}=\partial_{\mu}A_{\nu}-\partial_{\nu}A_{\mu}.
\end{equation}
Also spontaneous symmetry breaking can be explicitly studied if we
replace $\mu^{2}_{\sigma}\phi^{\star}\phi$ by an effective potential $U(\phi^{\star}\phi)$
which may have a minimum for $\phi\neq 0$. This approach has been followed in
previous studies \cite{Schaefer:em,Kodama:1999if,Jungnickel:1995fp,Bergerhoff:1999hr}.
We remark that for those terms to be present in the effective action it is not necessary for
them to be present in the initial (bosonized) action. They naturally receive non-vanishing corrections
by loop diagrams. E.g. the kinetic terms \eqref{equ::kinetic} get a correction from the diagram
depicted in Fig. \ref{fig::mass} with non-zero external momentum. Nevertheless, it is instructive
to investigate
how such terms would look like in the unbosonized language. For the potential terms this has already
been discussed in Sect. \ref{sec::toy}, i.e. Eqs. \eqref{equ::higherorder}, \eqref{equ::moregeneral},
for the kinetic (derivative) terms we will do this in the next section.

\subsection{Beyond Pointlike Interactions\protect\footnotemark[0]}\label{sec::momentumdep}
\footnotetext[0]{This section discusses
some details needed in Sect. \ref{sec::beyondlpa}, and can also be read then.}
So far our bosonization procedure seems relatively simple, and it is. However,
we should mention that above we have bosonized only the very special case of a
\emph{pointlike}, i.e. \emph{local} four-fermion
interaction\footnote{In this section we suppress all internal indices. In particular indices
distinguishing between $\bar{\psi}$ and $\psi$. $\lambda$ is a matrix with four such indices. All
problems connected with the internal indices are completely analogous to the previous sections.
Therefore, we allow ourselves to be somewhat sloppy concerning the internal indices, simplifying
the notation.},
\begin{equation}
\label{equ::fourfermion}
\int_{x}\lambda\,\Psi(x)\Psi(x)\Psi(x)\Psi(x)
=\int_{p_{1},p_{2},p_{3},p_{4}}\!\!\!\!\!\!\!\!\!\!\!\!\!\!\!\!\lambda
\,\Psi(p_{1})\Psi(p_{2})\Psi(p_{4})\Psi(p_{3})
\delta(p_{1}+p_{2}+p_{3}+p_{4}).
\end{equation}
This is by no means the most
general form of a four-fermion interaction.
Giving up locality, $\lambda$ can become an arbitrary\footnote{Of course, it must have the
right transformation properties under Lorentz transformations.} function of the four
momentum variables,
\begin{equation}
\label{equ::generalization}
\lambda\rightarrow\lambda(p_{1},p_{2},p_{3},p_{4}).
\end{equation}

\begin{figure}[t]
\begin{center}
\scalebox{0.7}[0.7]{\fbox{
\begin{picture}(190,140)(-10,0)
\SetOffset(3,10)
\LongArrow(13,10)(27,10)
\Text(20,2)[c]{$p_{2}$}
\LongArrow(147,10)(133,10)
\Text(140,2)[c]{$p_{1}$}
\LongArrow(133,110)(147,110)
\Text(140,118)[c]{$p_{3}$}
\LongArrow(27,110)(13,110)
\Text(20,118)[c]{$p_{4}$}
\ArrowLine(40,100)(0,100)
\ArrowLine(120,100)(40,100)
\ArrowLine(160,100)(120,100)
\ArrowLine(0,20)(40,20)
\ArrowLine(40,20)(120,20)
\ArrowLine(120,20)(160,20)
\Vertex(40,100){2}
\Vertex(120,100){2}
\Vertex(40,20){2}
\Vertex(120,20){2}
\Photon(40,100)(40,20){-5}{7.5}
\Photon(120,100)(120,20){5}{7.5}
\LongArrow(70,10)(90,10)
\Text(80,0)[c]{$q$}
\LongArrow(130,50)(130,70)
\Text(138,60)[l]{$q+p_{1}$}
\LongArrow(90,110)(70,110)
\Text(80,120)[c]{$q+p_{1}-p_{3}$}
\LongArrow(30,70)(30,50)
\Text(22,60)[r]{$q-p_{2}$}
\end{picture}}}
\end{center}
\caption{Typical diagram contributing to an effective four-fermion interaction
like in Eq. \eqref{equ::fourfermion} in QCD. The solid lines denote fermions with propagator $P^{-1}_{F}$,
the wiggled lines gluons with propagator $P^{-1}_{B}$. The labelled
arrows denote the momentum flow. The diagram suggests that the
contribution $\sim\int_{q}P^{-1}_{F}(q)P^{-1}_{F}(q+p_{1}-p_{3})P^{-1}_{B}(q+p_{1})P^{-1}_{B}(q-p_{2})$
to the effective four-fermion interactions is not a constant but depends, at least on some combination of
the external momenta $p_{1}$, $p_{2}$, $p_{3}$, $p_{4}$.}
\label{fig::qcd}
\end{figure}
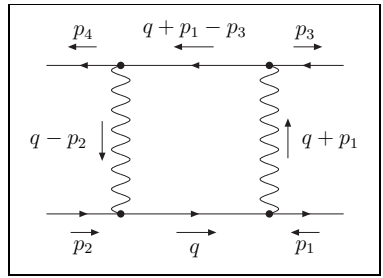

At first, giving up locality sounds like a big step not to be treated lightly. However,
we should remember, that we frequently use those four- and multi-fermion interactions
not as a fundamental interaction but to
model an effective interaction at some intermediate scale.
An example are the four-fermion interactions
used to model QCD at low energies. As an example, a diagram contributing to lowest order is depicted in
Fig. \ref{fig::qcd}.

So, what can we do about the bosonization of those awfully complicated interactions? In the previous section
we have considered local operators of the form ($C$ is a constant matrix in the space of internal indices),
\begin{equation}
\label{equ::bound}
O[\Psi](x)=\Psi^{\textrm{T}}(x)C\Psi(x)=\int_{x,y}\Psi^{\textrm{T}}(y)C\Psi(z)\delta(x-y)\delta(x-z)
=\int_{p}O[\Psi](p)\exp(ipx),
\end{equation}
with
\begin{equation}
\label{equ::composite}
O[\Psi](p)=\int_{q_{1},q_{2}}\Psi^{\textrm{T}}(q_{1})C\Psi(q_{2})\delta(q_{1}+q_{2}-p).
\end{equation}
Keeping in mind our physical picture of a bound state, we find that \eqref{equ::bound} is very restrictive. Indeed,
for a physical bound state we would expect that the particles of which the bound state is composed
are usually not at the same place, but smeared out over a certain region of space. Therefore,
we replace \mbox{$\delta(x-y)\delta(x-z)\rightarrow\tilde{g}(x-y,x-z)$} in Eq. \eqref{equ::bound}, i.e.
the ``elementary particles'' need no longer be located exactly at $x$ but they can be somewhat
distributed around $x$. In a momentum space formulation we find,
\begin{equation}
\label{equ::compoperator}
O[\Psi](p)=hG(p)\int_{p_{1},p_{2}}\Psi^{\textrm{T}}(p_{1})C\Psi(p_{2})g(p_{1},p_{2})\delta(p_{1}+p_{2}-p).
\end{equation}
establishing that the momentum $p$ of the composite operator is the sum, $p=p_{1}+p_{2}$, of the momenta
of the ``elementary particles'', as it must be for a bound state. The function $g(p_{1},p_{2})$ is
the so called (amputated) Bethe-Salpeter wave function \cite{Haymaker:1990vm,Ellwanger:wy}.
The function $hG(p)$ is a generalization of the factor $\frac{h}{m^2}$ in Eq. \eqref{equ::op}, and it
serves the same purpose namely, to obtain a simple form with Yukawa coupling $\sim h$
while keeping the direct relation $\phi\hat{=}O[\Psi]$ between the
bosonic fields $\phi$ and the composite operator $O[\Psi]$, as we will see below.

Proceeding along the lines of the previous sections we insert a functional integral
\begin{equation}
1={\mathcal{N}}\int {\mathcal{D}}\phi\exp(-\int_{p}\frac{1}{2}\phi^{\textrm{T}}(-p)G^{-1}(p^2)\phi(p)).
\end{equation}
Shifting the functional integral by the operator \eqref{equ::compoperator} we find that we can absorb
a four-fermion interaction of the following form,
\begin{equation}
\label{equ::absorb}
\lambda(p_{1},p_{2},p_{3},p_{4})\sim h^{2}g(p_{1},p_{2})G(s)g(-p_{3},-p_{4})C^{\textrm{T}}\otimes C,
\quad s=(p_{1}+p_{2})^2=(p_{3}+p_{4})^2,
\end{equation}
in a contribution
\begin{equation}
\!S_{\lambda}[\Psi,\phi]\!=\!\!\int_{p_{1},p_{2}}\!\!\!\!\!\!\!\!\left[
hg(p_{1},p_{2})\Psi^{\textrm{T}}(p_{1})C\phi^{\textrm{T}}(-p_{1}-p_{2})\Psi(p_{2})
+\phi^{\textrm{T}}(p_{1})G^{-1}(p^{2}_{1})\phi(p_{2})\delta(p_{1}+p_{2})\right],
\end{equation}
to the partially bosonized action. It is clear that $G(s)$ plays the role of the bosonic propagator.
Hence, $G(s)$ should have an appropriate pole structure in the complex $s$-plane,
e.g.
\begin{equation}
\label{equ::pole}
G(s)\sim\frac{1}{m^2+s}.
\end{equation}

Let us summarize this in the following properties:
\begin{enumerate}
\item{Eq. \eqref{equ::absorb} with the pole structure given by \eqref{equ::pole}
is the most general momentum structure
for the four-fermion interaction $\sim\lambda$ we can absorb
in a single bosonic field and an action quadratic in the bosons. We can only bosonize
four-fermion interactions \emph{factorizing} into two pairs of momenta. Usually, contributions
to $\lambda$ like the one depicted in Fig. \ref{fig::qcd} do not factorize completely, therefore
bosonization is usually only an approximation. On the other hand factorization
of the four-fermion interaction signals the onset of physical bound states and can be checked
numerically \cite{Ellwanger:mw}.}
\item{Permutation of the fields (corresponding to Fierz transformations) permute the momentum
variables on the RHS of Eq. \eqref{equ::absorb}. This allows us to absorb momentum structures with
poles in the $t$- respectively $u$-channels (\mbox{$t=(p_{1}-p_{3})^{2}=(p_{2}-p_{4})^{2}$,}
$u=(p_{1}-p_{4})^{2}=(p_{2}-p_{3})^{2}$, $s$, $t$, $u$ are the Mandelstam variables).}\label{item::deter}
\item{Turning the argument of \ref{item::deter}. around, we can determine the ``correct'' Fierz transformation
by an examination of the momentum structure (poles!).}\label{item::deter2}
\end{enumerate}

Since it might help us to resolve the whole mess of the Fierz ambiguity, let us illustrate the third point
by calculating an example. In addition this will also demonstrate how a momentum dependence of
the four-fermion interaction and a wave function renormalization for the composite
bosons are connected.

Starting from the action \eqref{equ::faction} extended by the kinetic terms \eqref{equ::kinetic} let
us calculate the corresponding purely fermionic action. For simplicity we take
$Z_{V}=Z_{A}=h_{V}=h_{A}=0$, i.e. we have no vector and axial vector bosons.
In momentum
space we then have,
\begin{eqnarray}
\label{equ::compboso}
S_{\textrm{B}}=
\int_{q_{1},q_{2}}\delta(q_{1}+q_{2})\!\!\!&\bigg[&(\mu^2_{\sigma}+Z_{\sigma}q^{2}_{1})\phi^{\star}(q_{1})\phi(q_{2})
\\\nonumber
&+&\!\!\!\int_{p_{1},p_{2}}\!\!\!h_{\sigma}\bar{\psi}(-p_{1})\left(\frac{1+\gamma^{5}}{2}\right)\psi(p_{2})\phi(p_{2})
\delta(q_{1}-p_{1}-p_{2})
\\\nonumber
&-&\!\!\!\int_{p_{3},p_{4}}\!\!\!h_{\sigma}
\phi^{\star}(q_{1})\bar{\psi}(p_{4})\left(\frac{1-\gamma^{5}}{2}\right)\psi(-p_{3})
\delta(q_{2}-p_{3}-p_{4})\bigg]
\\\nonumber
=\int_{q_{1},q_{2}}\delta(q_{1}+q_{2})\!\!\!&&\!\!\!\!\!\!\!\!\!(\mu^2_{\sigma}+Z_{\sigma}q^{2}_{1})
\\\nonumber
&&\!\!\!\!\!\!\!\!\!\!\!\!\!\!\!\!\!\!\!\!\!\!\!\!\!\!\!\!\!\!\!\!\!\!\!\!\left(\phi^{\star}(q_{1})+
\int_{p_{1},p_{2}}\!\frac{h_{\sigma}}{\mu^{2}_{\sigma}+Z_{\sigma}q^{2}_{1}}
\bar{\psi}(-p_{1})\left(\frac{1+\gamma^{5}}{2}\right)\psi(p_{2})\delta(q_{1}-p_{1}-p_{2})\right)
\\\nonumber
\times\quad\quad\quad\quad&&\!\!\!\!\!\!\!\!\!\!\!\!\!\!\!\!\!\!\!\!\!\!\!\!\!\!\!\!\!\!\!\!\!\!\!\!
\left(\phi(q_{2})-\int_{p_{3},p_{4}}\!\frac{h_{\sigma}}{\mu^{2}_{\sigma}+Z_{\sigma}q^{2}_{2}}
\bar{\psi}(p_{4})\left(\frac{1-\gamma^{5}}{2}\right)\psi(-p_{3})
\delta(q_{2}-p_{3}-p_{4})\right)
\\\nonumber
+\frac{1}{2}\int_{p_{1},p_{2},p_{3},p_{4}}\delta(p_{1}\!\!\!&+&\!\!\!p_{2}-p_ {3}-p_{4})
\frac{h^2_{\sigma}}{2(\mu^{2}_{\sigma}+Z_{\sigma}(p_{1}+p_{2})^{2})}
\\\nonumber
\times\!\!\!
&\big[&\!\!\!\!\!\bar{\psi}(-p_{1})\psi(p_{2})\bar{\psi}(p_{4})\psi(-p_{3})
-\bar{\psi}(-p_{1})\gamma^{5}\psi(p_{2})\bar{\psi}(p_{4})\gamma^{5}\psi(-p_{3})\big].
\end{eqnarray}
After the usual shift in the integration variable we can perform the Gaussian integration
over the bosonic fields, removing the first term on the RHS of Eq. \eqref{equ::compboso}.
From the second term we can read of the four-fermion interaction,
\begin{equation}
\label{equ::depcoup}
\lambda_{\sigma}(p_{1},p_{2},p_{3},p_{4})=\frac{h^{2}_{\sigma}}{2(\mu^{2}_{\sigma}+Z_{\sigma}(p_{1}+p_{2})^{2})}
=\frac{h^{2}_{\sigma}}{2(\mu^{2}_{\sigma}+Z_{\sigma}(p_{3}+p_{4})^{2})}
=\frac{h^{2}_{\sigma}}{2(\mu^{2}_{\sigma}+Z_{\sigma}s)}.
\end{equation}
In particular, the four-fermion interaction $\lambda$ only depends on the Mandelstam \mbox{variable $s$,}
while it is constant in $t$ and $u$.

An FT permutes $-p_{2}$ and $p_{3}$. After relabelling the integration indices we find,
\begin{eqnarray}
\lambda_{\sigma}(p_{1},p_{2},p_{3},p_{4})&=&0,
\\\nonumber
\lambda_{V}(p_{1},p_{2},p_{3},p_{4})&=&-\lambda_{A}(p_{1},p_{2},p_{3},p_{4})
=-\frac{h^{2}_{\sigma}}{4(\mu^{2}_{\sigma}+Z_{\sigma}(p_{1}-p_{3})^{2})}
=-\frac{h^{2}_{\sigma}}{4(\mu^{2}_{\sigma}+Z_{\sigma}t)}.
\end{eqnarray}
First of all these coupling do not factorize in functions depending only on $p_{1},p_{2}$ and
$p_{3},p_{4}$ respectively. Secondly, they have a pole in $t$ which cannot be directly absorbed by
bosonization. Therefore, this is not the ``right'' FT. A similar calculation for the vector and
axial vector bosons would have resulted in four-fermion interactions depending on $s$ in the
vector and axial vector channels, respectively. While after an FT we would have interactions
in all channels, scalar, vector and axial vector, but again with the ``wrong''
dependence on $t$ which cannot be absorbed into bosons. Roughly speaking we have the following
recipe, if the four-fermion interaction depends on $s$, bosonize, else if it depends on $t$
FT exactly once and then bosonize.
\section{Bosonic RG flow} \label{sec::bosoflow}
Having talked at length about how we can obtain the partially bosonized action Eq. \eqref{equ::baction}
it is time to put it to some use. Let us start with an RG calculation in a very simple truncation.
The flow equations in the bosonic language are obtained in complete analogy with the
fermionic formulation. In this section we restrict the discussion to a ``pointlike''
truncation as given by Eq. \eqref{equ::baction} with $k$-dependent couplings.
We will see (Chap. \ref{chap::boso2}) that we reproduce the result of the last section
in this approximation if
we take care of the fact that new fermionic interactions are generated by the flow and
have to be absorbed by an appropriate $k$-dependent redefinition of the bosonic fields.

\begin{figure}[t]
\begin{center}
\scalebox{0.7}[0.7]{\fbox{
\begin{picture}(160,140)
\SetOffset(-75,10)
\DashLine(90,60)(160,60){2}
\LongArrow(132,50)(118,50)
\Text(125,40)[c]{$p_{1}$}
\LongArrow(203.5,117.5)(192.5,106.5)
\Text(195,123)[c]{$p_{2}$}
\LongArrow(203.5,2.5)(192.5,13.5)
\Text(195,-3)[c]{$p_{3}$}
\Vertex(160,60){2}
\ArrowLine(160,60)(190,90)
\ArrowLine(190,90)(220,120)
\ArrowLine(190,30)(160,60)
\ArrowLine(220,0)(190,30)
\DashLine(190,90)(190,30){2}
\Vertex(190,90){2}
\Vertex(190,30){2}
\end{picture}}}
\caption{Vertex correction diagram in the bosonized model. Solid lines are fermions, dashed lines are bosons.
There exist several diagrams of this type since we have different species of bosons. The momentum
configuration indicated by the arrows is such that it gives a
contribution $\sim \phi(-p_{1})\bar{\psi}(-p_{2})\psi(p_{3})$. The pointlike limit
employed in this section corresponds to an evaluation for $p_{i}=0$.}
\label{fig::vertex}
\end{center}
\end{figure}
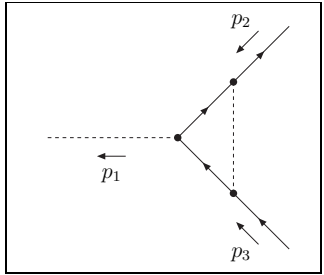

It is instructive to neglect in a first step all bosonic fluctuations by setting
all bosonic entries in the propagator matrix ${\mathcal{P}}^{-1}$ equal to zero. This removes all
diagrams with internal bosonic lines. Among other things this neglects the vertex
correction Fig. \ref{fig::vertex}
and therefore the running of the Yukawa couplings.
Indeed, Fig. \ref{fig::mass} is the only contributing diagram and we recover MFT.
One obtains the flow equations
\begin{eqnarray}
\label{equ::simple}
\partial_{t}\mu^{2}_{\sigma,k}&=&8h^{2}_{\sigma,k}v_{4}k^2l^{(F),4}_{1}(s),
\quad\partial_{t}\mu^{2}_{V,k}=8h^{2}_{V,k}v_{4}k^{2}l^{(F),4}_{1}(s),
\\\nonumber
\partial_{t}\mu^{2}_{A,k}&=&8h^{2}_{A,k}v_{4}k^{2}l^{(F),4}_{1}(s),
\\\nonumber
\partial_{t}h_{\sigma,k}&=&0,\quad\partial_{t}h_{V,k}=0,\quad\partial_{t}h_{A,k}=0.
\end{eqnarray}
As long as we do not consider the wave function renormalization \eqref{equ::kinetic} for the bosons,
the flow can completely be described in terms of the dimensionless combinations
\begin{equation}
\label{equ::convenient}
\tilde{\epsilon}_{\sigma,k}=\frac{\mu^{2}_{\sigma,k}}{h^{2}_{\sigma,k}k^{2}},
\quad \tilde{\epsilon}_{V,k}=\frac{\mu^{2}_{V,k}}{h^{2}_{V,k}k^{2}},
\quad \tilde{\epsilon}_{A,k}=\frac{\mu^{2}_{A,k}}{h^{2}_{A,k}k^{2}}.
\end{equation}

Due to the constant Yukawa couplings we can integrate Eq. \eqref{equ::simple}. We find
critical couplings:
\begin{equation}
\label{equ::simplecrit}
\frac{\mu^{2}_{\sigma}}{h^{2}_{\sigma}\Lambda^{2}}\mid_{\textrm{crit}}=4v_{4},
\quad \frac{\mu^{2}_{V}}{h^{2}_{V}\Lambda^{2}}\mid_{\textrm{crit}}=4v_{4},
\quad \frac{\mu^{2}_{A}}{h^{2}_{A}\Lambda^{2}}\mid_{\textrm{crit}}=4v_{4}.
\end{equation}
These are, of course, the results of MFT, Eq. \eqref{equ::crit}.
We note that in Eq. \eqref{equ::simple} the equations for the different species of bosons are completely decoupled.
The mass terms do not turn negative at the same scale for the different species. Indeed it is
possible that the mass of one boson species turns negative while the others do not. Such a behavior is
expected for the full theory, whereas for the fermionic RG of \mbox{Sect. \ref{sec::fermion}} all couplings
diverge simultaneously due to their mutual coupling. However, no real conclusion can be taken
from Eq. \eqref{equ::simplecrit} because of the strong dependence on
the unphysical parameter $\gamma$.

Now, let us also take into account the bosonic fluctuations. This includes the vertex correction
Fig. \ref{fig::vertex} and the flow of the Yukawa couplings does not vanish anymore
\begin{eqnarray}
\partial_{t}h^{2}_{\sigma,k}&=&
-32v_{4}l^{(F),4}_{1}(s)k^{2}h^{2}_{\sigma,k}
\left[\frac{h^{2}_{V,k}}{\mu^{2}_{V,k}}
-\frac{h^{2}_{A,k}}{\mu^{2}_{A,k}}\right],
\\\nonumber
\partial_{t}h^{2}_{V,k}&=&
-4v_{4}l^{(F),4}_{1}(s)k^{2}h^{2}_{V,k}\left[\frac{h^{2}_{\sigma,k}}{\mu^{2}_{\sigma,k}}
+2\left(\frac{h^{2}_{V,k}}{\mu^{2}_{V,k}}+\frac{h^{2}_{A,k}}{\mu^{2}_{A,k}}\right)\right],
\\\nonumber
\partial_{t}h^{2}_{A,k}&=&
-4v_{4}l^{(F),4}_{1}(s)k^{2}h^{2}_{A,k}
\left[-\frac{h^{2}_{\sigma,k}}{\mu^{2}_{\sigma,k}}
+2\left(\frac{h^{2}_{V,k}}{\mu^{2}_{V,k}}
+\frac{h^{2}_{A,k}}{\mu^{2}_{A,k}}\right)\right].
\end{eqnarray}
Using the
dimensionless $\tilde{\epsilon}$'s we now find:
\begin{eqnarray}
\label{equ::pubosonic}
\partial_{t}\tilde{\epsilon}_{\sigma,k}&=&-2\tilde{\epsilon}_{\sigma,k}
+8\left[1+4\tilde{\epsilon}_{\sigma,k}\left(\frac{1}{\tilde{\epsilon}_{V,k}}
-\frac{1}{\tilde{\epsilon}_{A,k}}\right)\right]v_{4}l^{(F),4}_{1}(s),
\\\nonumber
\partial_{t}\tilde{\epsilon}_{V,k}&=&-2\tilde{\epsilon}_{V,k}
+8\left[2+\tilde{\epsilon}_{V,k}\left(\frac{1}{2\tilde{\epsilon}_{\sigma,k}}
+\frac{1}{\tilde{\epsilon}_{A,k}}\right)\right]v_{4}l^{(F),4}_{1}(s),
\\\nonumber
\partial_{t}\tilde{\epsilon}_{A,k}&=&-2\tilde{\epsilon}_{A,k}
+8\left[2-\tilde{\epsilon}_{A,k}\left(\frac{1}{2\tilde{\epsilon}_{\sigma,k}}
-\frac{1}{\tilde{\epsilon}_{V,k}}\right)\right]v_{4}l^{(F),4}_{1}(s).
\end{eqnarray}
The onset of spontaneous symmetry breaking is
indicated by a vanishing of $\tilde{\epsilon}$ for at least one species of bosons.
Large $\tilde{\epsilon}$ means that the
corresponding bosonic species becomes very massive and therefore effectively
drops out of the flow.

For initial couplings larger than the critical values (see Tabs. \ref{tab::crit} and \ref{tab::crit2}) both
$\tilde{\epsilon}_{\sigma,k}$ and
$\tilde{\epsilon}_{V,k}$ reach zero for finite $t$. Due to the coupling between the different channels they reach zero
at the same $t$. At this point $\tilde{\epsilon}_{A,k}$ reaches infinity and drops out of the flow. This is quite different
from the flow without the bosonic fluctuations where the flow equations for the different species
were decoupled. The breakdown of all equations at one point resembles\footnote{This is an artefact
of the pointlike approximation.} now the case of the fermionic
model discussed in Sect. \ref{sec::fermion}.
The $\gamma$-dependence of the critical couplings  is
reduced considerably, as compared to MFT. This shows that the inclusion of the bosonic fluctuations
is crucial for any quantitatively reliable result. Nevertheless,
the difference between the bosonic and the fermionic flow remains of the order of $10\%$.

\section{Gap equation in the Bosonized Language} \label{sec::gapboso}
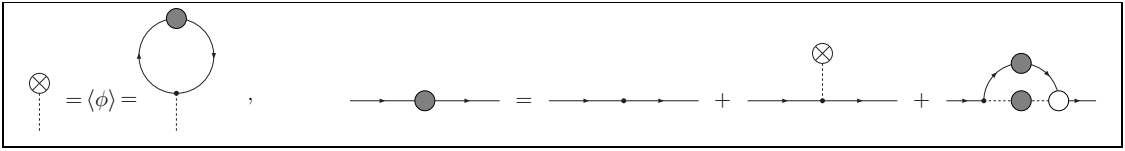
\begin{figure}[t]
\begin{center}
\scalebox{0.47}[0.47]{\fbox{
\begin{picture}(890,110)(-10,0)
\SetOffset(12,10)
\DashLine(0,0)(0,30){2}
\CArc(0,38)(8,0,360)
\Line(-5.66,43.66)(5.66,32.34)
\Line(-5.66,32.34)(5.66,43.66)
\Text(28,23)[]{\scalebox{1.5}[1.5]{$=$}}
\Text(50,24)[]{\scalebox{1.5}[1.5]{$\langle\phi\rangle$}}
\Text(72,24)[]{\scalebox{1.5}[1.5]{$=$}}
\SetOffset(62,10)
\DashLine(60,0)(60,30){2}
\Vertex(60,30){2}
\ArrowArc(60,60)(-30,90,270)
\ArrowArc(60,60)(-30,270,450)
\GOval(60,90)(8,8)(0){0.5}
\Text(120,24)[]{\scalebox{1.5}[1.5]{$,$}}
\SetOffset(262,34)
\ArrowLine(0,0)(52,0)
\ArrowLine(68,0)(120,0)
\GOval(60,0)(8,8)(0){0.5}
\Text(140,-1)[]{\scalebox{1.5}[1.5]{$=$}}
\SetOffset(422,34)
\ArrowLine(0,0)(60,0)
\ArrowLine(60,0)(120,0)
\Vertex(60,0){2}
\Text(140,0)[]{\scalebox{1.5}[1.5]{$+$}}
\SetOffset(582,34)
\ArrowLine(0,0)(60,0)
\ArrowLine(60,0)(120,0)
\Vertex(60,0){2}
\DashLine(60,0)(60,30){2}
\CArc(60,38)(8,0,360)
\Line(54.34,43.66)(65.66,32.34)
\Line(54.34,32.34)(65.66,43.66)
\Text(140,0)[]{\scalebox{1.5}[1.5]{$+$}}
\SetOffset(742,34)
\Vertex(30,0){2}
\ArrowLine(0,0)(30,0)
\ArrowLine(90,0)(120,0)
\DashLine(30,0)(90,0){2}
\ArrowArc(60,0)(-30,180,270)
\ArrowArc(60,0)(-30,270,360)
\GOval(60,0)(8,8)(0){0.5}
\GOval(90,0)(8,8)(0){1.0}
\GOval(60,30)(8,8)(0){0.5}
\end{picture}}}
\caption{Diagrammatic representation of the lowest-order SDE
for the partially bosonized model (Eq. \eqref{equ::baction}).
The shaded circles depict the full propagator, the circle with the cross
is the expectation value of the bosonic field and the empty circle is the full Yukawa vertex.}
\end{center}
\label{fig::schwingerb}
\end{figure}

Next, we turn to the SDE for the bosonized model \eqref{equ::baction}. They
are depicted in Fig. \ref{fig::schwingerb}. We will make here two further approximations
by replacing in the last graph of Fig. \ref{fig::schwingerb} the full fermion-fermion-boson
vertex by the classical Yukawa coupling and the full bosonic propagator by $\mu^{-2}_{\textrm{B}}$.
We remain with two coupled equations.

In a first step we approximate this equations even further by neglecting
the last diagram in \mbox{Fig. \ref{fig::schwingerb}} altogether. Then
no fermionic propagator appears on the right hand side of the equation for the fermionic propagator
which only receives a mass correction for $\langle\phi\rangle\neq 0$.
Without loss of generality we take $\phi$ real such that $M_{\textrm{F}}=h_{\sigma}\phi$ and
\begin{equation}
G^{-1}_{\textrm{F}}(q)=-\fss{q}+h_{\sigma}\phi\gamma^{5}.
\end{equation}
Inserting this into the equation for the expectation value $\phi$ we find
\begin{equation}
\label{equ::schMFT}
\phi=\frac{4v_{4}}{\mu^{2}_{\sigma}}\int^{\Lambda^2}_{0} dx\,x
\frac{h^{2}_{\sigma}\phi}{x+h^{2}_{\sigma}\phi^{2}}.
\end{equation}
For the onset of nontrivial solutions we now find the critical value
\begin{equation}
\left[\frac{h^{2}_{\sigma}}{2\mu^{2}_{\sigma}}\right]_{\textrm{crit}}=\frac{1}{8v_{4}\Lambda^{2}}
=\left[\bar{\lambda}_{\sigma}+2\gamma\bar{\lambda}_{V}\right]_{\textrm{crit}}
\end{equation}
which is the (ambiguous) result from MFT given in Eqs. \eqref{equ::crit} and \eqref{equ::MFT}. This is not
surprising since this \emph{exactly is} MFT from the viewpoint of Schwinger-Dyson equations.
Indeed, Eq. \eqref{equ::schMFT} is precisely the field equation which follows by differentiation of the
MFT effective action \eqref{equ::extremum} with respect to $\phi$.
\begin{equation}
\label{equ::MFTpot}
\Gamma^{\textrm{(F)}}=\mu^{2}_{\sigma}\phi^{2}
-4v_{4}\int^{\Lambda^2}_{0}dx\,x\ln(x+h^{2}_{\sigma}\phi^{2}).
\end{equation}

In a next step we improve our approximation and include the full set of diagrams shown in
\mbox{Fig. \ref{fig::schwingerb}}. Using the same ansatz as before, the self-energy
$\Sigma_{\textrm{F}}$ now has two contributions
\begin{equation}
\label{equ::sum}
\Sigma_{\textrm{F}}=M_{\textrm{F}}\gamma^{5}=h_{\sigma}\phi\gamma^{5}+\Delta m_{\textrm{F}}\gamma^{5}.
\end{equation}
The first one is the contribution from the expectation value of the bosonic
field whereas $\Delta m_{\textrm{F}}$
is the contribution from the last diagram in Fig. \ref{fig::schwingerb}, given by
an integral which depends on $M_{\textrm{F}}$. Both in the equation
for $\langle\phi\rangle$ and in the equation for the fermionic propagator only $M_{\textrm{F}}$ appears
on the RHS. Inserting $\langle\phi\rangle$ in the graph Fig. \ref{fig::schwingerb} one finds
a gap equation which determines $M_{\textrm{F}}$:
\begin{equation}
\label{equ::full}
M_{\textrm{F}}=8v_{4}\left[\frac{h^{2}_{\sigma}}{2\mu^{2}_{\sigma}}
+\frac{h^{2}_{V}}{\mu^{2}_{V}}-\frac{h^{2}_{A}}{\mu^{2}_{A}}\right]
\int^{\Lambda^2}_{0}dx\,x\frac{M_{\textrm{F}}}{x+M^{2}_{\textrm{F}}}.
\end{equation}
Once more, this can be expressed in terms of the invariant couplings and again we arrive at
\mbox{Eq. \eqref{equ::complete}}.

We point out that, in order to recover the result of the fermionic SDE we have started with MFT and added diagrams. Therefore,
the fermionic SDE (or the bosonic SDE with the extra contribution from the fermionic mass shift diagram)
sums over a larger class of diagrams which contains MFT as a subset. This is evident in the language of
statistical physics: MFT is the Hartree approximation, while the fermionic SDE is Hartree-Fock.

Looking more closely at the two contributions to $M_{\textrm{F}}$ we find that alone
neither the contribution $\sim\phi$ (which amounts to MFT as we have discussed above) nor
the ``fermionic contribution'' $\Delta m_{\textrm{F}}$ are invariant under FT's.
Only the combination $M_{\textrm{F}}$, which
is the fermion mass and therefore a physical quantity, is invariant. Indeed, changing the
FT amounts to a redefinition of the bosonic fields. This allows us to choose bosonic fields
such that $\Delta m_{\textrm{F}}=0$. Taking $\gamma=1/2$ gives us such a choice of the bosonic fields. This
explains why MFT gives the the same result as the purely fermionic calculation
in this special case.

In the next chapter we want to adapt this idea of a redefinition of the bosonic fields to the RG
calculation, i.e. we want to do it continuously during the flow.

\chapter{Partial Bosonization II: Scale Dependent Degrees of Freedom}\label{chap::boso2}
In the last Sect. \ref{sec::gapboso} we found that in the SDE formulation we can complete
MFT by adding the mass-shift diagram for the fermions (cf. Fig. \ref{fig::schwingerb}). The mass-shift diagram
is a contribution to the purely fermionic part of the effective action. Therefore, it makes sense
to look for purely fermionic contributions in the RG, too.
\section{New Four-Fermion Interactions} \label{sec::redef}
In our truncation of Chap. \ref{chap::boso1}
the bosonic propagators are approximated by
constants $\mu^{-2}_{k}$. The exchange of bosons therefore
produces effective pointlike four-fermion interactions. One
would therefore suspect that this approximation should contain the
same information as the fermionic formulation with pointlike four-fermion
interactions. An inspection of the results in Tabs.
\ref{tab::crit}, \ref{tab::crit2} shows, however, that this is not
the case for the formulation in the RG context. In
particular, in contrast to the fermionic language the results of
the bosonic flow equations still depend on the unphysical
parameter $\gamma$.

\begin{figure}[t]
\begin{center}
\subfigure[]{\scalebox{0.7}[0.7]{\fbox{
\begin{picture}(170,140)
\label{subfig::boxa} \SetOffset(3,10) \LongArrow(13,10)(27,10)
\Text(20,2)[c]{$p_{2}$} \LongArrow(147,10)(133,10)
\Text(140,2)[c]{$p_{1}$} \LongArrow(133,110)(147,110)
\Text(140,118)[c]{$p_{3}$} \LongArrow(27,110)(13,110)
\Text(20,118)[c]{$p_{4}$} \ArrowLine(40,100)(0,100)
\ArrowLine(120,100)(40,100) \ArrowLine(160,100)(120,100)
\ArrowLine(0,20)(40,20) \ArrowLine(40,20)(120,20)
\ArrowLine(120,20)(160,20) \Vertex(40,100){2} \Vertex(120,100){2}
\Vertex(40,20){2} \Vertex(120,20){2} \DashLine(40,100)(40,20){2}
\DashLine(120,100)(120,20){2}
\end{picture}}}}
\subfigure[]{\scalebox{0.7}[0.7]{\fbox{
\begin{picture}(170,140)
\label{subfig::boxb} \SetOffset(3,10) \LongArrow(13,10)(27,10)
\Text(20,2)[c]{$p_{2}$} \LongArrow(147,10)(133,10)
\Text(140,2)[c]{$p_{1}$} \LongArrow(133,110)(147,110)
\Text(140,118)[c]{$p_{3}$} \LongArrow(27,110)(13,110)
\Text(20,118)[c]{$p_{4}$} \ArrowLine(40,100)(0,100)
\ArrowLine(120,100)(40,100) \ArrowLine(160,100)(120,100)
\ArrowLine(0,20)(40,20) \ArrowLine(40,20)(120,20)
\ArrowLine(120,20)(160,20) \Vertex(40,100){2} \Vertex(120,100){2}
\Vertex(40,20){2} \Vertex(120,20){2} \DashLine(40,100)(120,20){2}
\DashLine(120,100)(40,20){2}
\end{picture}}}}
\end{center}
\caption{Box diagrams for the bosonized model. Again, solid lines
are fermions, dashed lines bosons and vertices are marked with a
dot. The diagrams generate new four-fermion $\sim \bar{\psi}(-p_{1})\psi(p_{2})\bar{\psi}(p_{4})\psi(-p_{3})$
interactions even for
the model \eqref{equ::baction} without direct four-fermion
interactions.} \label{fig::box}
\end{figure}
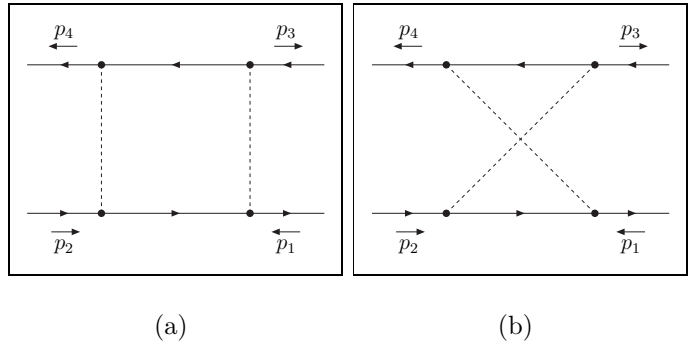

In fact, even for small couplings $\lambda$ the bosonic flow
equations of sect. \ref{sec::bosoflow} do not reproduce the
perturbative result. The reason is that at the one-loop level new
quartic fermion interactions are generated by the box diagrams
shown in Fig. \ref{fig::box}. A straightforward inspection shows that they
contribute to the same order $\lambda^{2}$ as the diagrams in
Figs. \ref{fig::mass} and \ref{fig::vertex}. Even if we start from
vanishing quartic couplings after partial bosonization, such
couplings are generated by the flow. The diagrams in Fig.
\ref{fig::box} yield
\begin{eqnarray}
\label{equ::lambdaflow}
\partial_{t}\lambda_{\sigma,k}&=&\beta_{\lambda_{\sigma}}=
-8v_{4}l^{(F),4}_{1}(s)k^{2}\frac{h^{2}_{\sigma,k}}{\mu^{2}_{\sigma,k}}\frac{h^{2}_{A,k}}{\mu^{2}_{A,k}}
+4k^{-2}\tilde{\gamma}(k),
\\\nonumber
\partial_{t}\lambda_{V,k}&=&\beta_{\lambda_{V}}=
24v_{4}l^{(F),4}_{1}(s)k^{2}\frac{h^{2}_{V,k}}{\mu^{2}_{V,k}}\frac{h^{2}_{A,k}}{\mu^{2}_{A,k}}
-2k^{-2}\tilde{\gamma}(k),
\\\nonumber
\partial_{t}\lambda_{A,k}&=&\beta_{\lambda_{A}}=
-v_{4}l^{(F),4}_{1}(s)k^{2}\left[
\frac{h^{4}_{\sigma,k}}{\mu^{4}_{\sigma,k}}
-12\frac{h^{4}_{V,k}}{\mu^{4}_{V,k}}
-12\frac{h^{4}_{A,k}}{\mu^{4}_{A,k}}\right]+2k^{-2}\tilde{\gamma}(k).
\end{eqnarray}
Here, $\tilde{\gamma}(k)$ is an in principle arbitrary function of
scale determining the choice of FT for the generated four-fermion
interactions. In other words, $\tilde{\gamma}(k)$ allows for the fact that we can choose
a different Fierz representation at every scale. We will make a special choice of this function
(similar to the one made in sects. \ref{sec::perturbation} and
\ref{sec::fermion}) namely we require
\begin{equation}
\label{equ::fix}
\frac{\tilde{\epsilon}_{V,k}}{\tilde{\epsilon}_{A,k}}=\frac{\gamma}{1-\gamma}\quad
\forall \,k,
\end{equation}
with $\tilde{\epsilon}$ given in Eq. \eqref{equ::convenient}. The
resulting equation
$\partial_{t}(\tilde{\epsilon}_{V,k}/\tilde{\epsilon}_{A,k})=0$
fixes $\tilde{\gamma}(k)$. An improved choice of
$\tilde{\gamma}(k)$ can be obtained once the momentum dependence
of vertices is considered more carefully (cf. \cite{Gies:2002nw} and Sects. \ref{sec::momentumdep}, \ref{sec::beyondlpa}).

\section{Adapted Flow Equation: Solving the Ambiguity} \label{sec::solving}
An inclusion of the couplings $\lambda_{k}$ into the truncation of
the effective average action does not seem very attractive.
Despite the partial bosonization we would still have to deal with
multi-fermion interactions and the bosonic formulation would
be of even higher algebraic complexity than the fermionic
formulation. A way out of this has been proposed in
\cite{Gies:2002nw}. There, it has been shown that it is possible
to reabsorb all four-fermion interactions generated during the
flow by a redefinition of the bosonic fields. In the following
brief description of this method we use a very symbolic
notation\footnote{We refrain from explicitly returning to the $0$-dimensional toy model of
Sect. \ref{sec::toy}, as we hope it is clear from Sect. \ref{sec::largerzero} that
for pointlike interactions the case $d>0$ involves no additional difficulties. Nevertheless,
let us note for completeness, that the replacements $\delta\rightarrow\partial$, $\phi_{k}\rightarrow y_{k}$,
$\bar{\psi}\psi\rightarrow x^{\textrm{T}}Sx$ would bring us back to the toy model.}.
In Sect. \ref{sec::beyondlpa} we will add some
details on the momentum dependence of field redefinitions.

Introducing an explicit $k$-dependence for the definition of the
bosonic fields in terms of fermion bilinears, the flow equation
\mbox{Eq. \eqref{equ::flow1}} is
modified\footnote{It has been pointed out by Jan Pawlowski that after the appropriate
modification of the infrared cutoff for the scale-dependent fields \cite{Gies:2002nw}, the
flow equation Eq. \eqref{equ::flow1} does not give the exact flow for $\partial_{t}\Gamma_{k}|$.
However, in the simple approximation of this section the bosonic fields do
not yet have an infrared cutoff. Therefore, we can still use \eqref{equ::flow1}.}:
\begin{equation}
\partial_{t}\Gamma_{k}=\partial_{t}\Gamma_{k}|_{\phi_{k}}
+\frac{\delta\Gamma_{k}}{\delta\phi_{k}}\partial_{t}\phi_{k}.
\end{equation}
Here
$\partial_{t}\Gamma_{k}|_{\phi_{k}}\equiv\partial_{t}\Gamma_{k}|$
is the flow of the effective average action at fixed
fields. Shifting $\phi$ by
\begin{equation}
\partial_{t}\phi_{k}=\left(\bar{\psi}\psi\right)\partial_{t}\omega_{k}
\end{equation}
we find
\begin{equation}
\label{equ::influence}
\partial_{t}\mu^{2}=\partial_{t}\mu^{2}|,
\quad\partial_{t}h=\partial_{t}h|+\mu^{2}\partial_{t}\omega_{k},
\quad\partial_{t}\lambda=\partial_{t}\lambda|-h\partial_{t}\omega_{k}
\end{equation}
and we can choose $\omega_{k}$ to establish:
\begin{equation}
\partial_{t}\lambda=0.
\end{equation}
Instead of including running four-fermion couplings explicitly we
therefore have to use only adapted flow equations for the
couplings contained in Eq. \eqref{equ::baction}.

Let us now apply this method explicitly to our model. Shifting
\begin{eqnarray}
\label{equ::shift}
\partial_{t}\phi&=&-\bar{\psi}\left(\frac{1-\gamma^{5}}{2}\right)\psi\partial_{t}\omega_{\sigma,k},
\quad
\partial_{t}\phi^{\star}=\bar{\psi}\left(\frac{1+\gamma^{5}}{2}\right)\psi\partial_{t}\omega_{\sigma,k},
\\\nonumber
\partial_{t}V^{\mu}&=&-\bar{\psi}\gamma^{\mu}\psi\partial_{t}\omega_{V,k},
\quad
\partial_{t}A^{\mu}=-\bar{\psi}\gamma^{\mu}\gamma^{5}\psi\partial_{t}\omega_{A,k}
\end{eqnarray}
we have
\begin{eqnarray}
\label{equ::lambda}
\partial_{t}\lambda_{\sigma,k}&=&\partial_{t}\lambda_{\sigma,k}|-h_{\sigma,k}\partial_{t}\omega_{\sigma,k},
\\\nonumber
\partial_{t}\lambda_{V,k}&=&\partial_{t}\lambda_{V,k}|-2h_{V,k}\partial_{t}\omega_{V,k},
\quad
\partial_{t}\lambda_{A,k}=\partial_{t}\lambda_{A,k}|-2h_{A,k}\partial_{t}\omega_{A,k}.
\end{eqnarray}
Requiring $\partial_{t}\lambda=0$ for all $\lambda$'s we can
determine the functions $\omega$:
\begin{eqnarray}
\label{equ::omegas}
\partial_{t}\omega_{\sigma,k}=\frac{\beta_{\lambda_{\sigma}}}{h_{\sigma,k}},
\quad
\partial_{t}\omega_{V,k}=\frac{\beta_{\lambda_{V}}}{2h_{V,k}},
\quad
\partial_{t}\omega_{A,k}=\frac{\beta_{\lambda_{A}}}{2h_{A,k}}
\end{eqnarray}
with the $\beta$-functions given in Eq. \eqref{equ::lambdaflow}.
This yields the adapted flow equations for the Yukawa couplings
\begin{eqnarray}
\label{equ::effect}
\partial_{t}h_{\sigma,k}&=&\partial_{t}h_{\sigma,k}|+\mu^{2}_{\sigma,k}\partial_{t}\omega_{\sigma,k},
\\\nonumber
\partial_{t}h_{V,k}&=&\partial_{t}h_{V,k}|+\mu^{2}_{V,k}\partial_{t}\omega_{V,k},
\quad
\partial_{t}h_{A,k}=\partial_{t}h_{A,k}|+\mu^{2}_{A,k}\partial_{t}\omega_{A,k}.
\end{eqnarray}
Combining Eqs. \eqref{equ::pubosonic}, \eqref{equ::lambdaflow},
\eqref{equ::fix}, \eqref{equ::omegas}, \eqref{equ::effect}
determines $\tilde{\gamma}(k)$
\begin{equation}
\tilde{\gamma}(k)=2v_{4}l^{(F),4}_{1}(s)\left[-\frac{3}{\tilde{\epsilon}^{2}_{V,k}}
+\frac{1}{\tilde{\epsilon}_{V,k}\tilde{\epsilon}_{A,k}}
+\frac{(4\tilde{\epsilon}_{\sigma,k}-\tilde{\epsilon}_{A,k})^{2}}
{4\tilde{\epsilon}^{2}_{\sigma,k}\tilde{\epsilon}_{A,k}(\tilde{\epsilon}_{V,k}+\tilde{\epsilon}_{A,k})}
\right].
\end{equation}

Having fixed the ratio between $\tilde{\epsilon}_{V,k}$ and
$\tilde{\epsilon}_{A,k}$ we only need two equations to describe
the flow. We will use the ones for $\tilde{\epsilon}_{\sigma,k}$
and $\bar{\epsilon}_{V,k}=(1-\gamma)\tilde{\epsilon}_{V,k}$
\begin{eqnarray}
\label{equ::invariantbos} \nonumber
\partial_{t}\tilde{\epsilon}_{\sigma,k}\!\!\!\!&=&\!\!\!\!-2\tilde{\epsilon}_{\sigma,k}
\!+\!4\!\left[(1+\gamma)-4(-2+\gamma+2\gamma^2)
\frac{\tilde{\epsilon}_{\sigma,k}}{\bar{\epsilon}_{V,k}}
+4(3-7\gamma+4\gamma^3)
\frac{\tilde{\epsilon}^{2}_{\sigma,k}}{\bar{\epsilon}^{2}_{V,k}}\right]\!l^{(F),4}_{1}(s)v_{4},
\\
\partial_{t}\bar{\epsilon}_{V,k}\!\!\!\!&=&\!\!\!\!-2\bar{\epsilon}_{V,k}
\!+\!4\!\left[\frac{\bar{\epsilon}_{V,k}}{2\tilde{\epsilon}_{\sigma,k}}-(2\gamma-1)\right]^{2}\!l^{(F),4}_{1}(s)v_{4}.
\end{eqnarray}
These equations are completely equivalent to the fermionic flow
Eq. \eqref{equ::fermionflow}. In order to see this we recall
that the simple truncation of the form \eqref{equ::baction} is at
most quadratic in the bosonic fields. We can therefore easily
solve the bosonic field equations as a functional of the fermion
fields. Reinserting the solution into the effective average action
we obtain the form \eqref{equ::faction} with the $k$-dependent
quartic couplings
\begin{equation}
\bar{\lambda}_{\sigma,k}=\frac{1}{2k^{2}\tilde{\epsilon}_{\sigma,k}}-2\gamma\frac{1}{k^{2}\bar{\epsilon}_{V,k}},
\quad \bar{\lambda}_{V,k}=\frac{1}{k^{2}\bar{\epsilon}_{V,k}}.
\end{equation}
Inserting this into Eq. \eqref{equ::invariantbos} we find Eq.
\eqref{equ::fermionflow}, establishing both the exact equivalence
to the fermionic model and the $\gamma$-independence of physical
quantities. On a numerical level, we can see the equivalence from
the critical couplings listed in \mbox{Tabs. \ref{tab::crit}, \ref{tab::crit2}.}

\begin{figure}[t]
\scalebox{0.99}[0.99]{\includegraphics{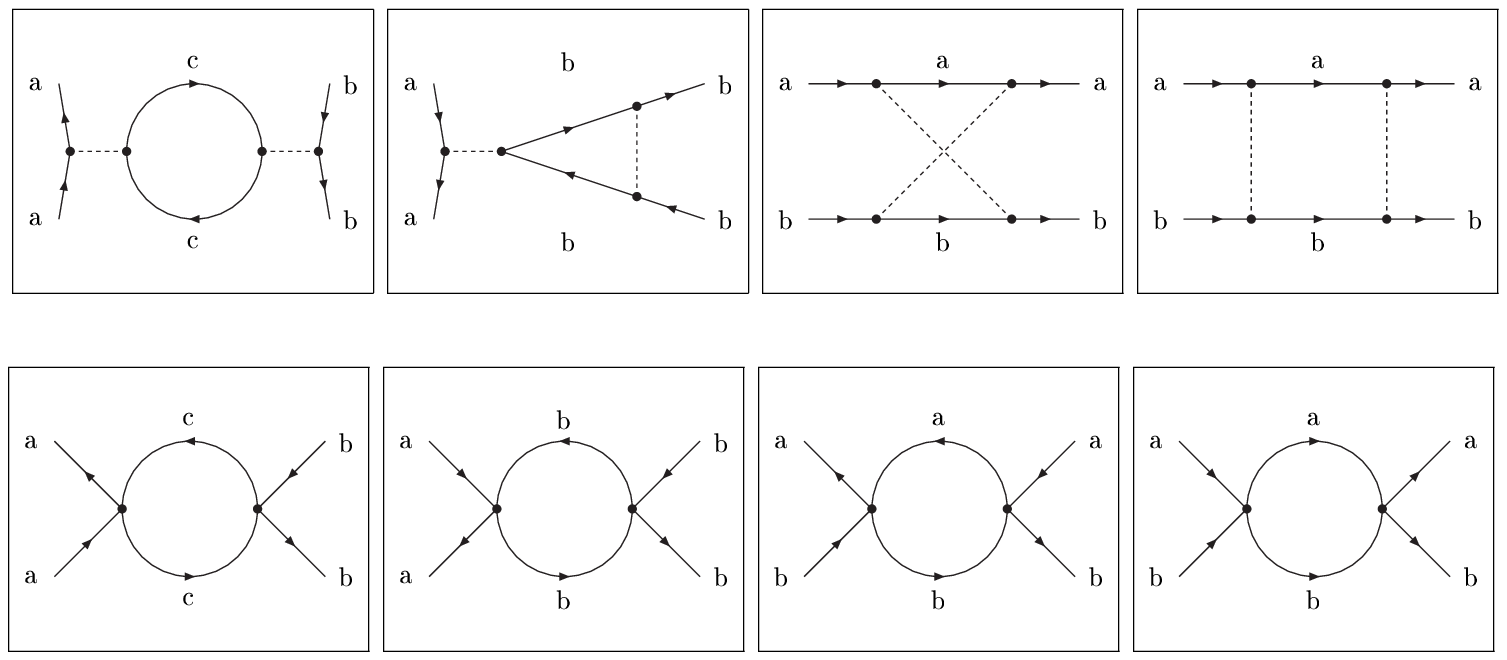}}
\caption{Summary of all diagrams encountered in the previous
sections. There is a one to one correspondence between the
diagrams of the bosonized model (first row) and the purely
fermionic model (second row). Solid lines with an arrow denote
fermionic lines. The letters in the diagrams are given for
visualizing the ways in which the fermionic operators are
contracted, e.g. the first diagram in the second row results from
a term $[(\bar{\psi}_{a}\psi_{a})
(\protect\Wwick{1}{<*{\bar{\psi}_{c}}\psi_{c})][({\bar{\psi}_{c}}>*\psi_{c}}
{1}{{\bar{\psi}_{c}}<+\psi_{c})][(>+{\bar{\psi}_{c}}\psi_{c}})(\bar{\psi}_{b}\psi_{b})]$.
Shrinking bosonic lines (dashed) to points maps the diagrams in
the first row to the second row. In the approximations of sect.
\ref{sec::bosoflow} only the first or the first two diagrams are
taken into account.} \label{fig::summ}
\end{figure}

On this level of truncation the equivalence between the fermionic
and the adapted bosonic flow can also be seen on a diagrammatic
level. As long as we do not have a kinetic term for the bosons the
internal bosonic lines shrink to points. On the one-loop level we
find an exact correspondence between the diagrams for the
bosonized and the purely fermionic model summarized in Fig.
\ref{fig::summ}. This demonstrates again that one-loop accuracy
cannot be obtained without adaption of the flow.
\section{Trouble With the LPA, an Example} \label{sec::trouble}
So far, everything seems quite satisfactory. In our simple truncation we have
been able to solve the problem of the Fierz ambiguity for the partially bosonized
language. Moreover, the adaption of the bosonic flow is quite intuitive as it
implements the idea of scale dependent degrees of freedom.
But, not everything is as rosy as it seems. Looking a little bit more closely
we notice, that although the critical coupling is independent of $\gamma$, the values
of the individual bosonic couplings $\tilde{\epsilon}_{\sigma}$, $\tilde{\epsilon}_{V}$,
$\tilde{\epsilon}_{A}$ are \emph{not}. Only two of them are fixed by the flow equation
\eqref{equ::invariantbos}, while the third one can be chosen freely. In particular, we cannot
determine from this truncation which type of boson will condense. This is completely analogous
to the purely fermionic description, and in view of the equivalence of both
descriptions not too surprising. It seems evident that this is a shortcoming of our present truncation.

Thinking about enlarging the truncation, two possibilities come to mind immediately, a more
complicated bosonic potential and a non-zero kinetic terms for the bosons.
In this section we will consider the first possibility. As it turns out this does not solve the
problem, i.e. we still cannot decide which type of bosons will condense. Hence, we turn to
the alternative of kinetic terms for the bosons in Sect. \ref{sec::beyondlpa}.
\subsection{The Gross-Neveu Model}
To get a first impression it is instructive to study an even simpler model than Eq. \eqref{equ::faction},
the so called $N=1$\footnote{N is the number of different fermion species.}
Gross Neveu model \cite{Gross:jv} in three spacetime dimensions,
\begin{equation}
\label{equ::grossaction}
S_{\textrm{F}}[\bar{\psi},\psi]=\int d^{3}x \left\{i\bar{\psi}\fss{\partial}\psi+\frac{G}{2}(\bar{\psi}\psi)^{2}\right\}
\end{equation}
Actually, it is indeed the same model, only the number of spacetime dimensions has been reduced by one.
The simplification as compared to Eq. \eqref{equ::faction} lies in the fact that in three
dimensions we can use spinors with
only two components (s. App. \ref{app::fermion}).

Despite its simplicity it is still
interesting in its own right. In particular it has a parity-like symmetry $\psi(x)\rightarrow-\psi(-x)$,
$\bar{\psi}(x)\rightarrow\bar{\psi}(-x)$ which is expected to be spontaneously broken by a non-vanishing
vacuum expectation value $\langle\bar{\psi}\psi\rangle$ for large enough $G$ \cite{Hoefling:2002hj}.

Following the lines of Chap. \ref{chap::boso1} it is straightforward to obtain the
equivalent partially bosonized action,
\begin{equation}
\label{equ::yukawagross}
S_{\textrm{B}}[\bar{\psi},\psi]=\int d^{3}x\left\{ i\bar{\psi}\fss{\partial}\psi
+ih\phi\bar{\psi}\psi+\frac{m^2}{2}\phi^{2}\right\}, \quad G=\frac{h^2}{m^2}.
\end{equation}
We now want to study this model in a truncation which includes an arbitrary local potential $V(\phi)$
but no kinetic term $\sim\partial_{\mu}\phi\partial^{\mu}\phi$
(in the following we will suppress the integration over the spacetime coordinates),
\begin{equation}
\label{equ::trunc}
\Gamma_{k}=i\bar{\psi}\fss{\partial}\psi+ih_{k}\phi\bar{\psi}\psi+V_{k}(\phi).
\end{equation}
The parity like symmetry translates into $\phi(x)\rightarrow-\phi(-x)$ in the bosonic language,
restricting any bosonic potential $V(\phi)$ to even powers of $\phi$.

Let us, for the moment, assume that during the flow we generate fermionic $\sim(\bar{\psi}\psi)^{n}$,
bosonic $\sim\phi^{n}$ and mixed
$\sim\phi^{n}(\bar{\psi}\psi)^{m}$ interactions. We neglect all other contributions
as they lie outside of our present truncation.
Hence, we write for the flow at fixed fields,
\begin{equation}
\partial_{t}\Gamma_{k}|\equiv\partial_{t}U_{k}(\phi_{k},\bar{\psi}\psi)
=\sum_{n}a^{n}_{k}(\phi_{k})\left(\bar{\psi}\psi+\frac{V^{\prime}_{k}(\phi_{k})}{ih_{k}}\right)^{n},
\end{equation}
where we have expanded the RHS in powers of $\bar{\psi}\psi$ about the point $-\frac{V^{\prime}(\phi)}{ih}$,
and $V^{\prime}_{k}(\phi_{k})=\frac{\partial V_{k}(\phi_{k})}{\partial\phi_{k}}$.
Allowing for a scale dependence of the field $\phi_{k}$ we obtain,
\begin{eqnarray}
\label{equ::grossflow}
\partial_{t}\Gamma_{k}&=&\partial_{t}\Gamma_{k}|+\frac{\delta\Gamma_{k}}{\delta\phi_{k}}\partial_{t}\phi_{k}
\\\nonumber
&=&\partial_{t}U_{k}(\phi_{k},\bar{\psi}\psi)+(ih_{k}\bar{\psi}\psi+V^{\prime}_{k}(\phi_{k}))\partial_{t}\phi_{k}
\\\nonumber
&=&a^{0}_{k}(\phi_{k})+\sum_{n\geq 1}a^{n}_{k}(\phi_{k})
\left(\bar{\psi}\psi+\frac{V^{\prime}_{k}(\phi_{k})}{ih_{k}}\right)^{n}
+(ih_{k}\bar{\psi}\psi+V^{\prime}_{k}(\phi_{k}))\partial_{t}\phi_{k}.
\end{eqnarray}
Now, let us make a non-linear field redefinition\footnote{One may wonder if such a field
redefinition has the right symmetry properties. A thorough inspection
tells us that the given field redefinition is of the
type $\partial_{t}\phi_{k}=\frac{\partial X(\phi_{k},\bar{\psi}\psi,\ldots)}{\partial\phi_{k}}$, where $X$ is
a singlet under all symmetries (the dots denote any type of additional fields).
A derivative of this type belongs to the conjugate representation
of $\phi_{k}$ and thereby to the same as $\phi_{k}$ since $\phi_{k}$ is self-conjugate.
This also provides us with a recipe
how we can obtain field redefinitions respecting the
symmetries for more complicated fields, e.g. vector fields.},
\begin{equation}
\partial_{t}\phi_{k}=-\frac{1}{ih_{k}}\sum_{n\geq 1}a^{n}_{k}(\phi_{k})
\left(\bar{\psi}\psi+\frac{V^{\prime}_{k}(\phi_{k})}{ih_{k}}\right)^{n-1}.
\end{equation}
Inserting this into Eq. \eqref{equ::grossflow} yields a simplified flow which only affects the
potential in $\phi_{k}$,
\begin{equation}
\partial_{t}\Gamma_{k}=a^{0}_{k}(\phi_{k})=\partial_{t}U_{k}
\left(\phi_{k},i\frac{V^{\prime}_{k}(\phi_{k})}{h_{k}}\right).
\end{equation}
At this point one might wonder why $h_{k}$ does not
receive any corrections as in \eqref{equ::effect}. The reason for this is that we have used the
additional freedom to rescale $\phi$ with a contribution $\sim\phi$ in $\partial_{t}\phi$, transforming
any change in $h_{k}$ into a change of $V(\phi)$.

So far this looks quite appealing, as we succeeded in including all the complicated interactions
contained in $\partial_{t}U_{k}(\phi,\bar{\psi}\psi)$ into a simple truncation
which includes a potential depending only on $\phi$. However, let us now show that we can simplify our
result even further by adding a suitably written $0$ to $\partial_{t}U_{k}(\phi_{k},\bar{\psi}\psi)$.
Indeed, starting from the action \eqref{equ::yukawagross} it is possible to reduce
$V_{k}(\phi_{k})$ to a mass term $\frac{m^{2}_{k}}{2}\phi^{2}_{k}$.
If $V_{k}(\phi_{k})=\frac{m^{2}_{k}}{2}\phi^{2}_{k}$ at a given point, we have
$V^{\prime}(\phi_{k})=m^{2}_{k}\phi_{k}$, and the flow of the potential reads,
\begin{equation}
\label{equ::totalflow}
\partial_{t}V_{k}(\phi_{k})=a^{0}_{k}(\phi_{k})
=\partial_{t}U_{k}\left(\phi_{k},i\frac{m^{2}_{k}}{h_{k}}\phi_{k}\right)
=\sum_{n}b^{n}_{k}\phi^{n}.
\end{equation}
Since we do not consider gravity, field independent terms in the effective action are of no
physical significance. Dropping those and remembering that $V(\phi)$ can contain only even
powers of $\phi$, it is no restriction to write
\begin{equation}
\partial_{t}U_{k}\left(\phi,i\frac{m^{2}_{k}}{h}\phi_{k}\right)=\sum_{n\geq 1}b^{2n}_{k}\phi^{2n}.
\end{equation}
Now, let us use that the spinors have only two components, i.e. at any given point $x$ we have only
four different Grassmann variables, $\bar{\psi}_{1,2}(x)$, $\psi_{1,2}(x)$. Consequently, we obtain
\begin{equation}
\label{equ::grassmannzero}
(\bar{\psi}(x)\psi(x))^{n}=0, \quad \forall n\geq 3,
\end{equation}
by use of the anticommutation relations. In particular, we have,
\begin{equation}
\label{equ::predesaster}
\partial_{t}\hat{U}_{k}(\phi_{k},\bar{\psi}\psi)=\partial_{t} U_{k}(\phi_{k},\bar{\psi}\psi)
-\sum_{n\geq 2}b^{2n}_{k}\left(\frac{ih}{m^{2}_{k}}\right)^{2n}(\bar{\psi}\psi)^{2n}
=\partial_{t}U_{k}(\phi_{k},\bar{\psi}\psi).
\end{equation}
Performing field redefinitions as above, but for $\partial_{t}\hat{U}$ instead of
$\partial_{t}U$ yields,
\begin{equation}
\label{equ::desaster}
\partial_{t}V_{k}(\phi_{k})=\partial_{t}\hat{U}\left(\phi_{k},i\frac{m^2_{k}}{h_{k}}\phi_{k}\right)
=b^{2}_{k}\phi^{2}_{k},
\end{equation}
i.e. our potential remains a mass term for all $k$ as we have claimed above.

At first, this seems very strange. Yet, as we know from Chap. \ref{chap::boso1} we can, at least formally,
remove the bosonic fields by performing the appropriate functional integral. Since we have no kinetic term
for the bosons, and $V(\phi)$ is local, this results in a completely local fermionic interaction.
However, Eq. \eqref{equ::grassmannzero} tells us that the highest order local and purely fermionic interaction
is $(\bar{\psi}\psi)^{2}$. Therefore, any action of the form \eqref{equ::trunc} is equivalent to
\eqref{equ::grossaction} and in consequence also to Eq. \eqref{equ::yukawagross}, as long as
we choose $G$ and $\frac{h^2}{m^2}$ correctly.
In other words, if we do not consider a non-vanishing kinetic term for the bosons,
an inclusion of a full bosonic potential does not give us any more physical information
than a simple truncation to a mass term or a purely fermionic calculation with a local four-fermion interaction.
In particular, we cannot proceed into the SSB regime.
\subsection{General Discussion}
It is straightforward to extend Eq. \eqref{equ::totalflow} to more general cases
with several different bosonic composite operators and corresponding bosonic fields, e.g., induced
by fermions with more components. In fact,
using our symbolic notation of Chap. \ref{chap::effectiveaction} the generalization
looks like \mbox{Eq. \eqref{equ::totalflow}}. The discussion leading to Eq. \eqref{equ::desaster}
can be generalized, too, but it is somewhat more complicated, as we have to use
the inverse function $V^{\prime\,-1}$ of $V^{\prime}$ to write down $\partial_{t}\hat{U}$. Assuming
the existence of $V^{\prime\,-1}$ we find,
\begin{equation}
\partial_{t}\hat{U}_{k}(\phi,\bar{\psi}O\psi)=\partial_{t}U_{k}(\phi,\bar{\psi}O\psi)
-\sum_{|n|> N}b^{n}_{k}\left[V^{\prime\,-1}(-ih_{k}\bar{\psi}O\psi)\right]^{n},
\end{equation}
where, for simplicity, we have employed a symbolic notation with the components $\phi_{i}$ corresponding
to the operators $O_{i}$. The $b$'s are
defined as in Eq. \eqref{equ::predesaster}, but $n=(n_{1},\ldots)$ is now a multi index. $N$ is the number of
components of the spinor $\psi$ and determines the highest-order monomial in $\psi$
which does not vanish by the anticommutation relations.
E.g. for a Dirac fermion in four dimensions (four spin indices)
with three colors we have
$N=4\times 3=12$. Using $\partial_{t}\hat{U}$ to define the field redefinitions we can force
all contributions $\phi^{n}$, $|n|>N$ to vanish as in Eq. \eqref{equ::desaster}.

Typically, a potential up to order $\phi^{4}$ is sufficient to describe at least the basic features of SSB.
As most models have $N\geq 4$ we might be tempted to conclude that, with the exception of
some special cases, the LPA works. However, this is not the case. We have already seen that
we can modify $\partial_{t}V_{k}(\phi)$ by adding a conveniently written
zero to $\partial_{t}U(\phi,\bar{\psi}\psi)$.
But, Eq. \eqref{equ::grassmannzero} and its generalizations to the case of more spinor
components are not the only way we can write a zero. Fierz identities like Eq. \eqref{equ::fierz}
provide another one. As we can see from the example of Eq. \eqref{equ::fierz} they allow us to find
non-vanishing $c^{n}$ such that
\begin{equation}
\label{equ::allgfierz}
\sum_{|n|=m}c^{n}(\bar{\psi}O\psi)^{n}=0,\quad m\leq N.
\end{equation}
An addition of this to $\partial_{t}U(\phi,\bar{\psi}O\psi)$ can be used to eliminate terms with $\phi^{n}$ and
$|n|\leq N$ in the potential. The $c^{n}$ in Eq. \eqref{equ::allgfierz} are not all independent,
but typically we can eliminate at least one species of bosons completely from the potential.
In our model and truncation this freedom is reflected by the arbitrariness of
$\tilde{\gamma}(k)$ in Eq. \eqref{equ::lambdaflow} and $\gamma$ in Eq. \eqref{equ::invariantbos}, respectively,
e.g. choosing $\gamma=0$ in Eq. \eqref{equ::invariantbos} yields $h_{A}=0$ and effectively removes
the axial vector bosons.

Physically, our findings in this section tell us that in the LPA without any kinetic terms
for the bosons
we simply cannot decide which type of boson will condense. To do this
we need additional information. Therefore, we will investigate
(simple) momentum dependent terms in the effective action in the next section.

\section{Going beyond the LPA}\label{sec::beyondlpa}
In Sect. \ref{sec::momentumdep} we have seen that a simple kinetic
term in the bosonized action gives a momentum dependent four-fermion interaction
in the purely fermionic language. Moreover, we could absorb this momentum dependent four-fermion interaction
into a boson only if we chose a certain Fierz transformation.
This provided us with the information to decide which FT is the ``right'' one.
Therefore, let us add the kinetic terms specified in Eq. \eqref{equ::kinetic} to
our pointlike action \eqref{equ::baction}.
Finally, let us impose for simplicity
one more restriction, $\alpha_{V}=\alpha_{A}=1$ on our effective action.
This simplifies the expressions for the vector-boson propagators, similar
to Feynman gauge in gauge theories.

Let us first outline the following complicated calculation,  using a
symbolic notation. In a next step we then discuss some points hidden in the notation and some
details of the employed approximations. Finally, we comment on a few properties and give
numerical results, while the whole set of equations and a more explicit and step-by-step
calculation is given in App. \ref{app::flowbeyond}.

\subsection{Adapting the Flow}
We restrict our truncation to lowest non-trivial order in $p^{2}$, and expand
all couplings up to this order. We write,
\begin{eqnarray}
\label{equ::def}
\mu^{2}(p)=\mu^{2}+p^2Z+{\mathcal{O}}(p^{4}),&&
\pt \mu^{2}(p)=\pt\mu^{2}+p^2\pt Z+\ldots,
\\\nonumber
h(p)=h+p^{2}h^{(2)}+{\mathcal{O}}(p^{4}),&&
\pt h(p)=\beta^{(0)}_{h}+p^2k^{-2}\beta^{(2)}_{h}+\ldots,
\\\nonumber
\lambda(p)=\lambda^{(0)}+p^{2}\lambda^{(2)}+{\mathcal{O}}(p^{4}),&&
\pt\lambda(p)=k^{-2}\beta^{(0)}_{\lambda}+p^2k^{-4}\beta^{(2)}_{\lambda}+\ldots.
\end{eqnarray}

At first sight, in the partially bosonized language it seems reasonable to take $\mu^{2}(p)$ as
in \eqref{equ::def} but restrict $h(p)=h$ and $\lambda(p)=0$.
However, in Sect. \ref{sec::solving}
we have seen that at least the latter is not a good approximation because the term $\lambda^{(0)}$ is crucial
for restoring the invariance under FT of the initial (pointlike) interaction.
Furthermore, we found that the flows of $\mu^{2}$, $h$ and $\lambda^{(0)}$ all contribute to the same order
to the effective four-fermion interaction (after integrating out the bosons).
It seems natural that this is also true
for the terms of order $p^{2}$: $Z$, $h^{(2)}$, $\lambda^{(2)}$.
Hence, we consider all these terms on equal footing.

Having chosen our truncation, we can calculate the flow equations. Since
we did not add higher powers of field operators, the remaining task is to evaluate
the diagrams depicted in Figs. \ref{fig::mass}, \ref{fig::vertex}, \ref{fig::box}.
The only difference to our previous calculations is that we have to consider non-trivial external momenta.

As in Sect. \ref{sec::solving} we want to keep the desired simple form of the effective
action, i.e. $h(p)=h$
and $\lambda(p)=0$, by choosing appropriate field redefinitions and
neglecting terms of order ${\mathcal{O}}(p^{4})$. To do so, we
shift
\begin{equation}
\label{equ::redefmom}
\partial_{t}\phi_{k}(q)=(\bar{\psi}\psi)(q)\partial_{t}\omega_{k}(q)+\partial_{t}\alpha_{k}(q)\phi_{k}(q),
\end{equation}
and, as in Sect. \ref{sec::solving}, we employ\footnote{As mentioned
in Sect. \ref{sec::solving}, after the appropriate modification of the cutoff,
the flow equation \eqref{equ::flow1} does not give
the exact flow $\partial_{t}\Gamma_{k}|$.
Since we now have a cutoff for the bosonic fields, the use of Eq. \eqref{equ::flow1} is really an approximation.
However, numerical tests performed in the appendix of \cite{Gies:2002nw} for a very similar case
suggest that it is a very good approximation. So we will use it without further comment.},
\begin{equation}
\label{equ::prinzi}
\partial_{t}\Gamma_{k}=\partial_{t}\Gamma_{k}|
+\int_{q}\frac{\delta\Gamma_{k}}{\delta\phi_{k}(q)}\partial_{t}\phi_{k}(q).
\end{equation}
This results in the following changes of the flow equations,
\begin{eqnarray}
\partial_{t}\mu^{2}(q)&=&\partial_{t}\mu^{2}(q)|+2\mu^{2}(q)\partial_{t}\alpha_{k}(q),
\\\nonumber
\partial_{t}h(q)&=&\partial_{t}h(q)|+h(q)\partial_{t}\alpha_{k}(q)+\mu^{2}(q)\partial_{t}\omega_{k}(q),\quad
\\\nonumber
\partial_{t}\lambda(q)&=&\partial_{t}\lambda(q)|-h(q)\partial_{t}\omega_{k}(q).
\end{eqnarray}
We can now use the freedom in choosing the field redefinitions to enforce,
\begin{equation}
\label{equ::enforce}
\partial_{t}\lambda(q)=0,\quad\partial_{t}h(q)=\partial_{t}h^{(0)}
,\quad\partial_{t}\mu^{2}(q)=\partial_{t}\mu^{2}+{\mathcal{O}}(p^4).
\end{equation}
Roughly speaking we absorb
the four-fermion interactions in the masses and Yukawa-couplings, and the momentum
dependence of the latter ones in the wave function renormalizations for the
bosonic fields ($\partial_{t}Z=0$).
In particular, we keep the simple form with a momentum independent Yukawa coupling
and no four-fermion interaction.

Using the fact that we start with a constant Yukawa coupling and $\lambda=0$ we can solve the
equations \eqref{equ::enforce} and find,
\begin{equation}
\eta\equiv2\partial_{t}\alpha(0)=-\frac{\partial_{t}Z|}{Z}+\frac{2\mu^2}{h}\left[\partial_{t}h^{(2)}|
+\mu^{2}\partial_{t}\lambda^{(2)}|+Z\partial_{t}\lambda^{(0)}|\right].
\end{equation}
We use this equation to define the anomalous dimension. From Eq. \eqref{equ::redefmom}
it is clear that $\partial_{t}\alpha(0)$ modifies the overall normalization of $\phi$,
hence the wave function renormalization.
Moreover, setting $\partial_{t}h^{(2)}|=0$, $\partial_{t}\lambda(q)|=0$,
it coincides with the original definition $\eta=-\frac{\partial_{t}Z}{Z}$.
Using this, we get pretty much the standard equations for the
flow of the mass and Yukawa coupling,
\begin{eqnarray}
\label{equ::modflow}
\partial_{t}\mu^2&=&\eta\mu^2+\partial_{t}\mu^2|,
\\\nonumber
\partial_{t}h&=&\frac{1}{2}\eta h+\partial_{t}h|+\frac{\mu^2}{h}\partial_{t}\lambda^{(0)}|.
\end{eqnarray}
We remark that together with the initial condition $Z=1$ the conditions \eqref{equ::enforce} automatically
guarantee that the couplings are renormalized.
\subsection{Choosing the Momentum Configurations}\label{sec::momconf}
So far everything seemed relatively straightforward. However, looking more
closely, we soon find that $\partial_{t}h(p)$ can actually depend
on two and $\partial_{t}\lambda(p)$ even on three momentum
variables. This is in contrast to $\mu^{2}(p)$ which depends only
on $p^2$. As we ultimately want to absorb those momentum dependencies in
$\partial_{t}\mu^{2}(p)$ it is clear that we have to make an approximation
such that $\partial_{t}h(p)$ and $\partial_{t}\lambda(p)$ depend
only on one momentum squared. To decide which of the
possible momenta to choose, we look at our example of Sect. \ref{sec::momentumdep}, in particular
at Eq. \eqref{equ::compboso}. From this we can read of, that the bosons
have the form,
\begin{equation}
\phi(q)=f(q)\int_{p_{1},p_{2}}\bar{\psi}(p_{1})\psi(-p_{2})\delta(q-p_{1}-p_{2}).
\end{equation}
Since we want to keep the simple form of the effective action, it is clear that the
bosons have to keep this form, too.
\begin{equation}
\label{equ::mostgeni}
\partial_{t}\phi_{k}(q)=\partial_{t}\alpha(q)\phi_{k}(q)+\partial_{t}\omega_{k}(q)
\int_{p_{1},p_{2}}\bar{\psi}(p_{1})\psi(-p_{2})\delta(q-p_{1}-p_{2}),
\end{equation}
is the most general form that accomplishes this. With Eq. \eqref{equ::mostgeni}
we can give meaning to our notation,
\begin{equation}
(\bar{\psi}\psi)(q)=\int_{p_{1},p_{2}}\bar{\psi}(p_{1})\psi(-p_{2})\delta(q-p_{1}-p_{2}).
\end{equation}
Inserting this into Eq. \eqref{equ::prinzi} we find that the most general
structures we can absorb are,
\begin{equation}
\label{equ::mostyukawa}
\int_{p_{1},p_{2},p_{3}}\!\!\!F(p^2_{1})\,\,\phi(-p_{1})\bar{\psi}(p_{2})\psi(-p_{3})\delta(p_{1}+p_{2}+p_{3}),
\end{equation}
\begin{equation}
\label{equ::mostfour}
\int_{p_{1},p_{2},p_{3},p_{4}}\!\!\!G((p_{1}+p_{2})^2)\,\,\bar{\psi}(-p_{1})\psi(p_{2})\bar{\psi}(p_{3})\psi(-p_{4})
\delta(p_{1}+p_{2}-p_{3}-p_{4}),
\end{equation}
where $F$ and $G$ are arbitrary functions which can then be expressed in terms of $\partial_{t}\alpha_{k}$
and $\partial_{t}\omega_{k}$.

Comparing the vertex correction $\sim \phi(-p_{1})\bar{\psi}(-p_{2})\psi(p_{3})$ depicted in Fig. \ref{fig::vertex}
with \eqref{equ::mostyukawa} it is clear that we have to restrict the momentum
dependence to $p_{1}$. A suitable
configuration for the evaluation then is, $(p_{1},p_{2},p_{3})=(p,\frac{1}{2}p,\frac{1}{2}p)$.

Recalling that a Fierz transformation for the four-fermion interaction exchanges $p_{2}$ and $-p_{3}$
we can absorb either a dependence on $s=(p_{1}+p_{2})^2$ or
one on \mbox{$t=(p_{1}-p_{3})^2$} (corresponding momentum configurations would
be e.g. $(p_{1},p_{2},p_{3},p_{4})=\frac{1}{2}(p,p,p,p)$ and
$(p_{1},p_{2},p_{3},p_{4})=\frac{1}{2}(p,-p,-p,p)$, respectively, \mbox{cf. Fig. \ref{fig::box}).}
Therefore, we have to ask which gives the better approximation.
In principle, we would have to calculate $\partial_{t}\lambda(s,t)$ (or even better
$\partial_{t}\lambda(p_{1},p_{2},p_{3},p_{4})$)
and test at every scale whether $\partial_{t}\lambda(s,t)\approx\partial_{t}\lambda(t)$ or
$\partial_{t}\lambda(s,t)\approx\partial_{t}\lambda(s)$ is a better approximation.
Analytically as well as numerically this is rather complicated. Therefore, we have adapted a much
simpler scheme: we have always absorbed the dependence on $t$, i.e.
we have evaluated the diagrams in Fig. \ref{fig::box} and Fierz transformed
the resulting interaction once. There are two reasons why we believe that this is a
reasonable approximation.
First of all, in the pointlike limit (at the beginning of the flow), i.e.
\begin{equation}
\label{equ::pointlimit}
\mu^{2}\rightarrow\infty, \quad h^{2}\rightarrow\infty,\quad \frac{h^2}{\mu^2}=\textrm{const},
\end{equation}
one finds $\partial_{t}\lambda(s,t)=\partial_{t}\lambda(t)$ exactly. Secondly, we have
checked for various combinations of $h$ and $\mu$ that in the vicinity of $(s,t)=(0,0)$ the dependence on
$t$ is usually (but not always) stronger than the dependence on $s$.

\subsection{Initial Flow and Numerical Results}
To get an impression of what we have achieved by all this let us take a look at the
effective four-fermion interaction at the beginning of the flow\footnote{Actually, in this section
we make one more
approximation: we have ignored the anomalous dimensions in the arguments of the threshold functions.
In the pointlike limit this neglects terms of order $\lambda^{3}$.
The flow equations in App. \ref{app::flowbeyond} include those terms, but they make the
evaluation much more difficult.}.

As we want to start with action \eqref{equ::faction} where the four-fermion interaction
is pointlike, the kinetic terms in the partially bosonized action vanish and
the renormalized couplings obey Eq. \eqref{equ::pointlimit} where the
constants are given by Eq. \eqref{equ::bosocouplings}.

In Sect. \ref{sec::momentumdep} we have already calculated,
\begin{equation}
\lambda(p)=\frac{h^2}{\mu^2+p^2}=\frac{h^2}{\mu^2}-\frac{h^2}{\mu^4}p^2+\cdots,
\end{equation}
where $\mu^2$ and $h^2$ are constants in momentum space.
Using our flow equations given in App. \ref{app::flowbeyond} and using the properties of the threshold
functions given in \ref{app::newthresh} we find (after some algebra) Eq. \eqref{equ::invariantbos}
for the flow of the $p^{0}$-terms.
For the flow of the $p^2$-term in $\lambda_{\sigma}$ we find,
\begin{equation}
\partial_{t}\left(\frac{h^2_{\sigma}}{\mu^4_{\sigma}}\right)
=16\vv \gamma^{(F),4}_{2}(0)(\bar{\lambda}_{\sigma}+\bar{\lambda}_{V})^2,
\end{equation}
and similar expressions for $\lambda_{V}$ and $\lambda_{A}$ which are
invariant under Fierz transformations, too.
This shows that at least at the beginning we have no Fierz ambiguity up to order $p^2$.

Now, let us come to the numerical results. Numerically it is impossible
to employ the pointlike limit exactly, therefore we have started with large values
of $\mu^2\sim 10^5$.
In addition to the results of the full set of flow equations we have
given some results for more simple approximations
in Tabs. \ref{tab::crit3}, \ref{tab::crit4}. The first approximation,
$(1)+(3)$, corresponds to the naive approach to the bosonized model, where all contributions
from the four-fermion interactions and the momentum dependence of
the Yukawa coupling are neglected. In the next step, $(1)-(3)$, we have included the box diagrams,
but only its constant parts, not the terms of order $p^2$. However, indirectly we have
included some knowledge of the momentum dependence as we have chosen the same FT as for
the truncation $(1)-(4)$ where we have included all terms of order $p^2$.

\begin{table}[!t]
\begin{center}
\scalebox{1.0}[1.0]{\begin{tabular}{|c|c|c|c|c|c|c|c}
  \hline
  Approximation & Chap.  & $\gamma=0.1$ & 0.25 & 0.5 & 0.75 & 0.9 \\
  \hline
  MFT &\ref{sec::mean} & 78.56 & 77.96 & 76.96 & 75.96 & 75.36 \\
  SD &\ref{sec::schwinger} &   76.96& 76.96 & 76.96 & 76.96 & 76.96 \\
  (1) & \ref{sec::bosoflow}&  76.32 & 76.36 & 76.42 & 76.49 & 75.54 \\
  Ferm. RG = (1) + (2) & \ref{sec::fermion}  & 86.15 & 86.15 & 86.15 & 86.15 & 86.15 \\
  \hline
  (1) + (3)  & \ref{sec::beyondlpa} & 76.43  & 76.45 &  76.49 &  76.55 & 76.58 \\
  (1) -- (3)  &\ref{sec::beyondlpa} & 83.97 & 83.95 & 83.92 & 83.89 & 83.87 \\
  (1) -- (4)  &\ref{sec::beyondlpa} & 86.17 & 86.18 & 86.20 & 86.21 & 86.22\\
  \hline
\end{tabular}}
\end{center}
\caption{Critical values $\bar{\lambda}^{\textrm{crit}}_{\sigma}$ for $\bar{\lambda}_{V}=2$ and for
various values of the unphysical
\mbox{parameter $\gamma$} (with $\Lambda=1$). To keep the table of manageable size we have abbreviated:
(1) the pointlike contibutions to the mass and the Yukawa coupling (Figs. \ref{fig::mass}, \ref{fig::vertex}),
(2) the pointlike contributions from the box diagrams (Fig. \ref{fig::box}),
(3) the contribution to the WFR from the purely bosonic diagram (Fig. \ref{fig::mass}) and
(4) the contribution to the WFR from the momentum dependence of the diagrams \ref{fig::vertex} and \ref{fig::box}.
We point out that differing from Tabs. \ref{tab::crit}, \ref{tab::crit2} we have employed
a UV regularization by the ERGE scheme
(cf. App. \ref{app::regularization}) with the linear cutoff Eq. \eqref{equ::cutoff} which
is better suited for numerical computations. This is why the
values for the critical coupling are roughly twice of
those given in Tabs. \ref{tab::crit}, \ref{tab::crit2}, since the critical
coupling is not a universal quantity, and therefore scheme dependent.}
\label{tab::crit3}
\end{table}

Moreover, we notice that the values for the critical coupling in the pointlike approximations are roughly twice of
those given in Tabs. \ref{tab::crit}, \ref{tab::crit2}. This is due to a difference
in the UV regularization. In this section we have employed the ERGE scheme described in App. \ref{app::regularization}.
Non-universal quantities can and do depend on the choice of UV regularization. In the
pointlike approximation this yields exactly a factor of two in our case (for the pair
of couplings $(\bar{\lambda}_{\sigma},\bar{\lambda}_{V})$). In the more
involved approximations this is not necessarily so, but the factor will still be somewhere around two.

\begin{table}[!t]
\begin{center}
\scalebox{1.0}[1.0]{\begin{tabular}{|c|c|c|c|c|c|c|c}
  \hline
  Approximation & Chap.  & $\gamma=0.1$ & 0.25 & 0.5 & 0.75 & 0.9 \\
  \hline
  MFT &\ref{sec::mean} & 74.96 & 68.96 & 58.96 & 48.96 & 42.96 \\
  SD &\ref{sec::schwinger} &   58.96 & 58.96 & 58.96 & 58.96 & 58.96 \\
  (1) & \ref{sec::bosoflow}& 53.16 & 52.93 & 53.32 & 54.64 & 55.88 \\
  Ferm. RG = (1) + (2)& \ref{sec::fermion}  & 58.83 & 58.83 & 58.83 & 58.83 & 58.83 \\
\hline
  (1) + (3) & \ref{sec::beyondlpa} & 53.89 &  53.66& 54.00 & 55.23  & 56.37 \\
  (1) -- (3)  &\ref{sec::beyondlpa} & 58.14 & 58.04 & 57.88 & 57.73 & 57.64 \\
  (1) -- (4)  &\ref{sec::beyondlpa} & 61.60 & 61.69 & 61.82 & 61.91 & 61.94 \\
\hline
\end{tabular}}
\end{center}
\caption{The same as in Tab. \ref{tab::crit3} but with $\bar{\lambda}_{V}=20$.}
\label{tab::crit4}
\end{table}

Aside from this, there is nothing new in the first four lines of Tabs. \ref{tab::crit3}, \ref{tab::crit4}.
Comparing the pointlike truncations for the RG with the improved approximations of this section we find
that the effect is of the order of $10\%$. Moreover, comparing the different non-pointlike
approximations we find that the differences between them are of the order of $5\%-10\%$, too.
While most of the Fierz ambiguity (more important for large values of $\bar{\lambda}_{V}$, Tab. \ref{tab::crit4})
is eliminated by including the pointlike contributions
of the boxes (in the ``right'' FT), $(2)$, for the absolute values of the critical coupling, the momentum dependence
of the Yukawa coupling and the box diagrams is not negligible.

On the more qualitative side we have checked for various values that for all truncations which include
kinetic terms (last three lines in the tables) and values of $\bar{\lambda}_{\sigma}$ slightly
larger than the critical $\bar{\lambda}^{\textrm{crit}}_{\sigma}$ only the renormalized scalar
boson mass turns negative, while the renormalized vector and axial vector boson masses remain positive.
This allows the conclusion that the scalar boson will condense first, and we
have a phase where only chiral symmetry is broken (at least in our approximation).

\chapter{Bosonic Effective Action (2PI)}\label{chap::bea}
In the last two chapters we have mainly worked on improving the RG description
in the partially bosonized language. It turned out, that it is necessary
to include a wave function renormalization (WFR) for the bosons. Without a WFR
we cannot determine the type of the bosonic
condensate (e.g. if it is a vector or scalar condensate).
Yet, consistent inclusion of a WFR leads to high algebraic complexity.
This might be appropriate for a quantitative description. However, if we
want to get a first, more qualitative, overview of a physical system this
seems to be a little bit excessive. For this purposes the SDE or MFT
approaches seem much more suitable.

Both methods allow for a computation of the order parameter in
systems which exhibit spontaneous symmetry breaking (SSB).
However, while the SDE approach leads directly to the gap equation
the MFT approach provides naturally a free-energy functional for
the bosonic composite degrees of freedom introduced by partial
bosonization via a Hubbard Stratonovich transformation (s. Chap. \ref{chap::boso1}).
The field
equation for this functional corresponds to the gap equation.
Knowledge of the free-energy functional becomes necessary if the
gap (or field) equation allows for solutions with different order
parameters and the free energy for the different solutions has to
be compared. The reconstruction of the free-energy functional from
the gap equation is not trivial and the method used in
\cite{Bowers:2002xr} for the case of color superconductivity may
not always work.

From this it seems that MFT is superior to SDE. Unfortunately,
as we have seen in Chaps. \ref{chap::njl}, \ref{chap::boso1},
it has a severe disadvantage: partial bosonization is not unique
and the results of the MFT calculation depend strongly on the
choice of the mean field. Moreover, MFT only includes
a subset of the SDE diagrams.

Hence, we want to find a functional which has the SDE
as its equation of motion, and which can be interpreted as a free energy.
Such a functional is given by the 2PI effective action
\cite{Baym:1961a,Baym:1961b,Cornwall:vz}.

In general, the 2PI effective action is a functional of
fields and propagators $\Gamma^{(2PI)}[\phi,G]$. However,
for a purely fermionic system, all the information
is already contained in $\Gamma^{(2PI)}[0,G]$.
$\Gamma^{(2PI)}[0,G]$ depends only on the bosonic variable $G$,
and therefore we will call it Bosonic Effective Action (BEA)
\cite{Wetterich:2002ky}.

In this chapter, we want to calculate a simple approximation of
the BEA for a general
local multi-fermion interaction. Already for a four-fermion interaction
the lowest non-trivial contribution to the BEA is of two-loop order.
For a general $n$-fermion interaction we have an $\frac{n}{2}$-loop structure.
However, we will show that this can be reduced
to a one-loop expression at the solution of the SDE, allowing for a comparison to MFT.

As an application we want to study an interaction resembling the six-fermion
interaction generated by instantons in the case of three flavors
and three colors
\cite{Callan:1977gz,Schafer:1996wv,Shuryak:1982hk,Shifman:uw,'tHooft:fv}.
In QCD this interaction is of special interest as it is $U(1)$-anomalous and solves the famous
$U(1)$-problem \cite{'tHooft:1986nc}.
In the simpler case of two flavors instantons mediate a four-fermion
interaction which has been investigated in works on chiral
symmetry breaking \cite{Diakonov:vw,Diakonov:1985eg,Carter:1999xb}
and color
superconductivity e.g. \cite{Berges:1998rc,Alford:1997zt}.

The effective interaction generated by the instantons does not
only lead to interactions between color singlet effective
quark-antiquark degrees of freedom ($\rightarrow$ chiral symmetry
breaking) but also between octets leading to the possibility of
octet condensation and spontaneous color symmetry breaking
\cite{Wetterich:2000ky,Wetterich:2000pp,Wetterich:1999vd}.
In the following we will consider both possibilities.

\section{Bosonic Effective Action (2PI)}\label{sec::bea}
To simplify the presentation we summarize all indices of the
fermionic field in $\tilde{\psi}_{\alpha}$. The index alpha contains
all internal indices (spin, color, flavor etc.) as well as
position or momentum. Furthermore
it also differentiates between $\psi$ and $\bar{\psi}$.

The partition function reads
\begin{equation}
Z[\eta,j]=\int{\mathcal{D}}\tilde{\psi}\exp(\eta_{\alpha}\tilde{\psi}_{\alpha}
+\frac{1}{2}j_{\alpha\beta}\tilde{\psi}_{\alpha}\tilde{\psi}_{\beta}-S_{\text{int}}[\tilde{\psi}])
\end{equation}
where we treat all quadratic terms as a bosonic source term.

We specify the interaction as
\begin{equation}
\label{equ::iniaction}
S_{\text{int}}[\tilde{\psi}]=\sum_{n}\frac{1}{n!}\lambda^{(n)}_{\alpha_{1}...\alpha_{n}}
\tilde{\psi}_{\alpha_{1}}\cdots\tilde{\psi}_{\alpha_{n}}.
\end{equation}
The usual generating functional of 1PI Greens functions in
presence of the bosonic sources $j$ is defined by a Legendre
transform with respect to the fermionic source term $\eta$:
\begin{equation}
\Gamma_{F}[\psi,j]=-W[\eta,j]+\eta_{\alpha}\psi_{\alpha}
\end{equation}
where
\begin{equation}
W=\ln Z[\eta,j], \quad
\psi_{\alpha}=\langle\tilde{\psi}_{\alpha}\rangle =\frac{\partial
W}{\partial\eta_{\alpha}}.
\end{equation}
$\Gamma_{F}$ can also be obtained by the following functional
integral\footnote{Note that in this formula $\tilde{\psi}$ is
shifted such that $\langle\tilde{\psi}\rangle=0$.}:
\begin{eqnarray}
\Gamma_{F}[\psi,j]&=&-\ln\int{\mathcal{D}}\tilde{\psi}\exp(\eta_{\alpha}\tilde{\psi}_{\alpha}-S_{j}[\tilde{\psi}+\psi]),
\\\nonumber
S_{j}[\tilde{\psi}]&=&-\frac{1}{2}j_{\alpha \beta}
\tilde{\psi}_{\alpha}\tilde{\psi}_{\beta}+S_{\text{int}}[\tilde{\psi}].
\end{eqnarray}
This form is especially useful to derive the SDE. Taking a
derivative with respect to $\psi$ we find
\begin{eqnarray}
\frac{\partial\Gamma_{F}}{\partial\psi_{\beta}}&=&-j_{\beta\alpha_{2}}\psi_{\alpha_{2}}
\\\nonumber
&&+\sum_{n}\frac{\lambda^{(n)}_{\beta\alpha_{2}\ldots\alpha_{n}}}{F_{n}}\psi_{\alpha_{2}}
\bigg\{
(\Gamma^{(2)}_{F})^{-1}_{\alpha_{3}\alpha_{4}}\cdots(\Gamma^{(2)}_{F})^{-1}_{\alpha_{n-1}\alpha_{n}}
+Z_{\alpha_{3}\ldots\alpha_{n}}+{\mathcal{O}}(\psi^{2})\bigg\}
\end{eqnarray}
where
\begin{equation}
F_{n}=(n-2)(n-4)\cdots 2
\end{equation}
and $Z$ summarizes all terms containing third and higher
derivatives of $\Gamma$. These are terms which have at least two
vertices. Taking another derivative with respect to
$\psi_{\alpha}$ and evaluating at $\psi=0$ we find the SDE:
\begin{eqnarray}
\label{equ::sdefermion}
(\Gamma^{(2)}_{F})_{\alpha\beta}&=&-j_{\alpha \beta}
+\sum_{n}\frac{\lambda^{(n)}_{\alpha\beta\alpha_{3}\ldots\alpha_{n}}}{F_{n}}
\bigg\{
(\Gamma^{(2)}_{F})^{-1}_{\alpha_{3}\alpha_{4}}\cdots(\Gamma^{(2)}_{F})^{-1}_{\alpha_{n-1}\alpha_{n}}
+Z^{\prime}_{\alpha_{3}\ldots\alpha_{n}}\bigg\}.
\end{eqnarray}
In this chapter we are only interested in the lowest order.
Therefore, from now on, we neglect $Z$, i.e. terms with at least
two vertices.

The ``Bosonic Effective Action'' (BEA) \cite{Wetterich:2002ky},
is defined by another Legendre transform with respect to $j$:
\begin{eqnarray}
\Gamma_{B}[G]&=&-W[0,j]+jG,
\\
\label{equ::grelation} G_{\alpha\beta}&=&\frac{\partial
W}{\partial
j_{\alpha\beta}}=(\Gamma^{(2)}_{F})^{-1}_{\alpha\beta}, \quad
\frac{\partial \Gamma_{B}}{\partial
G_{\alpha\beta}}=j_{\alpha\beta}.
\end{eqnarray}
Since $\Gamma_{F}$ is an even functional of $\psi$ the BEA
contains the same information as $\Gamma_{F}$. Indeed it is
related to $\Gamma_{F}$ by means of functional differential
equations like Eq. \eqref{equ::grelation}. Using this relation we
can conveniently write the SDE \eqref{equ::sdefermion} as
\begin{eqnarray}
\label{equ::sde} G^{-1}_{\alpha\beta}=-j_{\alpha \beta}
+\sum_{n}\frac{\lambda^{(n)}_{\alpha\beta\alpha_{3}\ldots\alpha_{n}}}{F_{n}}
G_{\alpha_{3}\alpha_{4}}\cdots G_{\alpha_{n-1}\alpha_{n}}.
\end{eqnarray}

Using \eqref{equ::grelation} we obtain a differential equation for
$\Gamma_{B}$
\begin{equation}
\frac{\partial\Gamma_{B}}{\partial G_{\alpha\beta}}=-G^{-1}_{\alpha\beta}
+\sum_{n}\frac{\lambda^{(n)}_{\alpha\beta\alpha_{3}\ldots\alpha_{n}}}{F_{n}}
G_{\alpha_{3}\alpha_{4}}\cdots G_{\alpha_{n-1}\alpha_{n}}.
\end{equation}
which we can integrate\footnote{Note that in our notation
$\frac{\partial G_{\alpha\beta}}{\partial
G_{\gamma\delta}}=\delta_{\alpha\gamma}\delta_{\beta\delta}
-\delta_{\alpha\delta}\delta_{\beta\gamma}$.} to obtain
\begin{equation}
\label{equ::1vertex} \Gamma_{B}=\frac{1}{2}\Tr\ln
G+\sum_{n}\frac{\lambda^{(n)}_{\alpha_{1}\ldots\alpha_{n}}}{nF_{n}}
G_{\alpha_{1}\alpha_{2}}\cdots G_{\alpha_{n-1}\alpha_{n}},
\end{equation}
the BEA at ``one-vertex order''.

It is sometimes convenient to introduce an auxiliary effective
action
\begin{equation}
\hat{\Gamma}[G,j]=\Gamma_{B}-\frac{1}{2}j_{\alpha\beta}G_{\alpha\beta}
\end{equation}
such that the physical propagator corresponds to the minimum of
$\hat{\Gamma}$ (cf. Eq. \eqref{equ::grelation}).
\section{BEA for Local Interactions}\label{sec::1vertex}
In the following we want to consider local interactions. For
clarity we now write $x$ (or momentum $p$) explicitly and use
latin letters for the remaining indices. The standard procedure
would be the insertion of the ansatz
$G^{-1}_{ab}(x,y)=-j_{ab}(x,y)+\Delta_{ab}(x)\delta(x-y)$ into Eq.
\eqref{equ::sde} to obtain the SDE for the local gap $\Delta$.
Since the BEA Eq. \eqref{equ::1vertex} is related to the SDE
\eqref{equ::sde} by differentiation with respect to $G$ it is not
clear that an effective action functional depending on $\Delta$
can be obtained by integration with respect to $\Delta$. Instead
we want to follow the construction presented in
\cite{Wetterich:2002ky} and start directly from the approximate BEA Eq.
\eqref{equ::1vertex}. With
\begin{equation}
\label{equ::localg} g_{ab}(x)=G_{ab}(x,x)
\end{equation}
we have
\begin{eqnarray}
\label{equ::localbea}
\hat{\Gamma}&=&\frac{1}{2}\Tr\ln G+\frac{1}{2}\Tr (G j)
+\int_{x}\sum_{n}\frac{\lambda^{(n)}_{a_{1}\ldots
a_{n}}}{nF_{n}} g_{a_{1}a_{2}}(x)\cdots g_{a_{n-1}a_{n}}(x).
\end{eqnarray}
For this relation it is essential that the interaction is strictly
local. Furthermore, we can use the locality of the interaction to
write \eqref{equ::sde} in the form of a local gap equation
\begin{eqnarray}
\label{equ::ansatz}
G^{-1}_{ab}(x,y)&=&-j_{ab}(x,y)+\Delta_{ab}(x)\delta(x-y).
\end{eqnarray}
We will evaluate the functional $\Gamma[G]$ for $G_{\alpha\beta}$
corresponding to Eq. \eqref{equ::ansatz}. This is actually a
restriction to a subspace of all possible $G$. However, locality
tells us that the extremum (solution of the SDE) is contained in
this subspace.

Using $j=-G^{-1}+\Delta$ we find (up to a shift in the irrelevant
constant and using $\Delta_{ab}(x,y)=\Delta_{ab}(x)\delta(x-y)$)
\begin{eqnarray}
\label{equ::vorstufe} \nonumber
\hat{\Gamma}[g,\Delta]\!\!&=&\!\!-\frac{1}{2}\Tr\ln (-j+\Delta)
-\frac{1}{2}\int_{x}\Delta_{ab}(x)g_{ab}(x)
\\&&\!\!
+\int_{x}\sum_{n}\frac{\lambda^{(n)}_{a_{1}\ldots
a_{n}}}{nF_{n}}g_{a_{1}a_{2}}(x)\cdots g_{a_{n-1}a_{n}}(x).
\end{eqnarray}
For the search of extrema of $\hat{\Gamma}$ it is actually
convenient to treat $\Delta$ and $g$ as independent variables. The
extremum of $\hat{\Gamma}[g,\Delta]$ then obeys
\begin{equation}
\label{equ::extremum}
\frac{\partial\hat{\Gamma}[g,\Delta]}{\partial\Delta}=0,\quad
\frac{\partial\hat{\Gamma}[g,\Delta]}{\partial g}=0.
\end{equation}
Evaluating the derivative with respect to $\Delta$ we recover the
inverse of Eq. \eqref{equ::ansatz} for $x=y$,
\begin{equation}
g_{ab}(x)=(-j+\Delta)^{-1}_{ab}(x,x)=g[\Delta(x)]
\end{equation}
Inserting this functional relation into Eq. \eqref{equ::sde} leads
to a gap equation for $\Delta$. In case of a six-fermion
interaction this takes, however, the form of a two-loop equation.

For n-fermion interactions with $n>4$ it is more appropriate to go
the other way around and first take a derivative with respect to
$g$. We obtain
\begin{eqnarray}
\nonumber
\Delta_{ab}(x)&=&\sum_{n}\frac{\lambda^{(n)}_{aba_{3}\ldots
a_{n}}}{F_{n}}g_{a_{3}a_{4}}(x)\cdots g_{a_{n-1}a_{n}}(x)
\\
\label{equ::equation} &=&\Delta_{ab}[g(x)],
\end{eqnarray}
which is precisely the value of the gap in Eq. \eqref{equ::sde}.
Inserting $\Delta[g]$ into \eqref{equ::vorstufe} we find the
effective action depending on $g$
\begin{eqnarray}
\label{equ::main} \nonumber \hat{\Gamma}[g]&=&-\frac{1}{2}\Tr\ln
(-j+\Delta[g]) -\frac{1}{2}\int_{x}\Delta_{ab}[g](x)g_{ab}(x)
\\
&&+\int_{x}\sum_{n}\frac{\lambda^{(n)}_{a_{1}\ldots
a_{n}}}{nF_{n}}g_{a_{1}a_{2}}(x)\cdots g_{a_{n-1}a_{n}}(x).
\end{eqnarray}
Searching for an extremum yields
\begin{equation}
\label{equ::mot} \frac{\partial{\hat{\Gamma}[g]}}{\partial
g}=\left\{(-j+\Delta[g])^{-1} -g\right\}\frac{d\Delta[g]}{d g}=0.
\end{equation}
For $\frac{d\Delta}{dg}\neq 0$ Eq. \eqref{equ::mot} indeed
corresponds to the SDE \eqref{equ::sde}, i.e.
\begin{equation}
g_{ab}(x)=(-j+\Delta[g])^{-1}_{ab}(x,x)
\end{equation}
This will be our central gap equation. We should point out that
possible extrema of $\hat{\Gamma}[g]$ corresponding to
$\frac{d\Delta}{dg}=0$ are not solutions of the gap equation
\eqref{equ::sde} and should be discarded. Finally, we also have
\begin{eqnarray}
\nonumber \frac{d \hat{\Gamma}[g]}{d
g}&=&\frac{d\hat{\Gamma}[g,\Delta[g]]}{d g}
=\frac{\partial\hat{\Gamma}[g,\Delta[g]]}{\partial
g}+\frac{\partial \hat{\Gamma}[g,\Delta[g]]}{\partial\Delta}
\frac{d \Delta[g]}{d g}
\\
&=&\frac{\partial \hat{\Gamma}[g,\Delta[g]]}{\partial\Delta}
\frac{d \Delta[g]}{d g}.
\end{eqnarray}
Only as long as $\frac{d\Delta[g]}{d g}\neq0$ is fulfilled we can
conclude that a solution of \eqref{equ::mot} fulfills both
extremum conditions \eqref{equ::extremum}.

Our procedure is quite powerful if $\Tr\ln(-j+\Delta)$ can be
explicitly evaluated as a functional of $\Delta$. Then
$\hat{\Gamma}[g]$ allows not only a search for the extremum
(discarding those with $\frac{d\Delta}{dg}=0$) but also a simple
direct comparison of the relative free energy of different local
extrema. This is crucial for the determination of the ground state
in the case of several ``competing gaps''.

This ``one-loop'' form of the equation of motion but also of the
effective action itself \eqref{equ::main} is very close to what we
would expect from MFT (cf. also the next section). In contrast to
the standard SDE, which is an equation of motion, we can use Eq.
\eqref{equ::main} to compare the values for the effective action
at different solutions of the equation of motion \eqref{equ::mot},
providing us with information about the stability.

Nevertheless, we have to be careful when considering
Eq. \eqref{equ::main} at points which are not solutions of Eq.
\eqref{equ::mot}. Going step by step through the procedure above,
we find that if we are not at a solution of \eqref{equ::mot} we do
not necessarily fulfill the ansatz \eqref{equ::ansatz}. Therefore,
at these points we are mathematically not allowed to insert the
ansatz into Eq. \eqref{equ::1vertex}. So, strictly speaking
\eqref{equ::main} only gives the value of the effective action at
the solution of the equation of motion\footnote{An alternative
would be to choose the gap $\Delta$ as the ``bosonic field''. Inserting Eq. \eqref{equ::ansatz}
into Eq. \eqref{equ::1vertex} we could calculate a functional $\Gamma[\Delta]$.
However, as one can check there are two drawbacks. First, even for four-fermion
interactions, $\Gamma[\Delta]$ is usually unbounded from below when considering
$\Delta\rightarrow\infty$. Second, in the case of a large four-fermion coupling the
``stable'' solution of the field equation is usually a local maximum.}. Although, this is already
more than we get from the standard SDE we would like to interpret
\eqref{equ::main} as a reasonable approximation in a small
neighborhood of the solution to the equation of motion.
Remembering $g(x)=\langle\tilde{\psi}(x)\tilde{\psi}(x)\rangle$ it
is suggestive to interpret $g$ as a bosonic field. Eq.
\eqref{equ::equation} gives the (non-linear) ``Yukawa coupling'' of
$g$ to the fermions, i.e. the relation between the gap and the bosonic field.
The $\Tr\ln$ is the contribution from the
fermionic loop in a background field $g$. The remaining terms can
then be interpreted as the cost in energy to generate the
background field $g$. This interpretation allows us to use
\eqref{equ::main} to calculate the mass and the couplings of the
bosonic field $g$.
\section{Comparison with MFT}\label{sec::mft}
From Chap. \ref{chap::boso1} we know that partial bosonization is not restricted to
four-fermion interactions. In particular Eq. \eqref{equ::higherorder} provides
us with the means to calculate a partially bosonized action for an arbitrary
local multi-fermion interaction. Associating
$\phi_{\alpha\beta}(x)=\langle\tilde{\psi}_{\alpha}(x)\tilde{\psi}_{\beta}(x)\rangle$,
the partially bosonized form of Eq. \eqref{equ::iniaction} becomes
\begin{eqnarray}
S_{\text{int}}[\phi,\tilde{\psi}]
&=&\int_{x}\sum_{n}m^{(n)}_{a_{1}b_{1}\ldots
a_{\frac{n}{2}}b_{\frac{n}{2}}} \bigg\{\phi_{a_{1}b_{1}}(x)
\cdots\phi_{a_{\frac{n}{2}}b_{\frac{n}{2}}}(x)
\\\nonumber
&&\!\!\!\!\!\!\!\!\!\!\!\!\!\!\!\!\!\!\!\!\!\!\!\!+\frac{1}{(\frac{n}{2}-1)!}\left[\psi_{a_{1}}(x)\psi_{b_{1}}(x)\phi_{a_{2}b_{2}}(x)
\cdots\phi_{a_{\frac{n}{2}}b_{\frac{n}{2}}}(x)+ \text{perm.}\,\,
\big\{1,\,\ldots\,,\frac{n}{2}\big\}\right]
+{\mathcal{O}}(\tilde{\psi}^{3}) \bigg\},
\end{eqnarray}
\begin{equation}
\label{equ::cond} m^{(n)}_{a_{1}\ldots
a_{n}}=-\frac{\lambda^{(n)}_{a_{1}\ldots a_{n}}}{n!}
+\Sigma_{a_{1}\ldots a_{n}}.
\end{equation}
Where $\Sigma$ is a sum of terms which are symmetric in at least one
pair of indices (cf. Sect. \ref{subsec::fierz}). The condition \eqref{equ::cond} ensures that the
partially bosonized action is equivalent to the original
fermionic one.

Neglecting the terms ${\mathcal{O}}(\tilde{\psi}^{3})$ and
performing the functional integral over the fermions provides us
with the MF effective action:
\begin{eqnarray}
\label{equ::mft}
\Gamma^{\text{MF}}[\phi]&=&-\frac{1}{2}\Tr\ln(-j+g[\phi])
+\int_{x}\sum_{n}m^{(n)}_{a_{1}\ldots
b_{\frac{n}{2}}}\phi_{a_{1}b_{1}}
\cdots\phi_{a_{\frac{n}{2}}b_{\frac{n}{2}}},
\\\nonumber
g[\phi]_{a_{1}b_{1}}&=&\frac{1}{(\frac{n}{2}-1)!}\left[m^{(n)}_{a_{1}\ldots
b_{\frac{n}{2}}}\phi_{a_{2}b_{2}}
\cdots\phi_{a_{\frac{n}{2}}b_{\frac{n}{2}}}+ \text{perm.}\,\,
\big\{1,\,\ldots\,,\frac{n}{2}\big\}\right].
\end{eqnarray}
By construction this is a one-loop result.
Moreover, it is strikingly similar to Eq. \eqref{equ::main}. However,
the coefficients differ. As discussed in Sect. \ref{subsec::fierz} the Fierz ambiguity is
reflected by the presence of a nearly arbitrary $\Sigma$
in Eq. \eqref{equ::cond}. Results usually depend on the
choice of $\Sigma$. Of course, further considerations as e.g. the
stability of the initial bosonic potential might reduce the
freedom of $\Sigma$ somewhat. But, sometimes this is not even enough
to get qualitatively unambiguous results \cite{Baier:2000yc,tobidoktor}.

Eqs. \eqref{equ::vorstufe}, \eqref{equ::main}, \eqref{equ::mot} do
not suffer from such an ambiguity since in the derivation of the
SDE \eqref{equ::sdefermion} the coefficients become antisymmetrized
and symmetric terms drop out. In Sect. \ref{sec::gapboso} we demonstrated
that the inclusions of certain diagrams cures the Fierz
ambiguity for four-fermion interactions and leads to the
SD result. We believe that this holds for higher fermion
interactions, too. Thus, we propose \eqref{equ::main} as a natural
generalization of \eqref{equ::mft}.

Finally, let us stress the similarity of both approaches
by noting that $g(x)=\langle
\tilde{\psi}(x)\tilde{\psi}(x)\rangle$ is exactly what we
had in mind as a ``mean field''.

\section{Wave Function Renormalization}\label{sec::wave}
In the partially bosonized language (cf. sect. \ref{sec::mft}) it
is possible to calculate a wave function renormalization for the
bosons. Allowing not only for constant but also for a slightly
varying $\phi$ ($p$ small)
\begin{equation}
\phi(x)=\phi(0)+\delta\phi\exp(ipx).
\end{equation}
Using this $\phi$ it is still possible to perform the fermionic
integral. Expanding in powers of the momentum up to the $p^2$-term
we can read off the wave function renormalization.

The same can be done for Eq. \eqref{equ::vorstufe}
\begin{equation}
g(x)=g(0)+\delta g\exp(ipx).
\end{equation}
Expanding again in powers of momentum we interpret the $p^2$-term as
the wave function renormalization for the boson corresponding to
the field $g$. As in MFT the only contribution to the wave
function renormalization comes from the $\Tr\ln$ and therefore
from a simple one-loop expression.

Knowledge of the wave function renormalization together with the
second derivative of the effective action for constant fields
allows us to compute the mass of the boson.

Again, this calculation is unambiguous. This is in contrast to a
calculation in the partially bosonized language where we
again have the problems with the Fierz ambiguity. On the other hand we do not
want to hide the fact that by considering values of $g$ which do not coincide
with the solution of the SDE we have left the solid ground of a direct computation
from the BEA given in Eq. \eqref{equ::localbea}.

\section{Chiral Symmetry Breaking from a 3-Flavor Instanton Interaction} \label{sec::instanton1}
In this section we want to use the method described above to study
chiral symmetry breaking in an NJL-type model with a six-fermion
interaction modelling the QCD-instanton interaction with three
colors and three flavors
\cite{Shuryak:1982hk,Shifman:uw,Callan:1977gz,'tHooft:fv,Schafer:1996wv}.
The three flavor instanton vertex can be written in the following
convenient form \cite{Wetterich:2000ky}
\begin{eqnarray}
\label{equ::instinteraction} \nonumber
S_{\text{inst}}[\psi]=
-\frac{\zeta}{6}&&\!\!\!\!\!\!\!\!\!\int_{x}\epsilon_{a_{1}a_{2}a_{3}}\epsilon_{b_{1}b_{2}b_{3}}
\\\nonumber
\bigg\{&&\!\!\!\!\!\!\!\!\! \bigg[(\bar{\psi}^{a_{1}}_{L}\psi^{b_{1}}_{R})
(\bar{\psi}^{a_{2}}_{L}\psi^{b_{2}}_{R})
(\bar{\psi}^{a_{3}}_{L}\psi^{b_{3}}_{R})
-\frac{1}{8}
(\bar{\psi}^{a_{1}}_{L}\lambda^{z}\psi^{b_{1}}_{R})
(\bar{\psi}^{a_{2}}_{L}\lambda^{z}\psi^{b_{2}}_{R})
(\bar{\psi}^{a_{3}}_{L}\psi^{b_{3}}_{R})
\\\nonumber
-&&\!\!\!\!\!\!\!\!\!\frac{1}{8}
(\bar{\psi}^{a_{1}}_{L}\lambda^{z}\psi^{b_{1}}_{R})
(\bar{\psi}^{a_{2}}_{L}\psi^{b_{2}}_{R})
(\bar{\psi}^{a_{3}}_{L}\lambda^{z}\psi^{b_{3}}_{R})
-\frac{1}{8} (\bar{\psi}^{a_{1}}_{L}\psi^{b_{1}}_{R})
(\bar{\psi}^{a_{2}}_{L}\lambda^{z}\psi^{b_{2}}_{R})
(\bar{\psi}^{a_{3}}_{L}\lambda^{z}\psi^{b_{3}}_{R})\bigg]
\\
&&\quad\quad-(R\leftrightarrow L)\bigg\}
\end{eqnarray}
where $\lambda^{z}$ are the Gell-Mann matrices corresponding to
the $SU(3)_{\text{c}}$ color group and the brackets $(\quad)$
indicate contractions over color and spinor indices.

The coupling constant $\zeta$ can be calculated in terms of the
gauge coupling. However, it involves an IR divergent integral over
the instanton size. Therefore, one needs to provide a physical
cutoff mechanism. To avoid this difficulty we treat $\zeta$ as a
parameter.

Inspection of \eqref{equ::instinteraction} tells us that this
interaction is $U(1)$ anomalous with a residual
${\mathcal{Z}}_{3}$-symmetry. This is important because we cannot
restrict ourselves to real condensates from the start.

In order to extract the interaction matrix $\lambda$ we have to antisymmetrize over
flavor indices $(a=1\ldots3)$, color indices $(i=1\ldots3)$
Weyl spinor indices $(\alpha=1,2)$, chirality indices
$(\chi=1,2=L,R)$ and the indices distinguishing between $\psi$ and
$\bar{\psi}$ $(s=1,2)$.
\begin{eqnarray}
\lambda_{m_{1}\ldots m_{6}}&=&P\{\tilde{\lambda}_{m_{1}\ldots
m_{6}}\},
\\\nonumber
\tilde{\lambda}_{m_{1}\ldots m_{6}}&=&
\frac{\zeta}{12}\epsilon_{a_{1}a_{2}a_{3}}\epsilon_{a_{4}a_{5}a_{6}}
\delta_{\alpha_{1}\alpha_{4}}\delta_{\alpha_{2}\alpha_{5}}\delta_{\alpha_{3}\alpha_{6}}
\\\nonumber
&&\times[
\delta_{i_{1}i_{4}}\delta_{i_{2}i_{5}}\delta_{i_{3}i_{6}}
-\frac{1}{8}\lambda^{z}_{i_{1}i_{4}}\lambda^{z}_{i_{2}i_{5}}\delta_{i_{3}i_{6}}
-\frac{1}{8}\lambda^{z}_{i_{1}i_{4}}\delta_{i_{2}i_{5}}\lambda^{z}_{i_{3}i_{6}}
-\frac{1}{8}\delta_{i_{1}i_{4}}\lambda^{z}_{i_{2}i_{5}}\lambda^{z}_{i_{3}i_{6}}]
\\\nonumber
&&\times[\delta_{\chi_{1}1}\delta_{\chi_{2}1}\delta_{\chi_{3}1}
\delta_{\chi_{4}2}\delta_{\chi_{5}2}\delta_{\chi_{6}2}
-\delta_{\chi_{1}2}\delta_{\chi_{2}2}\delta_{\chi_{3}2}
\delta_{\chi_{4}1}\delta_{\chi_{5}1}\delta_{\chi_{6}1}]
\\\nonumber
&&\times[
\delta_{s_{1}2}\delta_{s_{2}2}\delta_{s_{3}2}\delta_{s_{4}1}\delta_{s_{5}1}\delta_{s_{6}1}
+\delta_{s_{1}1}\delta_{s_{2}1}\delta_{s_{3}1}\delta_{s_{4}2}\delta_{s_{5}2}\delta_{s_{6}2}].
\end{eqnarray}
Here $P$ denotes the sum over all $6!$ permutations of the
multiindices $m_{j}=(a_{j},i_{j},\alpha_{j},\chi_{j},s_{j})$,
$j=1\ldots 6$, with minus signs appropriate for total
antisymmetrization.

As a first example we consider a flavor singlet, color singlet
scalar chiral bilinear
($\sigma=\frac{1}{3}\bar{\psi}^{a}_{L}\psi^{a}_{R}$,
\mbox{$\sigma^{\star}=-\frac{1}{3}\bar{\psi}^{a}_{R}\psi^{a}_{L}$})
\begin{eqnarray}
g_{mn}&=&g_{ai\alpha\chi s,bj\beta\tau t}
\\\nonumber
&=&\frac{1}{6}\delta_{ab}\delta_{ij}\delta_{\alpha\beta}
\bigg[\sigma(\delta_{\chi1}\delta_{\tau2}\delta_{s2}\delta_{t1}-\delta_{\chi2}\delta_{\tau1}\delta_{s1}\delta_{t2})
-\sigma^{\star}(\delta_{\chi2}\delta_{\tau1}\delta_{s2}\delta_{t1}
-\delta_{\chi1}\delta_{\tau 2}\delta_{s1}\delta_{t2})\bigg].
\end{eqnarray}
We evaluate
\begin{eqnarray}
\label{equ::delta}
\Delta[g]_{mn}&=&-\frac{\lambda^{(n)}_{mm_{2}m_{3}nm_{5}m_{6}}}{8}g_{m_{2}m_{5}}g_{m_{3}m_{6}}
\\\nonumber
&=&-6\bigg[\tilde{\lambda}^{(n)}_{mm_{2}m_{3}nm_{5}m_{6}}-\tilde{\lambda}^{(n)}_{mm_{2}m_{3}m_{5}nm_{6}}
+\tilde{\lambda}^{(n)}_{mm_{2}m_{3}m_{5}m_{6}n}\bigg]
\\\nonumber
&&\quad\quad\quad\quad\quad\quad\quad\quad\quad\quad\quad\quad
\times\bigg[g_{m_{2}m_{5}}g_{m_{3}m_{6}}-g_{m_{2}m_{6}}g_{m_{3}m_{5}}\bigg]
\end{eqnarray}
and
\begin{eqnarray}
\nonumber U&=&-\frac{1}{2}\Delta[g]_{mn}g_{mn}
+\frac{\lambda_{m_{1}\ldots
m_{6}}}{48}g_{m_{1}m_{2}}g_{m_{3}m_{4}}g_{m_{5}m_{6}}
\\\nonumber
&=&-\frac{1}{24}\lambda_{m_{1}\ldots
m_{6}}g_{m_{1}m_{2}}g_{m_{3}m_{4}}g_{m_{5}}g_{m_{6}}
\\
&=&-\frac{1}{3}\Delta[g]_{mn}g_{mn},
\end{eqnarray}
where we have used the fact that $\tilde{\lambda}$ is symmetric
under permutations of the three $\bar{\psi}\psi$ bilinears.
Exploiting the flavor, spin and color structure for
\eqref{equ::delta} yields
\begin{eqnarray}
\nonumber
\Delta[g]_{mn}&=&-\frac{10}{9}\zeta\delta_{ab}\delta_{ij}\delta_{\alpha\beta}
\bigg[\sigma^{2}(\delta_{\chi1}\delta_{\tau2}\delta_{s2}\delta_{t1}
-\delta_{\chi2}\delta_{\tau1}\delta_{s1}\delta_{t2})
\\
&&\quad\quad-\sigma^{\star
2}(\delta_{\chi2}\delta_{\tau1}\delta_{s2}\delta_{t1}
-\delta_{\chi 1}\delta_{\tau 2}\delta_{s1}\delta_{\tau2})\bigg]
\end{eqnarray}
and
\begin{eqnarray}
U[\sigma]=\frac{20}{9}\zeta(\sigma^{3}+\sigma^{\star 3}).
\end{eqnarray}
Evaluating for a $\sigma$ constant in space and pulling out a
volume factor we obtain the effective potential and the relation
between condensate and fermion mass,
\begin{eqnarray}
\nonumber \label{equ::potentialsigma}
\hat{\Gamma}[\sigma]&=&-36v_{4}\int dx \,\,x
[\ln(x+|m|^2_{\sigma})] +U[\sigma],
\\
m_{\sigma}&=&\frac{10}{9}\zeta \sigma^2+m^{0}_{\sigma}.
\end{eqnarray}
Here $m^{0}_{\sigma}$ is a current quark mass which we take to be
equal for all quarks. The integral in \eqref{equ::potentialsigma}
is, of course, divergent. Our UV regularization is simply to cut
it off at $\Lambda^2$. Measuring all quantities in units of
$\Lambda$ we can put $\Lambda=1$.
\subsection{The chiral limit $\msigma=0$}
Let us now look at the field equation or equivalently search for
extrema of $\hat{\Gamma}[\sigma]$. Since
$\frac{d\Delta[\sigma]}{d\sigma}\neq 0$ for all $\sigma\neq 0$ we
do not need to worry for non-trivial solutions to be spurious. In
addition $\sigma=0$ is always a solution in the chiral limit.

Inspection of $\hat{\Gamma}[\sigma]$ tells us that it is invariant
under the combined operation $\zeta\rightarrow-\zeta$,
$\sigma\rightarrow-\sigma$. This allows us to restrict our
analysis to positive $\zeta$.

In the chiral limit it is useful to parametrize
\begin{equation}
\sigma=|\sigma|\exp(i\alpha).
\end{equation}
Form this one finds
\begin{equation}
\hat{\Gamma}[|\sigma|,\alpha]=\frac{40}{9}\zeta|\sigma|^{3}\cos(3\alpha)+f(|\sigma|),
\end{equation}
where $f$ is a function determined by the integral in Eq.
\eqref{equ::potentialsigma}. We can see that the only $\alpha$-dependence
comes from $\cos(3\alpha)$ which is the explicit manifestation of the
${\mathcal{Z}}_{3}$-symmetry.

It is clear that extrema can only occur at
$\alpha=\frac{n\pi}{3}$, $n\in\mathbb{Z}$. Using the
${\mathcal{Z}}_{3}$-symmetry we can restrict ourselves to
$\alpha=0,\pi$ or restrict ourselves simply
to real $\sigma$.

\begin{figure}[t]
\begin{center}
\scalebox{0.8}[0.8]{
\begin{picture}(350,185)
\Text(70,147)[c]{\scalebox{1.5}[1.5]{$\Gamma[\sigma]$}}
\Text(249,35)[c]{\scalebox{1.5}[1.5]{$\sigma$}}
\includegraphics{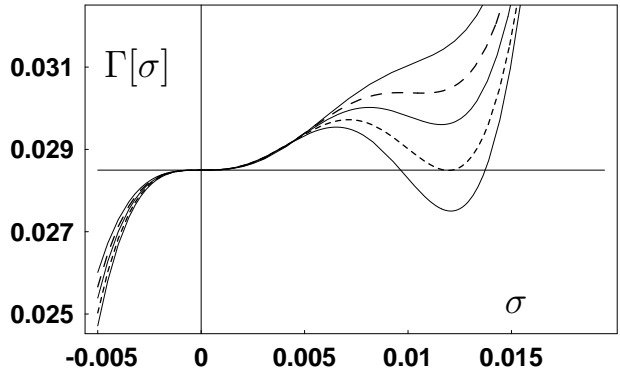}
\end{picture}
}
\end{center}
\caption{Plot of the BEA for various values of the coupling constant increasing from the topmost line
$\zeta=3000$ to the lowest line $\zeta=4200$. The second line (long dashed) is for
$\zeta_{\text{crit}}\approx 3350$, the critical coupling for the onset of non-vanishing solutions.
The third is for $\zeta=3600$ while the fourth (short dashed) is the for the onset
of SSB $\tilde{\zeta}_{\text{SSB}}\approx 3900$. The horizontal line indicates the value
of $\Gamma[\sigma]$ at the trivial solution $\sigma=0$.} \label{fig::beapotential}
\end{figure}

\begin{figure}[t]
\begin{center}
\scalebox{0.8}[0.8]{
\begin{picture}(350,185)
\Text(60,160)[c]{\scalebox{1.8}[1.8]{$|m_{\scalebox{0.5}[0.5]{$\sigma$}}|$}}
\Text(122,40)[c]{\scalebox{1.8}[1.8]{$\zeta_{\scalebox{0.5}[0.5]{\text{crit}}}$}}
\Text(208,40)[c]{\scalebox{1.8}[1.8]{$\zeta_{\scalebox{0.5}[0.5]{\text{SSB}}}$}}
\Text(155,53)[c]{\scalebox{1.8}[1.8]{$\tilde{\zeta}_{\scalebox{0.5}[0.5]{\text{SSB}}}$}}
\Text(260,40)[c]{\scalebox{1.8}[1.8]{$\zeta$}} \put(124.5,32.5)
{\vector(1,-1){10}} \put(198,32.5)  {\vector(-1,-1){10}}
\put(153,42.5) {\vector(1,-2){10}}
\includegraphics{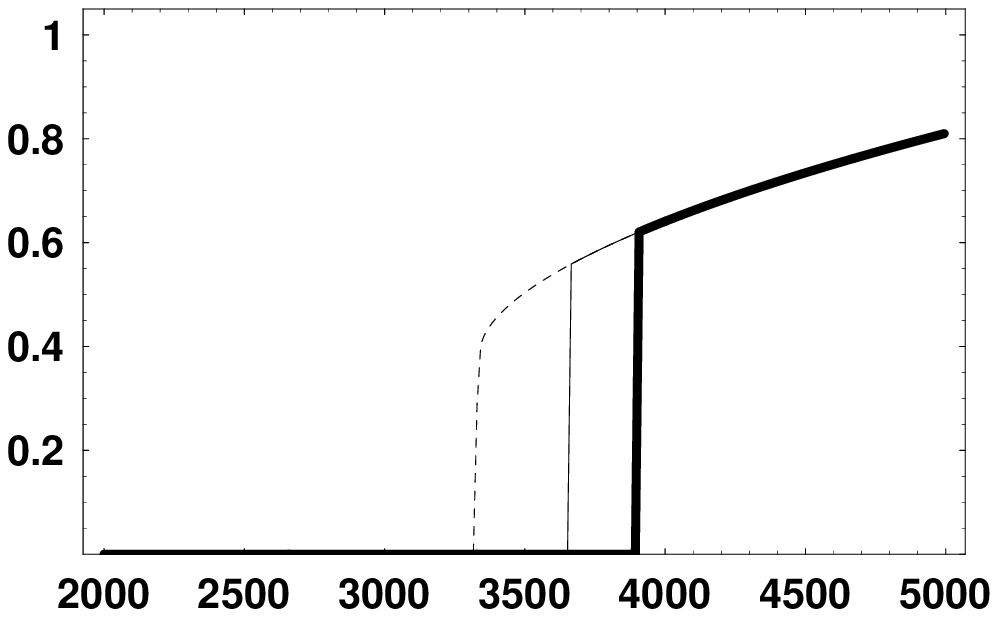}
\end{picture}
}
\end{center}
\caption{Fermion mass $|m|_{\sigma}$ (gap) versus the strength of
the six-fermion interaction $\zeta$ in the chiral limit
$\msigma=0$. The thick line corresponds to the solution with
smallest action. The dashed line is the largest non-trivial
solution. Finally, the thin line is obtained by minimizing an
``effective action'' $\tilde{\Gamma}[\Delta]$ obtained by direct
integration of the SDE with respect to the gap $\Delta$. We find
three special couplings, $\zeta_{\text{crit}}$ for the onset of
non-trivial solutions to the SDE, $\zeta_{\text{SSB}}$ for the
onset of SSB, i.e. a non-trivial solution has lower action than
the trivial solution and $\tilde{\zeta}_{\text{SSB}}$ where the
lowest extremum of $\tilde{\Gamma}[\Delta]$ becomes non-trivial.
We point out that these three ``critical'' couplings differ.
Moreover, to calculate $\zeta_{\text{SSB}}$ we need to know the
effective action. In our approximation we obtain a
first order phase transition.} \label{fig::firstorder}
\end{figure}

Taking all this into account we find up to three solutions (cf. Fig. \ref{fig::beapotential}).
As already mentioned $\sigma=0$ is a solution for all values of the
coupling. Going to larger couplings we encounter a point
$\zeta_{\text{crit}}$ where we have two solutions. For even larger
couplings there are three solutions
$0=\sigma_{0}<\sigma_{1}\leq\sigma_{2}$. We know that
$\hat{\Gamma}[\sigma_{1}]>\hat{\Gamma}[\sigma_{0}=0]$ therefore
$\sigma_{1}$ is not the stable solution. As can be seen from Figs.
\ref{fig::beapotential}, \ref{fig::firstorder} there is a
$\zeta_{\text{crit}}\leq\zeta\leq\zeta_{\text{SSB}}$ where there
exist non-trivial solutions to the SDE but there is still no SSB
because $\hat{\Gamma}[\sigma_{2}]\geq\hat{\Gamma}[\sigma_{0}=0]$.
We point out that in order to calculate $\zeta_{\text{SSB}}$ we
need to know the value of $\hat{\Gamma}$, i.e. information beyond
the SDE.

In Fig. \ref{fig::firstorder} we have plotted the mass gap versus
the six-fermion coupling strength.
Looking at Fig. \ref{fig::firstorder} we observe a first order
phase transition. From Eq. \eqref{equ::main} we actually expect
this quite generically as long as we have only $n$-fermion
interactions with $n\geq 6$. However, this might also be an
artifact of the ``one-vertex'' approximation.

Finally, we would like to remark that in general it is not enough
to simply integrate the SDE with respect to the gap $\Delta$
reconstruct the effective action functional. This functional, let
us call it $\tilde{\Gamma}[\Delta]$ is in general \emph{not}
equal to the the BEA $\hat{\Gamma}[G]$, not even at solutions of
the SDE. Indeed, as can be seen from Fig. \ref{fig::firstorder},
the results for physical quantities like the effective fermion
mass can differ. The underlying reason for this is that the gap
$\Delta$ is not the correct integration variable. The SDE is
obtained by a $G$-derivative of the BEA functional
$\hat{\Gamma}[G]$. Therefore, in order to reconstruct
$\hat{\Gamma}[G]$, we have to integrate with respect to $G$. As
can be seen from Eq. \eqref{equ::equation} $\Delta$ is, in
general, not even a linear function of $G$. Simple integration
with respect to $\Delta$ therefore neglects the Jacobi matrix,
which is a non-trivial function of $\Delta$ for interactions more
complicated than a four-fermion interaction.

\subsection{Non-vanishing current quark masses $\msigma\neq 0$}
The non-vanishing current quark mass explicitly breaks the
residual ${\mathcal{Z}}_{3}$-symmetry.
$\hat{\Gamma}[|\sigma|,\alpha]$ does no longer depend on
$\cos(3\alpha)$ only, and we have to look at the complete complex
$\sigma$-plane for possible extrema.

Moreover, for $\msigma=0$, $\hat{\Gamma}[\sigma]$ is completely
symmetric under $\zeta\rightarrow -\zeta, \sigma\rightarrow
-\sigma$. Therefore, we could restrict ourselves to $\zeta\geq 0$.
For $\msigma\neq 0$ we need to add the transformation
$\msigma\rightarrow -\msigma$. We can still restrict ourselves to
positive $\zeta$ but we need to consider both positive and
negative $\msigma$.

In the case of $\msigma\neq 0$ we still encounter an extremum of
$\hat{\Gamma}[\sigma]$ at $\sigma=0$. However, in this case it is
not a solution of the SDE. It is a spurious solution due to
$\frac{d\Delta[\sigma]}{d\sigma}=0$. The difference to the chiral
limit is that the derivative of the effective potential
$\hat{\Gamma}[\sigma]$ now has only a simple zero while in the chiral
limit it is a twofold zero. After dividing the field equation by
$\frac{d\Delta[\sigma]}{d\sigma}$ a simple zero remains, giving a
solution of the SDE in the chiral limit.

\begin{figure}[t]
\begin{center}
\scalebox{0.8}[0.8]{
\begin{picture}(350,185)
\Text(235,140)[c]{\scalebox{1.8}[1.8]{$\Delta|m_{\scalebox{0.5}[0.5]{$\sigma$}}|$}}
\Text(70,140)[c]{\scalebox{1.8}[1.8]{$\zeta_{\scalebox{0.5}[0.5]{\text{SSB}}}$}}
\Text(200,32)[c]{\scalebox{1.8}[1.8]{$m^{\scalebox{0.5}[0.5]{0}}_{\scalebox{0.5}[0.5]{$\sigma$}}$}}
\includegraphics{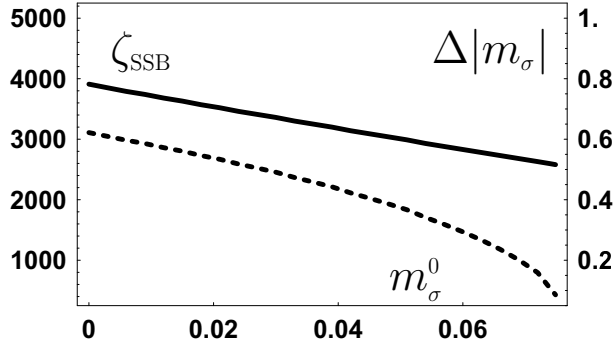}
\end{picture}
}
\end{center}
\caption{Dependence of the critical coupling on the current quark
mass $\msigma$ (thick solid line). Jump in the fermion mass at the
phase transition (thick dashed line). We observe that there exists
a critical $m^{0}_{\sigma}\approx 0.076$ above which there is no
phase transition.} \label{fig::currentquark}
\end{figure}
Although, chiral symmetry is now broken explicitly we can still
observe a first-order phase transition signaled by a jump in the
fermion mass. The critical coupling for the phase transition depends on
$\msigma$ as depicted in Fig. \ref{fig::currentquark}. The
critical line ends at
$m^{0}_{\sigma}=m^{0}_{\sigma,\textrm{crit}}\approx 0.076$, i.e.
for $m^{0}_{\sigma}>m^{0}_{\sigma,\textrm{crit}}$ we have no first
order phase transition in our approximation.

\section{Color-Octet Condensation} \label{sec::instanton2}
In the last section we have considered only one direction in the
space of all possible $g$ resulting in a phase diagram for chiral
symmetry breaking. Let us now consider the more general case where
we also allow for a non-vanishing expectation value in the
color-octet channel, more explicitly in the color-flavor locking
direction.
\begin{eqnarray}
g_{mn}&=&g_{ai\alpha\chi s,bj\beta\tau t}
\\\nonumber
&=&\frac{1}{6}\delta_{ab}\delta_{ij}\delta_{\alpha\beta}
\bigg[\sigma(\delta_{\chi1}\delta_{\tau2}\delta_{s2}\delta_{t1}-\delta_{\chi2}\delta_{\tau1}\delta_{s1}\delta_{t2})
-\sigma^{\star}(\delta_{\chi2}\delta_{\tau1}\delta_{s2}\delta_{t1}
-\delta_{\chi1}\delta_{\tau 2}\delta_{s1}\delta_{t2})\bigg]
\\\nonumber
&+&\frac{1}{12}\lambda^{z}_{ab}\lambda^{z}_{ij}\delta_{\alpha\beta}
\bigg[\chi(\delta_{\chi1}\delta_{\tau2}\delta_{s2}\delta_{t1}-\delta_{\chi2}\delta_{\tau1}\delta_{s1}\delta_{t2})
-\chi^{\star}(\delta_{\chi2}\delta_{\tau1}\delta_{s2}\delta_{t1}
-\delta_{\chi1}\delta_{\tau 2}\delta_{s1}\delta_{t2})\bigg]
\end{eqnarray}
Following the outline of the previous section we obtain
\begin{eqnarray}
\nonumber
\Delta[g]_{mn}&=&-\zeta\bigg[\frac{5}{1296}(288\sigma^{2}+6\sigma\chi-7\chi^{2})\delta_{ab}\delta_{ij}
-\frac{5}{432}\chi(6\sigma+\chi)\delta_{ai}\delta_{bj}\bigg]
\\\nonumber
&&\quad\quad\quad\quad\quad\quad\times\delta_{\alpha\beta}
\bigg[\delta_{\chi1}\delta_{\tau2}\delta_{s2}\delta_{t1}
-\delta_{\chi2}\delta_{\tau1}\delta_{s1}\delta_{t2}\bigg]
\\\nonumber
&&-(\chi\leftrightarrow\tau,\sigma\rightarrow\sigma^{\star},\chi\rightarrow\chi^{\star})
\end{eqnarray}
and
\begin{eqnarray}
U[\sigma,\chi]=\zeta\bigg[\frac{20}{9}\sigma^{3}-\frac{5}{27}\sigma\chi^{2}-\frac{5}{243}\chi^{3}+c.c.\bigg].
\end{eqnarray}
\begin{eqnarray}
\nonumber \hat{\Gamma}[\sigma,\chi]&=&-4v_{4}\int_{x}
dx\,\,x[8\ln(x+|m|^{2}_{\sigma}) +\ln(x+|m|^{2}_{\chi})]
+U[\sigma,\chi]
\\\nonumber
m_{\sigma}&=&\frac{5}{1296}\zeta(288\sigma^{2}+6\sigma\chi-7\chi^{2})+\msigma
\\
m_{\chi}&=&\frac{5}{81}\zeta(18\sigma^{2}-3\sigma\chi-\chi^{2})+\msigma.
\end{eqnarray}

In the chiral limit every point on the line $\chi=-6\sigma$
($m_{\sigma}=0$ and $m_{\chi}=0$) has the same value of
$\hat{\Gamma}$, and both derivatives with respect to $\sigma$ and
$\chi$ vanish. But, in this direction the
derivative of $\Delta$ with respect to the condensates vanishes,
too, and is indeed the null function in this direction. Therefore,
on this line only the point $(\sigma,\chi)=(0,0)$ is a true
solution to the SDE.

Restricting both $\sigma$ and $\chi$ to be real we have not found a
solution with $\chi\neq 0$. Thus, we have not identified a solution which
breaks color symmetry but not parity.

For the most general case of complex $\sigma$ and $\chi$ things
are considerably more difficult since we now have to search for an
extremum of a potential which depends on four real parameters. We
checked several values of the coupling constant. So far we have not
found a solution which has lower action than the lowest one with
$\chi=0$.

Still, we would like to point out that the potential is unbounded
from below. In various directions including those with $\chi\neq
0$, $\hat{\Gamma}\rightarrow -\infty$. Therefore, a physical cutoff mechanism
like the one discussed in \cite{Wetterich:2000ky} or a better approximation
which makes the potential bounded from
below may provide additional solutions.

\chapter{Outlook: Quest for a Renormalizable Standard Model\protect\footnotemark[0]} \label{chap::quest}
\footnotetext[0]{This is
work in progress in collaboration with Holger Gies and Christof Wetterich.}
The ``Standard Model'' (SM) of particle physics is probably one of the most widely studied
physical theories. It describes a wide range of physical situations with a satisfactory amount of
accuracy. Yet, there are still open questions. In particular, we want to concentrate
on ``Renormalizability'' and the ``Hierarchy Problem'', as those are problems
tightly connected to the
existence of an elementary scalar boson -- the Higgs -- in the SM. This
returns us to the speculation of a composite Higgs which we mentioned in the
introduction.

Before going into details, let us briefly outline those two problems.

\textbf{Renormalizability:} If asked whether the SM is renormalizable many physicists\footnote{A (not representative)
survey in the ``Graduiertenkolleg: Physical Systems with many Degrees of Freedom'' of the university
of Heidelberg
resulted in $\approx 70\%$ positive answers.} would answer this
question positively. So, why search for something we have already found? Well, while this answer is
certainly correct for most practical purposes, i.e. when the UV-cutoff scale is of
a reasonable size $\Lambda\lesssim 1\textrm{TeV}$, it is not
certain that this is so if we send the cutoff to infinity.

To get a grasp of
the problem let us look at the example of $\phi^{4}$-theory (a model
for the Higgs potential). Roughly speaking renormalizability means
that the physics at short distances does not really matter, and therefore we can send the
cutoff to infinity without changing the results.
Yet, a straightforward calculation gives the
(1-loop) running four-boson coupling as follows,
\begin{equation}
\label{equ::running}
\lambda_{\textrm{eff}}(q^2)=\frac{\lambda}{1-\frac{3\lambda}{4\pi^2}\ln\left(\frac{q^2}{\mu^{2}}\right)},
\end{equation}
where $q^2$ is the (Euclidean) momentum scale and $\lambda$ is the coupling at the scale $\mu^2$.
From this we can read off, that the coupling constant grows with increasing momentum scale. It is
plausible that this does not fit into our picture that long distance physics (small
momenta) decouples from short distance physics (large momenta), as the coupling is strong in the latter regime.

This gives us
an intuitive understanding that problems might arise when we want to send the UV cutoff to
infinity. Those theories are not renormalizable in a strict sense.
Requiring that a ``fundamental'' theory should be renormalizable such
theories cannot be ``fundamental'', i.e. valid at all scales.
They are valid only as effective theories up to a certain cutoff scale $\Lambda$ where ``new''
physics will set in.

This \cite{Callaway:ya,Lindner:1985uk,Frohlich:tw,Aizenman:du,Luscher:1987ay}
and a similar problem
in QED (Landau pole!) \cite{Broda:qj,Gockeler:1990bc},
hint to serious trouble in the Higgs respectively $U(1)$-sector of the SM.
This is often referred to as ``triviality'' because when starting with some
bare interactions and sending the cutoff to infinity, the renormalized
coupling will be strictly zero.

\textbf{Hierarchy Problem:} In contrast to chiral fermions, where chiral symmetry
prevents the fermion mass from acquiring large quantum correction, the mass
of a scalar boson is not protected against such corrections. Thus,
assuming that the RG flow of the SM is ``released'' at, say, the
GUT\footnote{Grand Unified Theory.} scale, enormous fine-tuning
of the scalar initial conditions is required to separate
the electroweak scale from the GUT scale by many orders of magnitude.

To illustrate this let us do a very simplified calculation.
In a crude approximation the Higgs mass runs as follows ($c^2=\textrm{const.}={\mathcal{O}}(1)$),
\begin{equation}
m^2_{H}(q^2_{1})-m^{2}_{H}(q^2_{2})=c^2(q^{2}_{1}-q^2_{2}).
\end{equation}
Defining the dimensionless Higgs mass $\mu^2(q^2)=\frac{m^{2}_{H}(q^2)}{q^2}$ this
can be rewritten as,
\begin{equation}
\label{equ::fine}
\mu^{2}(q^2_{1})=c^2-\frac{q^2_{2}}{q^{2}_{1}}\left(c^2+\mu^2(q^2_{2})\right).
\end{equation}
So far so good, but let us now consider two very different scales, e.g. the GUT scale
$q^{2}_{1}=M^2_{\textrm{GUT}}\sim 10^{30}(\Gev)^2$ and the electroweak
symmetry breaking scale \mbox{$q^{2}_{2}\sim 10^4 (\Gev)^2$.} Inserting values for the Higgs mass at the
electroweak scale we find that the brackets on the RHS are of order ${\mathcal{O}(1)}$. Hence,
\begin{equation}
\mu^2(M_{\textrm{P}})=c^2-{\mathcal{O}}(10^{-26}),
\end{equation}
the dimensionless mass at the GUT scale must be fine-tuned to $c^2$ by an incredible amount.
Although not excluded, such an amount of fine-tuning seems unnatural.
\section{UV Fixed Points and Renormalizability}
Let us now address the problem of renormalizability on a more formal level.

\subsection{Non-perturbative Renormalizability}
Commonly, theories are considered to be renormalizable if they have a (small coupling) perturbative expansion and
we can absorb the infinities by a finite number of counterterms. In consequence, they only
have a finite number of couplings (masses etc.) since it requires
physical information to fix a counterterm  unambiguously. For instance,
the value of a coupling or mass at some scale cannot be determined in the theory itself, but
must be measured. All other quantities can be calculated from these couplings. From this
renormalizable theories derive their predictive power. In particular, the renormalization
procedure allows us to take
the continuum limit, i.e. to send the UV cutoff to infinity. In this sense we can consider renormalizable theories
as fundamental. A prominent example of such a theory is QCD.

It comes as no surprise that not all theories have this property of
``perturbative renormalizability'' (PR). Gravity is probably the most well known
example, but there are many more: as theories which contain a coupling of negative
mass dimension are usually not PR.
The NJL model \eqref{equ::faction}
has a coupling $\sim(mass)^{-2}$ and consequently is not PR, either.
Moreover, naively renormalizable theories with dimensionless coupling constants
may also face difficulties spoiling PR, as discussed at the beginning of this chapter.
Usually, theories which are not PR
are thought to be effective field theories valid only up to
a finite UV scale $\Lambda$.

Yet, PR is not the end of the game. Using a little bit of imagination it is obvious
that a logical alternative is a ``non-perturbative'' renormalizable (NPR)
\cite{Weinberg:1976xy,Parisi:1975im,Gawedzki:uq,Gawedzki:ed,Gawedzki:1985jn,Rosenstein:1990nm,deCalan:km}
theory.
So far so good, but does such a thing exist and what is it? Fortunately, the answer
to the first question is yes. Among the examples are various
versions of the Gross-Neveu model. More recently,
it has been proposed that gravity is NPR, too \cite{Lauscher:2001ya,Lauscher:2001rz}.
But now, let us find out what hides behind NPR\footnote{In this brief description
of NPR we follow \cite{Lauscher:2001rz}.}.

The space of all possible action functionals can be parametrized by an infinite number
of dimensionless couplings (if necessary we use a suitable rescaling with the scale $k$).
As discussed in Sect. \ref{sec::rg} the RG describes a trajectory in this space
(it is parametrized by the scale $t=\ln(k/k_{0})$),
\begin{equation}
\partial_{t}g^{i}_{t}=\beta^{i}(g^{1},g^{2},\ldots).
\end{equation}
Starting point for the construction of a NPR is the existence of a non-Gaussian fixed point (FP)
$g_{\star}=(g^{1}_{\star},g^{2}_{\star},\ldots)$ with at least one $g^{i}_{\star}\neq 0$, in the RG flow,
\begin{equation}
\beta^{i}(g^{1}_{\star},g^{2}_{\star},\ldots)=0 \quad \forall i.
\end{equation}
In the setting of statistical physics such a FP is exactly what we would
associate with a second order phase transition. The dimensionless constants
do not change with the scale (typically the scale can be associated with
the difference from the critical temperature $\sim|T-T_{c}|$), accordingly the dimensionful quantities are
simply powers of the scale\footnote{It is often useful to include the wave function
renormalization into the couplings. This modifies the naive power laws by anomalous dimensions.}.
This gives the typical power laws
of the critical behavior near a second order phase transition.
Moreover, we would like to remark that in this context a PR is just the special case
of a Gaussian, i.e. $g_{\star}=0$,
FP.

In the vicinity of this FP we can linearize the RG equations,
\begin{equation}
\partial_{t}g^{i}=\sum_{j}B^{ij}(g^{j}_{\star}-g^{j}_{t})
\end{equation}
where
\begin{equation}
B^{ij}=\frac{\partial\beta^{i}}{\partial g^{j}}\bigg|_{g_{\star}}
\end{equation}
is the Jacobi matrix of the $\beta$-functions. This is a linear differential equation
and the general solution reads,
\begin{equation}
\label{equ::flowaway}
g^{i}_{t}=g^{i}_{\star}+\sum_{I}C^{I}V^{I}_{i}\left(\frac{k_{0}}{k}\right)^{\Theta^{I}},
\end{equation}
where the $V^{I}$ are right eigenvectors of $B$ with corresponding
eigenvalues\footnote{The $\Theta^{I}$ are not necessarily real as $B$ is not necessarily symmetric.
For simplicity let us pretend that they are real, anyway. The
general case can be recovered by replacing, $\Theta>0\rightarrow Re(\Theta)>0$ etc. in our argumentation.}
$-\Theta^{I}$,
\mbox{$BV^{I}=-\Theta^{I}V^{I}$},
and the $C^{I}$ are constants determined by the initial conditions.
Setting $C^{I}=0$ if the corresponding eigenvalue\footnote{For $\Theta^{I}=0$ it
depends on the details if the UV limit is finite or not. Correspondingly, it might
or might not be necessary to set $C^{I}=0$. Prominent examples for both cases
are the $\phi^{4}$-theory where the UV limit is not finite, and QCD where the
limit is finite and which is therefore strictly renormalizable.} $\Theta^{I}<0$ we can safely take the UV limit
$k\rightarrow\infty$. In other words this gives us an RG trajectory which ends in $g_{\star}$
for $k\rightarrow\infty$. Any such trajectory defines a theory with a meaningful UV limit.
The space of all such trajectories is a submanifold ${\mathcal{S}}$ of dimensionality
$\Delta$, given by the number of eigenvalues $\Theta^{I}>0$. We can specify
a trajectory in this space by giving the values of integration constants $C^{I}$.
Thus our theory has $\Delta$ ``renormalizable couplings'', which have to be taken from experiment. In particular,
$\Delta$ must be \emph{finite} (and preferably small) in order for our theory to have predictive power.

Stated differently, all trajectories in the submanifold ${\mathcal{S}}$ are \emph{attracted} toward
$g_{\star}$ for \emph{increasing} scale $k$, consequently, for \emph{decreasing} $k$ they are \emph{repelled}.
Therefore, the $\Delta$ parameters specifying the trajectory are the \emph{relevant} parameters to describe
physics in the fixed point regime. The remaining ``non-renormalizable couplings'' corresponding to $\Theta^{I}<0$ are \emph{irrelevant}
in the sense that starting with a finite value at some scale $k<\infty$ they are attracted
toward the submanifold ${\mathcal{S}}$ for decreasing $k$. Thus it does not really matter if we give them some finite
value at a large scale\footnote{We will see below that this statement
should be taken with some care. More precisely, this is true only if we are close enough to
the fixed point.} $k$.

Finally, let us give a naive argument why the submanifold ${\mathcal{S}}$ should be finite dimensional.
To obtain the dimensionless coupling we have to rescale the dimensionful coupling
constant $G^{i}$ by an appropriate power of $k$, $g^{i}=k^{-d^{i}}G$. The $\beta$-function
has now the form $\beta^{i}=-d^{i}g^{i}+\cdots$ where the dots denote the loop corrections.
In the Jacobi matrix this gives a contribution $B^{ij}=-d^{i}\delta^{ij}+\cdots$. The eigenvalues
get a contribution $\Theta^{i}=d^{i}+\cdots$. Constructing a higher order operator,
we usually add derivatives or field operators, thus decreasing $d^{i}$. Therefore, we should
only have a finite number of $\Theta^{i}<0$. Yet, this argument is based on
the assumption that the loop corrections are small. Hence, it might be not valid in
the non-perturbative regime with strong coupling and large loop corrections.
Still, it seems reasonable to first look for a possible FP and establish
that it is not an artifact of the approximation, and then worry about the dimensionality
of ${\mathcal{S}}$.

\subsection{A Toy Model}
Let us illustrate this idea for a simple NJL model in $d$-dimensions, a four-fermion
interaction with dimensionless (positive) coupling constant $\hat{\lambda}=k^{d-2}\lambda$,
and the flow equation (which, for the moment, is assumed to be exact),
\begin{equation}
\label{equ::toyflow}
\partial_{t}\hat{\lambda}=(d-2)\hat{\lambda}-C\hat{\lambda}^2.
\end{equation}
For $d>2$ the Gaussian FP $\hat{\lambda}=0$ is UV repulsive, i.e. $\hat{\lambda}$
is an eigenvector with eigenvalue $-\Theta=(d-2)>0$. The model
is not PR for a non-vanishing $\hat{\lambda}$.
Moreover, for any value $\hat{\lambda}<\frac{d-2}{C}$ we approach the Gaussian FP
in the IR, $\hat{\lambda}$ is ``irrelevant''.
By contrast, for $\hat{\lambda}\geq\frac{d-2}{C}$ the coupling $\hat{\lambda}$ grows and eventually diverges.
This typically signals some kind of symmetry breaking, $\hat{\lambda}$ is not so irrelevant
any more. Looking a little bit more closely we notice that $\hat{\lambda}_{c}=\frac{d-2}{C}$ is
a FP. The eigenvector $(\hat{\lambda}-\hat{\lambda}_{c})$ has an
eigenvalue $-\Theta^{\prime}=-(d-2)<0$, i.e.
it is a relevant parameter in the fixed point regime. On the other hand, the FP is UV attractive and we
have a meaningful UV limit even for a (small) non-vanishing $(\hat{\lambda}-\hat{\lambda}_{c})$.
The theory with this parameter is NPR.

Finally, let us remark that the ad hoc flow equation of our toy
model resembled a typical flow equation
for an NJL-type model in a truncation to four-fermion interactions,
e.g. with $C=4(N-2)v_{3}\lF$ and $d=3$ the flow equation for the Gross-Neveu model in
three dimensions.

\subsection{Manifestation of the Hierarchy Problem}
In this setting we also have the tools available for discussing the hierarchy problem.
So far we have already seen that a renormalizable coupling is relevant in the infrared, i.e.
starting with a small deviation from the fixed point in the far UV this deviation soon grows large.
This is quantified in Eq. \eqref{equ::flowaway}. Taking $k_{0}$ to be some large
UV scale $k_{0}=\Lambda\sim M_{\textrm{P}}$ and $\Theta\sim{\mathcal{O}}(1)$,
the deviation grows very fast with $(\frac{\Lambda}{k})^{\Theta}$
when $k$ is lowered. Turning the argument around,
we have to fine tune the initial conditions
(choose the initial $C$ very small) at the scale $\Lambda$ in order
to achieve a value comparable to some scale $k\ll\Lambda$ for the coupling at the scale $k$.

From this it may seem that renormalizability and a solution to the hierarchy problem
more or less exclude each other. However, there is a way out, we simply
need to make $\Theta$ small or zero. At first this may sound simply like another
type of fine-tuning, but it is not necessarily so, as the eigenvalues
are a prediction of our fixed point solution and not a parameter. To understand this
we can once more look at our simple toy model. For $d=2$ we
have exactly the case of a vanishing eigenvalue.
Solving the flow equation \eqref{equ::toyflow}, we find
\begin{equation}
\label{equ::toysol}
\hat{\lambda}(t=\ln(k))=\frac{\hat{\lambda}_{0}}{1+C\hat{\lambda}_{0}\ln\frac{k}{\Lambda}},
\end{equation}
where $\hat{\lambda}_{0}=\hat{\lambda}(t_{0}=\ln(\Lambda)$. As $\hat{\lambda}$ depends
only logarithmic on the scale we can have very different scales without fine-tuning.

Finally, let us remark, that Eq. \eqref{equ::toysol} also gives us an understanding of what can happen
when $\Theta=0$. Only for $C>0$ we approach the fixed point $\hat{\lambda}^{\star}=0$ in the far ultraviolet.
Only in this case our simple toy model is renormalizable. Moreover, our $d=2$ flow equation
for our toy model has the same form as the lowest order terms of the $d=4$ flow in QED and QCD around
their Gaussian fixed points. In this language QCD has $C>0$ and is renormalizable while
QED has $C<0$ and is not strictly renormalizable.
\section{One more NJL Model}
Inspired by our toy model let us once again investigate an NJL model but now with $\Nf$ fermion species
and an $SU(\Nc)$-gauge interaction.
\subsection{Truncation and Flow Equations}
 A simple truncation including all possible pointlike the four-fermion interaction reads,
\begin{eqnarray}
\label{equ::truncsym}
\Gk=\int \yb\I\fsl{D} \psi+\frac{1}{4}F^{\mu\nu}F_{\mu\nu}
+\frac{1}{2}\!\!\!\!\!\!\!\! && \Big[
  \bls \SP +\blm\VAm + \blp \VAp  \\\nonumber
&&+\blsf \SPN +\blsc \SPn +\blva \VAn \Big],
\end{eqnarray}
with the covariant derivative,
\begin{equation}
\fsl{D}=\fss{\partial}-ig\fss{A}.
\end{equation}
The color and flavor singlets are
\begin{eqnarray}
\SP&=&(\yb\psi)^2-(\yb\gamma_5\psi)^2,\nonumber\\
\VAm&=&(\yb\gamma_\mu\psi)^2 + (\yb\gamma_\mu\gamma_5\psi)^2,
\nonumber\\
\VAp&=&(\yb\gamma_\mu\psi)^2 - (\yb\gamma_\mu\gamma_5\psi)^2
\nonumber,
\end{eqnarray}
where color ($i,j,\dots$) and flavor ($a,b,\dots$) indices are
pairwise contracted, e.g., $(\yb\psi)\equiv (\yb_i^a \psi_i^a)$.
The operators of non-trivial color or flavor structure are denoted by,
\begin{eqnarray}
\SPn&=&(\yb_i\psi_j)^2-(\yb_i\gamma_5\psi_j)^2 \equiv
   (\yb_i^a\psi_j^a)^2-(\yb_i^a\gamma_5\psi_j^a)^2,\nonumber\\
\SPN&=&(\yb^a\psi^b)^2-(\yb^a\gamma_5\psi^b)^2 \equiv
   (\yb_i^a\psi_i^b)^2-(\yb_i^a\gamma_5\psi_i^b)^2,\nonumber\\
\VAn&=&(\yb_i\gamma_\mu\psi_j)^2 +
(\yb_i\gamma_\mu\gamma_5\psi_j)^2
   \equiv\VAm^{\text{N}}.
\nonumber
\end{eqnarray}
The last equation holds because of a Dirac Fierz identity (cf. App. \ref{app::fierz}).
The truncation \eqref{equ::truncsym} is symmetric under a simultaneous
exchange of $\Nc\leftrightarrow\Nf$, $\blsf\leftrightarrow\blsc$,
$(\dots)_{\text{N}} \leftrightarrow (\dots)^{\text{N}}$.
However, it is not invariant under $SU(\Nf)_{\text{L}}\times SU(\Nf)_{\text{R}}$.
We can obtain the part invariant under this additional symmetry by setting $\lsh=\lsc=0$. The full
action Eq. \eqref{equ::truncsym} is invariant only under the subgroup
$SU(\Nf)_{\text{V}}$ of simultaneous right and left handed
rotations.

Following along the lines of Sect. \ref{sec::fermion}, in particular using Eq. \eqref{equ::expansion},
the calculation of the flow equations is straightforward.
Using the dimensionless coupling constants $\hat{\lambda}=k^2\lambda$ we find,
\begin{eqnarray}
\pat\lsh\!\!\!&=&\!\!\!\left(2\lsh-12g^{2}\left(\frac{\Nc^2-1}{\Nc}\lsh+\lsc\right)v_{4}\right)
\\\nonumber
&-&\!\!\!8v_4\lF
  \Big[2\Nc\Nf \lsh^2 -2\lsh\big[\lm + 3\lp - 2(\Nf\lsc+\Nc\lsf)\big]
       +3\lsc\lsf\Big],
       \\\nonumber
\pat\lm\!\!\!&=&\!\!\!-\frac{1}{8\Nc^2}\left(12+\Nc^2(3+\alpha)^2\right)g^{4}v_{4}\lFB
+\left(2\lm-\frac{12}{\Nc}g^2\left(\lm-\Nc\lva\right)v_{4}\right)
\\\nonumber
&-&\!\!\!8 v_4\lF
  \Big[-\Nf\Nc(\lm^2+\lp^2) + \lm^2-2(\Nc+\Nf)\lm\lva  \\\nonumber
&&\qquad\qquad\qquad
       + \lp(\lsh+\Nc\lsc+\Nf\lsf) + 2\lva^2
        -\case{1}{4}\lsh^2-\case{1}{2}\lsc\lsf \Big],
        \\\nonumber
\pat\lp\!\!\!&=&\!\!\!\frac{1}{8\Nc^2}\left(12+\Nc^2(3-\alpha(6+\alpha))\right)g^{4}v_{4}\lFB+
\left(2+\frac{12}{\Nc}g^{2}v_{4}\right)\lp
\\\nonumber
&-&\!\!\!8 v_4 \lF
  \Big[ - 3\lp^2 - 2\Nc\Nf\lm\lp
        - 2\lp(\lm+(\Nc+\Nf)\lva)  \\\nonumber
&&\qquad\quad
        + \lm(\lsh+\Nc\lsc+\Nf\lsf) + \lva(\lsc+\lsf)
        +\case{1}{4}(\lsh^2+\lsc{}^2+\lsf{}^2) \Big],
  \\\nonumber
\pat\lsc\!\!\!&=&\!\!\!\left(2+\frac{12}{\Nc}g^2v_{4}\right) \lsc
\\\nonumber
&-&\!\!\! 8 v_4 \lF
  \Big[ 2\Nf \lsc{}^2 - 2 \lsh\lva - 2\lm\lsc - 2\Nc\lsc\lva
        - 6\lp\lsc-\lsh\lsf \Big],
        \\\nonumber
\pat\lsf\!\!\!&=&\!\!\!\frac{1}{4\Nc}\left(24-\Nc^2(3+\alpha)^2\right)g^{4}v_{4}\lFB
+\left(2\lsf+24g^2\left(\lp-\frac{\Nc^2-1}{2\Nc}\lsf\right)v_{4}\right)
\\\nonumber
&-&\!\!\! 8 v_4 \lF
  \Big[ 2\Nc \lsf{}^2 - 2 \lsh\lva - 2\lm\lsf - 2\Nf\lsf\lva
        - 6\lp\lsf -\lsh\lsc\Big],
        \\\nonumber
\pat\lva\!\!\!&=&\!\!\!\frac{1}{8\Nc}\left(24-\Nc^2(3-\alpha(6+\alpha))\right)g^{4}v_{4}\lFB
+\left(2\lva-\frac{12}{\Nc}(\lva-\Nc\lm) g^{2}v_{4}\right)
\\\nonumber
&-&\!\!\! 8 v_4 \lF
  \Big[ - (\Nc+\Nf)\lva^2 + 4\lm\lva
        - \case{1}{4} (\Nc \lsc{}^2 +\Nf \lsf{}^2)
 -\case{1}{2}\lsh(\lsc+\lsf) \Big],
\end{eqnarray}
where the gauge fixing parameter is denoted by $\alpha$.

\subsection{Many Fixed Points but no Solution to the Hierarchy Problem -- the
Case of Vanishing Gauge Coupling}\label{sec::vanishing}
To get a first insight we have numerically solved the FP equation,
\begin{equation}
\pt\hat{\lambda}=0,
\end{equation}
for vanishing gauge coupling $g$ and $\Nc=\Nf=3$. It turns out that not only do we find a solution, but
quite a few, $64=2^6$ to be exact (one is always the Gaussian FP).
Looking more closely, only $44$ are real.

Let us, for the moment postpone the question, if the FP are
physical or an artifact of the truncation.
Having found so many FP we could become kind of greedy and ask if among all those $44$
FP there is one which in addition
to providing us with a renormalizable theory, could solve the hierarchy problem.
However, calculating the eigenvalues of the stability matrix we find, that the
largest eigenvalue $\Theta^{\textrm{max}}\geq 2$ for all non-trivial fixed points. Or, stated more physically, one direction
is at least as unstable as a scalar boson mass.

Is this a very special property of our choice of $\Nf$ and $\Nc$?
No, but it is a property of our truncation to pointlike four-fermion interactions and vanishing
gauge coupling.
More precisely, in this truncation there is always an eigenvalue $\Theta=(d-2)$. This can be
seen by the following argument.
In a four-fermion truncation we can write the flow equations in the form,
\begin{equation}
\pt\lambda_{i}=(d-2)\lambda_{i}+\lambda_{k}A^{i}_{kl}\lambda_{l},
\end{equation}
where $A^{i}$ is a symmetric matrix i.e. $A^{i}_{kl}=A^{i}_{lk}$.
The stability matrix is then given by,
\begin{equation}
B_{ij}=\frac{\partial(\pt\lambda_{i})}{\partial\lambda_{j}}=(d-2)\delta_{ij}+2\lambda_{k}A^{i}_{kj}.
\end{equation}
Let $\lambda^{\star}$ be a solution of the fixed point equation,
\begin{equation}
\label{equ::fixed}
\pt\lambda^{\star}_{i}=(d-2)\lambda^{\star}_{i}+\lambda^{\star}_{k}A^{i}_{kl}\lambda^{\star}_{l}=0
\quad \forall i.
\end{equation}
Acting with $B_{ij}\mid_{\lambda^{\star}}$ on
$\lambda^{\star}_{j}\neq 0$ we have,
\begin{eqnarray}
B_{ij}\mid_{\lambda^{\star}}\lambda^{\star}_{j}&=&(d-2)\lambda^{\star}_{i}+2\lambda^{\star}_{j}A^{i}_{jk}\lambda^{\star}_{k}
\\\nonumber
&=&-(d-2)\lambda^{\star}_{i}+2(d-2)\lambda^{\star}_{i}+2\lambda^{\star}_{j}A^{i}_{jk}\lambda^{\star}_{k}
\\\nonumber
&=&-(d-2)\lambda^{\star}_{i}+2((d-2)\lambda^{\star}_{i}+\lambda^{\star}_{j}A^{i}_{jk}\lambda^{\star}_{k})
\\\nonumber
&=&-(d-2)\lambda^{\star}_{i},
\end{eqnarray}
where we have used the fixed point equation \eqref{equ::fixed} in the last step.

This shows that $\lambda^{\star}$ itself is an eigenvector of the
stability matrix with the eigenvalue $-(d-2)$, hence $\Theta=(d-2)$. Therefore, in this
truncation there cannot be an infrared stable fixed point beside $\lambda=0$.
\subsection{Non-vanishing Gauge Coupling}
Having not found the desired properties for the eigenvalues of the fixed point, let us look if a
non-vanishing gauge coupling can stabilize the fixed point.

\begin{figure}[!t]
\begin{center}
\begin{picture}(220,150)(10,10)
\Text(218,32)[]{\scalebox{1.3}[1.3]{$g$}}
\Text(50,110)[]{\scalebox{1.3}[1.3]{$\Theta^{\textrm{max}}_{<}$}}
\includegraphics[width=8cm]{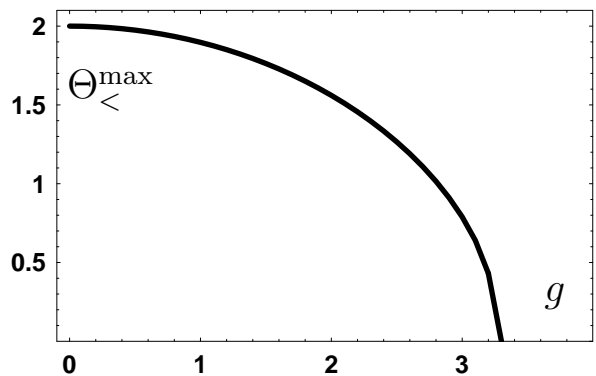}
\end{picture}
\end{center}
\caption{Largest eigenvalue $\Theta^{\textrm{max}}_{<}$ for the most ``stable'' fixed point (cf. Eq. \eqref{equ::minmax})
depending on the gauge coupling. At a realistic value $g\approx 1$
for the gauge coupling $\Theta^{\textrm{max}}_{<}>1.5$. Giving
us no solution to the hierarchy problem.}
\label{fig::eigengauge}
\end{figure}

A measure for the stability of a fixed point $\hat{\lambda}^{\star}_{x}$, $x=1\ldots 44$
is the size of its largest eigenvalue $\Theta^{\textrm{max}}_{x}$. When $\Theta^{\textrm{max}}_{x}$ is smaller
the fixed point $\hat{\lambda}^{\star}_{x}$ is more stable. Thus, we have searched for the smallest,
\begin{equation}
\label{equ::minmax}
\Theta^{\textrm{max}}_{<}=\min_{x}\Theta^{\textrm{max}}_{x}.
\end{equation}
We have plotted this eigenvalue as a function of the gauge coupling in Fig. \ref{fig::eigengauge}.
To get an impression of a reasonable value of the gauge coupling, let us take the strong
gauge coupling. At a scale of $\sim 100\Gev$ we have $g_{s}\approx 1$.
At larger scales the gauge coupling is even smaller.
As can be seen from Fig. \ref{fig::eigengauge}  $\Theta^{\textrm{max}}_{<}>1.5$ in this range, bringing
us nowhere near the desired $\Theta\approx 0$.
\section{The Future}
So far the results of our toy model can be summarized as follows. It looks as if there are many
possibilities to have fixed points, but it seems difficult to find one which has very small
positive eigenvalues $\Theta$. But, as we have already seen in Sect. \ref{sec::vanishing}
this may be an artifact of our truncation to four-fermion interactions. Therefore, a next step
is certainly to enlarge the truncation.

\begin{figure}[t]
\begin{center}
\scalebox{0.65}[0.65]{
\fbox{
\begin{picture}(160,80)
\SetOffset(-87,-20)
\ArrowLine(130,60)(95,95)
\ArrowLine(95,25)(130,60)
\Vertex(130,60){2}
\ArrowArc(160,60)(-30,0,180)
\ArrowArc(160,60)(-30,180,0)
\Vertex(190,60){2}
\Photon(190,60)(235,60){5}{4}
\end{picture}}}
\end{center}
\caption{Correction to the gauge-boson--fermion vertex (fermions solid with arrow, gauge boson wiggled)
in the presence of a four-fermion interaction.}
\label{fig::gauge}
\end{figure}
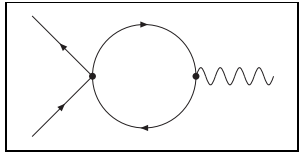

However, there is another interesting direction we can take. So far, in our simple
approximations we have neglected the influence of the four-fermion interaction
on the running of the gauge coupling. Instead we have treated the gauge coupling as coming
from the outside. Yet, at a strong-coupling non-Gaussian fixed point the flow of
the gauge-boson--fermion vertex is certainly influenced by the presence of the four-fermion
interaction. The lowest order contribution is depicted in Fig. \ref{fig::gauge}.
If we define the gauge coupling by this vertex,
the running is modified by a contribution $\sim g^2\hat{\lambda}$,
\begin{eqnarray}
\pt g^{2}&=&-8\vv g^2\left [
\lF(\ls-2\lm+\Nf\lsf-2\Nf\lva)+\beta_{0}g^2\right],
\end{eqnarray}
with
\begin{equation}
\beta_{0}=\frac{1}{2}\left(\frac{11}{3}N_{c}-\frac{2}{3}N_{f}\right).
\end{equation}
First of all this may have interesting consequences for
properties like asymptotic freedom, because at a non-Gaussian fixed point the terms
$\sim g^2\hat{\lambda}$ will dominate for very small gauge coupling. Therefore, the fixed
point in $\hat{\lambda}$ determines if the gauge interaction is asymptotically free or not\footnote{
As $\beta_{0}>0$ it is sufficient for asymptotic freedom that the sum of the four-fermion couplings
in the brackets is positive. First numerical checks indicate that regardless of which of the $44$
fixed points ($\Nc=\Nf=3$) we choose, asymptotic freedom is ensured. This even seems to apply to
various other combinations of $\Nf$, $\Nc$.}.

Secondly, in Fig. \ref{fig::gauge} we have concentrated on the gauge-boson--fermion vertex. But, in
non-abelian gauge theories alternative definitions of the gauge coupling are by
the three- and four-gauge-boson vertices. Some thought reveals that those do not get a direct contribution
from the four-fermion interaction. This raises the question how this can be reconciled
with the fact that, at least naively, gauge invariance tells us that both definitions agree.
These interesting possibilities are subject to future investigations.

\chapter{Summary and Conclusions}\label{chap::final}
We encounter strongly interacting fermions in many situations, ranging from
color superconductivity and chiral symmetry breaking ($\sim 100\textrm{MeV}$) to ordinary
superconductivity ($\sim \textrm{meV}$). Typical
features of such systems are the formation of bosonic bound states and spontaneous symmetry breaking (SSB).
Non-perturbative techniques are essential as SSB cannot be
described by (standard) perturbation theory in these systems (cf. Sect. \ref{sec::perturbation}).
Therefore, all methods discussed in the following correspond to non-perturbative
resummations of perturbative diagrams.

We have started this work with a comparison of various standard methods used
for non-perturbative calculations in this kind of systems. We have calculated the
critical coupling for the onset of spontaneous chiral symmetry breaking
in a simple NJL model. In particular, we have used mean field theory (MFT), a renormalization group (RG)
equation with a truncation to
pointlike four-fermion interactions (from now on referred to as fermionic RG) and the
lowest-order Schwinger-Dyson equation (SDE). The results for the critical
coupling $\bar{\lambda}^{\textrm{crit}}_{\sigma}$ and
two different values of $\bar{\lambda}_{V}$ can be found in
Tabs. \ref{tab::crit3}, \ref{tab::crit4}. Since the most characteristic
features and problems of the different methods are most
clearly seen when the couplings $\bar{\lambda}_{\sigma}$ and $\bar{\lambda}_{V}$ are of similar
size we concentrate on Tab. \ref{tab::crit4}.

Both MFT and the lowest order SDE sum only over fermionic fluctuations
in presence of a bosonic background. They include,
in principle the same type of diagrams, Fig. \ref{fig::mass}. The MFT result depends
strongly on the choice of the background field. This ''Fierz ambiguity'' (FA) is expressed by the
dependence on the unphysical parameter $\gamma$ in the tables. No such ambiguity appears
in the SDE approach which therefore seems more reliable.
We note that for a particular choice of $\gamma$ the MFT and the SD approaches give
identical results - in our case $\gamma=1/2$. This has led to widespread belief that MFT
and SD are equivalent if the basis for the Fierz ordering is appropriately chosen. However,
this is not the case, as can be seen by calculating also the critical coupling where
spontaneous symmetry breaking sets in in the vector channel (in absence of other order parameters).
There is again a value $\gamma=-(\bar{\lambda}_{\sigma}+\bar{\lambda}_{V})/(2\bar{\lambda}_{V})$
where MFT and SD give identical results, but it differs from $\gamma=1/2$ as encountered in
the scalar channel\footnote{Actually, $\gamma$ is negative and therefore outside the
range of strict validity of MFT.}. We conclude that there is no possible choice of $\gamma$ where
\emph{both} critical couplings for SSB in the scalar and vector channels are identical
in the MFT and SD approaches.

For a better understanding of the FA of MFT it is instructive to consider MFT
on a more formal basis as a simple approximation, taking into account only the fermionic
fluctuations, in a ``partially bosonized'' language.
Partial bosonization is a powerful tool for an understanding of strongly interacting fermionic systems
beyond the level of MFT or SDE. It allows us to treat the bosonic fluctuations in an explicit
manner, treating them on equal footing with those of the elementary particles,
and provides for a rather simple framework for the discussion of SSB.
Most importantly,
it permits the direct exploration of the ordered phase which is, in practice, almost
inaccessible for the fermionic RG.
Yet, already on the level of the classical action we can get a grasp of the
origin of the FA. Partial bosonization
is not unique. From one fermionic action we can obtain a whole family of equivalent
bosonized actions, related to each other by redefinitions of the bosonic fields.

In order to permit a simple comparison with the fermionic
RG we have used a rather crude approximation for the purely bosonic sector by retaining only
a mass term and neglecting bosonic interactions as well as the momentum dependence
of the bosonic propagator. In this approximation the effect of the boson exchange between fermions
does not go beyond pointlike fermionic interactions. Taking into
account only the running of the Yukawa couplings (Fig. \ref{fig::vertex}) in the bosonic RG
of Sect. \ref{sec::bosoflow}, we observe already a very substantial improvement as compared to MFT.
The dependence on $\gamma$ is greatly reduced and the numerical value of the critical coupling
comes already close to the result of the fermionic RG. These features can be compared to the inclusion
of higher loop effects in perturbation theory in particle physics: they often reduce the dependence
of the results on unphysical parameters, like the choice of the renormalization scale.

Using SDE in a partially bosonized language we fared even better.
Again, MFT appears as the approximation which includes only the fermionic fluctuations
in a bosonic background. Adding the mass-shift diagram (cf. Fig. \ref{fig::schwingerb}) we
recover the unambiguous result of the fermionic SDE -- MFT sums
only over a subset of the diagrams contained in the fermionic SDE.

The mass-shift diagram in the partially bosonized language has only fermionic external legs.
This has prompted us to look for similar purely fermionic contributions in our partially bosonized
RG description. The box diagrams (Fig. \ref{fig::box}) generate
new four-fermion interactions and contribute to the same order as the mass-shift and
vertex corrections. Those four-fermion interactions are included
in the adapted bosonic RG discussed in Sect. \ref{sec::solving}.
Here the relation between the bosonic composite fields and the fermion bilinears becomes scale
dependent. This formulation is well adapted to the basic idea of renormalization where
only effective degrees of freedom at a certain scale $k$ and their effective couplings
should matter for physics associated to momenta $q^2\lesssim k^2$. The system should loose all
memory of the detailed microscopic physics. In particular, the choice of an
optimal bosonic field for the long distance physics should not involve the parameters of the
microscopic theory, but rather the renormalized parameters at the scale $k$. In this
formulation it has also become apparent that the distinction between
''fundamental degrees of freedom'' and ''bound states'' becomes a matter of scale \cite{Gies:2002nw}.
The adapted bosonic RG reproduces in our crude approximation the results of the fermionic RG.
We argue that for precision estimates in the partially bosonized approach the
''adaption'' of the definition of the composite field seems mandatory.

For an improvement of the truncation two possibilities come to mind immediately. One is to
enlarge the bosonic potential beyond the mass term, the other is to add kinetic terms for the bosons.
Yet, our discussion of Sect. \ref{sec::trouble} shows that including only a bosonic potential
without kinetic terms for the bosons does not help us in deciding which type of boson will condense
in the SSB phase. The basic reason for this is, that the corresponding interactions
in the fermionic language are still pointlike and can be Fierz transformed in many ways.
Kinetic terms correspond to a momentum dependence of the interactions in the fermionic language,
greatly reducing the freedom to Fierz transform. For this reason we have turned
to the second possibility. To cut a long story short, for a consequent inclusion of all terms
with up to two derivatives on the bosonic fields, we have to take into account
not only the momentum dependence of the mass-shift diagram (cf. Fig. \ref{fig::mass}), but in addition
part of the momentum dependence of the vertex correction and the box diagrams
(cf. Figs. \ref{fig::vertex}, \ref{fig::box}).

Comparing the numerical values (cf. Tabs. \ref{tab::crit3}, \ref{tab::crit4})
with those of the pointlike approximations, we find that the inclusion of the kinetic terms affects
the critical coupling on a $10\%$ level. Moreover, the contributions from the different
diagrams are of similar size, confirming once more that an adaption of the flow is
necessary to achieve a high precision. The $\gamma$ dependence is small, as we would
expect for a systematic enlargement of the truncation. Nevertheless, it does not
completely vanish since we have been forced to make approximations when including the
momentum dependence of the vertex corrections and box diagrams. Moreover, if the ``right''
FT is known, inclusion of the pointlike contributions seems already sufficient for
reducing the $\gamma$ dependence (cf. the second to last line in Tab. \ref{tab::crit4}). In view of the high algebraic
complexity involved with including the full momentum dependence we would like to suggest this approximation for
more moderate demands on accuracy.
Finally, let us point out the following important feature (shared by all of the three considered
approximations which include kinetic terms): with kinetic terms for the bosons we can decide which
boson will condense. For values $\bar{\lambda}_{\sigma}$ (slightly) larger than
$\bar{\lambda}^{\textrm{crit}}_{\sigma}$, only the renormalized mass of the scalar boson turns negative, while
the renormalized masses of the vector and axial vector bosons remain positive. This confirms that
we have a phase where chiral symmetry (and nothing else) is broken.

In summary, the FA can be used as one possible test of approximation errors
for some of the methods proposed to deal with strongly interacting fermionic systems.
The spread of the results within an acceptable range of the unphysical parameter
$\gamma$ should be considered as a lower bound for the systematic uncertainty within a given
approximation. We find that MFT can have a very substantial ambiguity which should then be reduced by systematic
improvements. On the other hand, the FA
is, of course, not the only source of error. Several methods such as SDE or the fermionic RG
have no such ambiguity by construction. This holds similarly for the adapted bosonic RG (with or without
kinetic terms) which is constructed to reduce the Fierz ambiguity. Hence, in this cases
the uncertainty due to the Fierz ambiguity is small compared to other uncertainties which
can roughly be estimated from the spread of the results between the different approximations.
However, improving the truncation we should keep an eye on the FA as it is likely
to increase, when our ``improvement'' neglects something essential. Comparing the RG
and SD approaches the adapted bosonic RG sums over a larger class of diagrams and therefore
seems more reliable. Moreover, the RG accounts for the fact that physics at the scale $k$ should
involve only renormalized parameters at this scale, while the SDE involves ``bare'' couplings.
We think that with the adapted bosonic RG we have reached a promising starting point for future investigations along
the lines of \cite{Berges:1999eu,Jungnickel:1995fp}. In particular, a more elaborate bosonic
potential would allow us to proceed into the SSB phase.

Removing the FA in a partially bosonized setting has turned out to be quite tedious.
For a first investigation of the phase diagram, lowest order SDE seem to be a viable
alternative which allow for a description of the SSB phase without auxiliary fields.
However, at first sight there is one drawback:
SDE determine only the derivatives of the action, not the action itself, making
it difficult to compare different solutions with regard to stability.
This shortcoming is cured by the use of the 2PI effective action, or
its simplified form at vanishing fermionic sources, the bosonic effective action (BEA).
In this context the SDE is the field equation of the BEA.

Integrating the lowest-order SDE for a
multi-fermion interaction we have obtained the bosonic effective
action at ``one-vertex'' level. Within this approximation we computed
a simple ``one-loop'' expression for the BEA and the SDE.
In this form the BEA appears very similar to MFT but does not suffer from its ambiguities and,
as we have already seen, sums over a larger class of diagrams.

We have applied the BEA at one-vertex level to a six-fermion interaction
resembling the instanton vertex for three colors and flavors.
For vanishing current quark masses we find a first-order phase transition.
It turns out that the value $\zeta_{\textrm{crit}}$ for the onset of non-trivial
solutions for the SDE is not necessarily equal to the value
$\zeta_{\text{SSB}}$ for the onset of SSB. To calculate the latter
we need the value of the effective action in addition to the
solution of the SDE. For nonzero current quark masses and positive a coupling constant
we find a phase transition only
for current quark masses below a $m^{0}_{\textrm{crit}}\approx 0.076$. At this
point the gap in the effective mass vanishes and we expect a second-order phase transition.
We have also searched for solutions with a non-vanishing color-octet condensate.
Although they are not excluded by symmetry we have
not found a stable solution of the field equations with
non-vanishing octet condensate $\chi\neq 0$. Nevertheless, we would like
to point out that in our simple approximation
the BEA is unbounded from below in several directions including some
with $\chi\neq 0$. Hence, some sort of cutoff mechanism which renders the potential bounded might
yield additional solutions. Therefore, a more detailed
investigation of the instanton interaction in a color-octet background
would be of great value.

Leaving aside the more technical aspects of this work, we have turned to an intriguing speculation,
namely that the Higgs is not an elementary scalar boson, but a bound state of fermions,
more precisely a top-antitop bound state \cite{Nambu:bk,Miransky:1989ds,Miransky:1988xi,Bardeen:1989ds}.
To motivate a thorough investigation
we have briefly reviewed the ``Hierarchy Problem'' and the ``Triviality'' of the
$\phi^4$-theory which cause trouble for the Standard Model (SM) with its \emph{elementary} scalar boson.
In particular, the latter problem may prevent the SM from being renormalizable in a strict sense.
In a model with a bound state Higgs, renormalizability may be provided by a non-Gaussian
fixed point in the RG. A toy model with $\Nf$ fermions interacting via a four-fermion
interaction and a $SU(\Nc)$ gauge interaction implied that there might
be plenty of those. On the other hand it became apparent that a solution to the hierarchy problem is
beyond a simple approximation to a pointlike four-fermion interaction, as in this setting the
required fine-tuning to achieve a separation of scales is more or less as bad as for an elementary scalar.
However, already in this simple truncation another interesting effect turned up: at
a non-Gaussian fixed point, the flow of the gauge coupling might be crucially influenced
by the four-fermion-interactions. In particular, this aspect seems to be interesting for
future investigations.

It may well be that there are fundamental scalar fields, most candidates for
unifying theories have plenty of them, but so far not even one has been detected
with certainty. Hence, it might also be that nature has simulated once more
an elementary scalar with a bound state. Upcoming experiments (e.g. at Tevatron and LHC)
will help deciding this issue. But, independent of this, bound states of fermions
are still abundant in nature and we hope that our formalism with scale-dependent degrees of freedom
may be of help in understanding some of them.

\begin{appendix}
\chapter{Conventions, Abbreviations and Symbols}
\section{Conventions}
\begin{itemize}
\item{We use Euclidean conventions in flat spacetime, i.e. the metric is the $d$ dimensional unit matrix.}
\item{Greek indices $\mu,\nu\ldots=0,\ldots, d$ denote spacetime indices.}
\item{Latin indices $a,b\ldots=1,\ldots, \Nf$ are flavor indices, $i,j,\ldots=1,\ldots, \Nc$ color indices.}
\item{Our conventions for the Fourier transform are
\begin{eqnarray}
\phi(x)&=&\int\frac{d^dq}{(2\pi)^{d}}\phi(q)\exp(ipx)
\\\nonumber
\psi(x)&=&\int\frac{d^dq}{(2\pi)^{d}}\psi(q)\exp(ipx),\quad \bar{\psi}(x)=\int\frac{d^dq}{(2\pi)^{d}}\bar{\psi}(q)\exp(-ipx).
\end{eqnarray}}
\end{itemize}
\section{Mathematical Symbols}
\begin{tabular}{|l|l|}
  \hline
$\sim$ & Proportional to\\
\hline
  $\approx$ & approximately equal to\\
  \hline
$\otimes$ & Tensor product\\
\hline
$\dagger$ & Hermitian conjugation \\
\hline
$C$ & Charge Conjugation (operator)\\
\hline
$P$ & Parity (operator)\\
\hline
$R$ & Reflection (operator) in Euclidean spacetime \\
\hline
$T$ & Time Reversal (operator)\\
\hline
\end{tabular}
\section{Abbreviations}
\begin{tabular}{|l|l|}
  \hline
Ad. & Adapted \\
\hline
BBS & Bosonic Bound State\\
\hline
  BEA & Bosonic Effective Action \\
  \hline
  cf. & confer\\
  \hline
Chap. & Chapter \\
\hline
ERGE & Exact Renormalization Group Equation(s)\\
\hline
Eq. & Equation\\
\hline
FA & Fierz Ambiguity\\
\hline
FP & Fixed Point\\
\hline
FT & Fierz Transformation\\
\hline
IR & Infrared\\
\hline
LHS & Left Hand Side\\
\hline
LPA & Local Potential Approximation\\
\hline
MF & Mean Field\\
\hline
MFT & Mean Field Theory\\
\hline
RG & Renormalization Group \\
  \hline
RHS & Right Hand Side\\
\hline
s. & see\\
\hline
SD & Schwinger-Dyson\\
\hline
SDE & Schwinger-Dyson Equation(s)\\
\hline
Sect. & Section\\
\hline
SSB & Spontaneous Symmetry Breaking\\
  \hline
  UV & Ultraviolet\\
  \hline
  WFR & Wave Function Renormalization(s)\\
  \hline
\end{tabular}

\chapter{Fermion Conventions, Fierz Identities} \label{app::fermion}
\section{Dirac-Algebra in 4 Dimensions}
Throughout this work we use an Euclidean metric,
$g^{\mu\nu}=\delta^{\mu\nu}$ and
\mbox{$|x|^2=x^{2}_{0}+x^{2}_{1}+\cdots+x^{2}_{d-1}$.} With the
exception of one tiny excursion in Chap. \ref{chap::boso2} the
number of spacetime dimensions will be $d=4$.

Accordingly the Dirac-algebra is
\begin{eqnarray}
\label{equ::algebra}
\left\{\gamma^{\mu},\gamma^{\nu}\right\}&=&\gamma^{\mu}\gamma^{\nu}+\gamma^{\nu}\gamma^{\mu}
=2\delta^{\mu\nu}\mathbf{1},
\\\nonumber
\left(\gamma^{\mu}\right)^{\dagger}&=&\gamma^{\mu},
\\\nonumber
\gamma^{5}&=&\gamma^{1}\gamma^{2}\gamma^{3}\gamma^{0},
\\\nonumber
\sigma^{\mu\nu}&=&\frac{i}{2}[\gamma^{\mu},\gamma^{\nu}]
=\frac{i}{2}(\gamma^{\mu}\gamma^{\nu}-\gamma^{\nu}\gamma^{\mu}).
\end{eqnarray}
We use a chiral basis $\psi=\left(\begin{array}{c}
\psi_{\textrm{L}} \\ \psi_{\textrm{R}} \\\end{array}\right)$,
$\bar{\psi}=(\bar{\psi}_{\textrm{R}},\bar{\psi}_{\textrm{L}})$,
with the projection operators $P_{\textrm{L,R}}=\frac{1\pm\gamma^{5}}{2}$ on the chiral
components. An explicit representation is then
given by,
\begin{equation}
\gamma^{\mu}=\left(
\begin{array}{cc}
  0 & -i\sigma^{\mu} \\
  i\sigma^{\mu} & 0 \\
\end{array}
\right),\quad\gamma^{5}=\left(
\begin{array}{cc}
  \mathbf{1} & 0 \\
  0 & -\mathbf{1} \\
\end{array}
\right),
\end{equation}
with $\sigma^{\mu}=(i\mathbf{1},\sigma^{i})$. $\sigma^{i}$ are the
standard Pauli-matrices
\begin{equation}
\sigma^{1}=\left(%
\begin{array}{cc}
  0 & 1 \\
  1 & 0 \\
\end{array}
\right),\quad \sigma^{2}=
\left(%
\begin{array}{cc}
  0 & -i \\
  i & o \\
\end{array}
\right),\quad \sigma^{3}=\left(
\begin{array}{cc}
  1 & 0 \\
  0 & -1 \\
\end{array}
\right),
\end{equation}
and $\mathbf{1}$ is here the $2\times 2$ unit matrix.

Using Eq. \eqref{equ::algebra} it is quite easy to derive
relations to simplify expressions containing several $\gamma$
matrices, e.g.
$\gamma^{\mu}\gamma^{\nu}\gamma^{\mu}=-2\gamma^{\nu}$. This can be
automated and we use the \emph{Tracer}-package \cite{Jamin:1991dp}
for \emph{Mathematica} to do this.

\section{Dirac-Algebra in 3 Dimensions}
In Chap. \ref{chap::boso2} we use the Gross-Neveu model in 3
dimensions to demonstrate a shortcoming of partial bosonization.
The Dirac-algebra in odd dimensions is somewhat different from the
case of even dimensions.

Nevertheless, it is quite easy to find an explicit realization as
the Pauli matrices already fulfill the requirements for the
Dirac-algebra,
\begin{equation}
\left\{\sigma^{i},\sigma^{j}\right\}=2\delta^{ij}.
\end{equation}
Consequently, we can use spinors with only two components. Since
we do not need any more subtle properties, let us leave it at that
and return to the case of four dimensions.
\section{Fierz Identities} \label{app::fierz}
Defining $O_{S}=\mathbf{1}$, $O_{P}=\gamma^{5}$,
$O_{V}=\gamma^{\mu}$, $O_{A}=\gamma^{\mu}\gamma^{5}$ and
$O_{T}=\sigma^{\mu\nu}$ we obtain the following Fierz identities,
\begin{equation}
\label{equ::fierz1}
(\bar{\psi}_{a}O_{X}\psi_{d})(\bar{\psi}_{c}O_{X}\psi_{a})
=\sum_{Y}C_{XY}
(\bar{\psi}_{a}O_{Y}\psi_{b})(\bar{\psi}_{c}O_{Y}\psi_{d}),
\end{equation}
with
\begin{equation}
\label{equ::fierz2}
C_{XY}=\frac{1}{4}\left(%
\begin{array}{ccccc}
-1  & -1 & -1 & 1 & -1 \\
-1   & -1 & 1 &-1  &-1  \\
-4   & 4 & 2 & 2 & 0 \\
4  &  -4 & 2 & 2 & 0 \\
-6   & -6  & 0 & 0 & 2 \\
\end{array}%
\right).
\end{equation}
From the indices $a,b,c,d$ which combine all but spin-indices we
can see that we have done nothing but exchanged the two $\psi$'s
appearing in the four-fermion term.

For the special case of only one fermionic species we find that
the structure $(\bar{\psi}O_{V}\psi)^2+(\bar{\psi}O_{A}\psi)^2$ is
invariant. Moreover, we can use Eqs. \eqref{equ::fierz1},
\eqref{equ::fierz2} to obtain the identity (actually this is
exactly Eq. \eqref{equ::fierz}),
\begin{equation}
(\bar{\psi}O_{V}\psi)^2-(\bar{\psi}O_{A}\psi)^2=-2[(\bar{\psi}O_{S}\psi)^2-(\bar{\psi}O_{P}\psi)],
\end{equation}
which allows us to transform
$(\bar{\psi}O_{V}\psi)^2-(\bar{\psi}O_{A}\psi)^2$ completely into
scalar and pseudoscalar channels.

If we have more than one fermion species, e.g. several flavors
and/or colors, the Fierz transformations turns singlets into
non-singlets and vice versa. This can be used to reduce the
number of possible couplings as we do in Chap. \ref{chap::quest}.

Finally, let us mention two possible generalizations of the above.
First, the same idea of permuting the $\psi$'s can, of course,
also be applied to higher order interactions like a 6- or
8-fermion interaction. Second, the four-fermion operators
considered above are invariant under the discrete transformations $C$,
$P$ and $T$ (charge conjugation, parity and time reversal).
However, as we know the weak interactions violate parity.
Therefore, we might also want to consider interactions which are
only invariant under $CP$ and $T$. The set of parity violating
operators can be obtained by multiplying one $O_{X}$ in the four-fermion
operator by $\gamma^{5}$. This yields
$(\bar{\psi}O_{X}\psi)(\bar{\psi}O_{X}\gamma^{5}\psi)$. Of course
there is a set of equations similar to Eqs. \eqref{equ::fierz1},
\eqref{equ::fierz2}.

\chapter{Infrared and Ultraviolet Regularization} \label{app::infanduv}
One of the central building blocks of the flow equation Eqs. \eqref{equ::flow1}, \eqref{equ::flow2}
is the IR cutoff $R_{k}$. In
this appendix we want to give explicit examples of the cutoff
functions and define the threshold functions which appear in explicit calculations
as a consequence of the trace over momentum space. Furthermore, we briefly comment
on some technical points concerning the
correspondence between UV and IR regularization.
\section{Cutoff Functions}
To derive the flow equation \eqref{equ::flow2} in Chap. \ref{chap::effectiveaction}
we added an additional term to the effective action providing for
an IR regularization,
\begin{equation}
\Delta S_{k}[\phi]=\frac{1}{2}\int_{p} \phi^{\textrm{T}}(-p)R_{k}(p)\phi(p).
\end{equation}

The inverse massless average (i.e. regularized) propagator $P_{B}$ for bosons and
the corresponding squared quantity $P_{F}$ for fermions are given
by (cf. Sect. \ref{sec::rg}, Eqs. \eqref{equ::optcutoff}, \eqref{equ::fermioncutoff}),
\begin{eqnarray}
P_{B}&=&q^2+Z^{-1}_{\phi,k}R_{k}(q)=q^2(1+r_{B}(q^2)),\quad
\\\nonumber
P_{F}&=&q^2(1+r_{F}(q^2))^2,
\end{eqnarray}
where $r_{B}$ and $r_{F}$ reflect the presence of the IR cutoff.
The dimensionless functions $r_{B}$ and $r_{F}$ depend only on
$y=q^2/k^2$.

Expressed in terms of the $r$'s the linear cutoff, Eqs. \eqref{equ::optcutoff}, \eqref{equ::fermioncutoff})
is as follows,
\begin{eqnarray}
\label{equ::cutoff}
r_{B}(y)&=&\left(\frac{1}{y}-1\right)\Theta(1-y),
\\\nonumber
r_{F}(y)&=&\left(\frac{1}{\sqrt{y}}-1\right)\Theta(1-y).
\end{eqnarray}
Other typical examples are the sharp momentum cutoff,
\begin{eqnarray}
\label{equ::sharpcutoff}
r_{B}(y)&=&\frac{\Theta(1-y)}{1-\Theta(1-y)},
\\\nonumber
r_{F}(y)&=&\frac{\Theta(1-y)}{1-\Theta(1-y)},
\end{eqnarray}
and the popular exponential cutoff
\begin{eqnarray}
r_{B}(y)&=&\frac{1}{1-\exp(-y)}-1,
\\\nonumber
r_{F}(y)&=&\sqrt{\frac{1}{1-\exp(-y)}}-1.
\end{eqnarray}

\section{Threshold Functions}\label{app::cutoff}
The (super-)trace in the flow equations \eqref{equ::flow1}, \eqref{equ::flow2} includes
a momentum space integral, reminiscent of a one-loop expression. Typically,
the integral kernels are products of the IR regularized propagators and their derivatives.
In most parts of this work, we use standard threshold functions as defined in \cite{Berges:2000ew},
which we evaluate below explicitly for a finite UV cutoff $\Lambda$ and for the linear cutoff \eqref{equ::cutoff}.
In Sect. \ref{sec::beyondlpa} we need several additional threshold functions, not defined in the
standard literature. To facilitate the notation we define a new,
enlarged set of threshold functions in Sect. \ref{app::newthresh}, and label the threshold functions
somewhat differently.
\subsection{Evaluation with Finite UV Cutoff}\label{app::finite}
To adapt the IR regulator to a finite UV cutoff one can modify
the cutoff functions by a term which becomes infinite
for all $y\geq\frac{\Lambda^{2}}{k^2}$.
For our purposes it is simpler ´to restrict the range of
$x$ to $[0,\Lambda^{2}]$. This has the same effect.
In absence of
mass terms the threshold functions can only depend on the ratio
$s=k^{2}/\Lambda^{2}$. With
\begin{equation}
\tilde{\partial}_{t}=\frac{q^2}{Z_{\phi,k}}\frac{\partial[Z_{\phi,k}r_{B}]}{\partial t}\frac{\partial}{\partial P_{B}}
+\frac{2}{Z_{\psi,k}}\frac{P_{F}}{1+r_{F}}
\frac{\partial[Z_{\psi,k}r_{F}]}{\partial t}\frac{\partial}{\partial P_{F}}
\end{equation}
we find for bosons ($x=q^2$)
\begin{eqnarray}
\label{equ::bosonthresh}
l^{(B),d}_{0}(\omega,\eta_{\phi},s)
&=&\frac{1}{2}k^{-d}\int^{\Lambda^{2}}_{0}dx\,x^{\frac{d}{2}-1}\tilde{\partial_{t}}\ln(P_{B}(x)+\omega
k^{2})
\\\nonumber
&=&\frac{2}{d}\left[1-\frac{\eta_{\phi}}{d+2}\right]\frac{1}{1+\omega}\Theta(1-s)
\\\nonumber
&&+\frac{2}{d}s^{-\frac{d}{2}}\left[1-\frac{\left(2+d(1-s^{-1})\right)\eta_{\phi}}{2(d+2)}\right]\frac{1}{1+\omega}
\Theta(s-1)
\end{eqnarray}
and for fermions
\begin{eqnarray}
\label{equ::fermionthresh}
l^{(F),d}_{0}(\omega,\eta_{\psi},s)
&=&\frac{1}{2}k^{-d}\int^{\Lambda^{2}}_{0}dx\,x^{\frac{d}{2}-1}\tilde{\partial_{t}}\ln(P_{F}(x)+\omega
k^{2})
\\\nonumber
&=&\frac{2}{d}\left[1-\frac{\eta_{\psi}}{d+1}\right]\frac{1}{1+\omega}\Theta(1-s)
\\\nonumber
&&+\frac{2}{d}s^{-\frac{d}{2}}\left[1-\frac{(d+1-ds^{-1})\eta_{\psi}}{d+1}\right]\frac{1}{1+\omega}\Theta(s-1).
\end{eqnarray}
Higher threshold functions can be obtained simply by
differentiating with respect to $\omega$:
\begin{equation}
l^{d}_{n+1}(\omega,\eta,s)=-\frac{1}{n+\delta_{n,0}}\frac{d}{d\omega}l^{d}_{n}(\omega,\eta,s).
\end{equation}
For a finite value of the UV cutoff $\Lambda$ the threshold
functions are explicitly $s$- and therefore $k$-dependent. Taking
$\Lambda\rightarrow\infty$ we have $s=0$ for any value of $k$.
This renders the threshold functions $k$-independent.

In the pointlike truncations of Chaps. \ref{chap::njl}-\ref{chap::boso2} we neglect the anomalous
dimensions $\eta_{\phi}$,
$\eta_{\psi}$ and effectively only consider a fermionic cutoff
since $Z_{\phi,k}=0$. Moreover, the fermions are massless
and we abbreviate for $\omega=0$
\begin{equation}
l^{d}_{n}(0,0,s)=l^{d}_{n}(s).
\end{equation}
This yields explicitly
\begin{equation}
l^{(F),4}_{1}(s)=\frac{1}{2}\left[\Theta(1-s)+s^{-2}\Theta(s-1)\right].
\end{equation}
To obtain the perturbative result from the fermionic RG equation
we used
\begin{equation}
\label{equ::universalint} \int^{\infty}_{-\infty}dt
k^{2}l^{(F),4}_{1}(s)=\int^{\infty}_{0}dk\, k
l^{(F),4}_{1}(s)=\frac{\Lambda^{2}}{2}.
\end{equation}
As long as we keep the sharp momentum cutoff at $q=\Lambda$ this
integral is universal, i.e. it does not depend on the precise
choice of the IR cutoff. Indeed the universality is necessary to
reproduce perturbation theory for every choice of the IR cutoff.

\subsection{Cutoff Independence for Pointlike Truncations}
We have also used other cutoff functions $R_{k}$ different from
the linear cutoff. Within the local interaction approximation we
have found that the value of the critical coupling comes out
independent of the choice of $R_{k}$. The basic reason is that a
multiplicative change of $l^{(F),4}_{1}$ due to the use of another
threshold function can be compensated by a rescaling of $k$ (cf.
Eq. \eqref{equ::fermionflow}). The rescaling is simply
multiplicative for $s<1$, with a suitable generalization for
$s>1$. Critical values of the flow which are defined for
$k\to\infty$ are not affected by the rescaling. Let us demonstrate
this for $\bar{\lambda}_{V}$. Writing Eq. \eqref{equ::fermionflow}
in the scale variable $k$ we have
\begin{eqnarray}
\partial_{k}\bar{\lambda}_{V,k}
=-4v_{4}l^{(F),4}_{1}(s)k(\bar{\lambda}_{\sigma,k}+\bar{\lambda}_{V,k})^{2}.
\end{eqnarray}
Rescaling to
\begin{equation}
\tilde{k}(k)=\int^{k}_{0} dk\, k l^{(F),4}_{1}(s)
\end{equation}
we find
\begin{equation}
\partial_{\tilde{k}}\bar{\lambda}_{V,\tilde{k}}
=-4v_{4}
(\bar{\lambda}_{\sigma,\tilde{k}}+\bar{\lambda}_{V,\tilde{k}})^{2}.
\end{equation}
Due to the universality of Eq. \eqref{equ::universalint} the
domain for $k$, $[0,\infty]$, is now mapped to
$[0,\frac{\Lambda^{2}}{2}]$, giving the domain for $\tilde{k}$
independent of the IR cutoff. Having obtained identical
differential equations for every choice of IR cutoff without any
rescaling of $\bar{\lambda}$ establishes the above claim for the
critical couplings.

Note however, that this would not hold if we would start the
integration of the flow equation at $k=\Lambda$. In this case the
domain $[0,\Lambda]$ for $k$ is mapped into an interval for
$\tilde{k}$ that depends on the threshold function and therefore
on $R_{k}$. Actually, the $R_{k}$ dependence in this case is not
very surprising because different IR cutoffs then correspond to
different UV regularizations. Since our model is naively
non-renormalizable results can depend on the choice of UV
regularization (cf. Sect. \ref{app::regularization}).
\subsection{Threshold Functions for Sect. \ref{sec::beyondlpa}} \label{app::newthresh}
Similar to Eq. \eqref{equ::bosonthresh}, \eqref{equ::fermionthresh} we define our modified
threshold functions by integrals over $x=q^2$. Threshold functions with derivatives of
the fermion propagator are denoted by a Greek letter, all others by Latin letters.
With
\begin{equation}
F^{-1}(x)=\frac{1+r_{F}(x)}{P_{F}(x)},
\end{equation}
and suppressing the arguments
$(\omega_{F},\eta_{F},\omega_{1},\eta_{1},\omega_{2},\eta_{2})=(F,1,2)$ of the threshold
functions and the argument $(x)$ of the inverse propagators we write,
\begin{eqnarray}
a^{(FBB),d}_{n,m_{1},m_{2}}\!\!
&=&\!\!-\frac{1}{2}k^{2(n+m_{1}+m_{2}-1)-d}\,\tilde{\partial}_{t}\,\int dx x^{\frac{d}{2}}F^{-n}P^{-m_{1}}_{1}P^{-m_{2}}_{2},
\\
\beta^{(FBB),d}_{n,m_{1},m_{2}}\!\!&=&\!\!-\frac{1}{2}k^{2(n+m_{1}+m_{2})-d}\,\tilde{\partial}_{t}\,\int dx x^{\frac{d}{2}}
\dot{F}F^{-(n+1)}P^{-m_{1}}_{1}P^{-m_{2}}_{2},
\\
b^{(FBB),d}_{n,m_{1},m_{2}}\!\!&=&\!\!-\frac{1}{2}k^{2(n+m_{1}+m_{2})-d}\,\tilde{\partial}_{t}\,\int dx x^{\frac{d}{2}}
F^{-n}\dot{P}_{1}P^{-(m_{1}+1)}_{1}P^{-m_{2}}_{2},
\\
\gamma^{(FBB),d}_{n,m_{1},m_{2}}\!\!&=&\!\!-\frac{1}{2}k^{2(n+m_{1}+m_{2})-d}\,\tilde{\partial}_{t}\,\int dx x^{\frac{d}{2}+1}
(\dot{F})^{2}F^{-(n+2)}P^{-m_{1}}_{1}P^{-m_{2}}_{2},
\\
c^{(FBB),d}_{n,m_{1},m_{2}}\!\!&=&\!\!-\frac{1}{2}k^{2(n+m_{1}+m_{2})-d}\,\tilde{\partial}_{t}\,\int dx x^{\frac{d}{2}+1}
F^{-n}(\dot{P}_{1})^{2}P^{-(m_{1}+2)}_{1}P^{-m_{2}}_{2},
\\
\delta^{(FBB),d}_{n,m_{1},m_{2}}\!\!&=&\!\!-\frac{1}{2}k^{2(n+m_{1}+m_{2})-d}\,\tilde{\partial}_{t}\,\int dx x^{\frac{d}{2}+1}
\dot{F}F^{-(n+1)}\dot{P}_{1}P^{-(m_{1}+1)}_{1}P^{-m_{2}}_{2},
\\
d^{(FBB),d}_{n,m_{1},m_{2}}\!\!&=&\!\!-\frac{1}{2}k^{2(n+m_{1}+m_{2})-d}\,\tilde{\partial}_{t}\,\int dx x^{\frac{d}{2}+1}
F^{-n}\dot{P}_{1}P^{-(m_{1}+1)}_{1}\dot{P}_{2}P^{-(m_{2}+1)}_{2},
\\
\epsilon^{(FBB),d}_{n,m_{1},m_{2}}\!\!&=&\!\!-\frac{1}{2}k^{2(n+m_{1}+m_{2})-d}\,\tilde{\partial}_{t}\,\int dx x^{\frac{d}{2}+1}
\ddot{F}F^{-(n+1)}P^{-m_{1}}_{1}P^{-m_{2}}_{2},
\\
e^{(FBB),d}_{n,m_{1},m_{2}}\!\!&=&\!\!-\frac{1}{2}k^{2(n+m_{1}+m_{2})-d}\,\tilde{\partial}_{t}\,\int dx x^{\frac{d}{2}+1}
F^{-n}\ddot{P}_{1}P^{-(m_{1}+1)}_{1}P^{-m_{2}}_{2}.
\end{eqnarray}
If possible, i.e. if in the threshold function $t$ no $x$-derivatives
act on $P_{2}$, respectively $P_{1}$ and $P_{2}$, we abbreviate
\begin{equation}
t^{(FBB),d}_{n,m,0}=t^{(FB),d}_{n,m} \,\,\,\textrm{and}\,\,\ t^{(FB),d}_{n,0}=t^{(F),d}_{n}.
\end{equation}

For the linear cutoff Eq. \eqref{equ::cutoff} the integrals can be done explicitly
as in Sect. \ref{app::finite}. This is useful for numerical computations but
not very enlightening.

Some of the threshold functions defined above are straightforwardly related to the standard threshold functions
used in the literature. Particularly noteworthy are the following relations,
\begin{eqnarray}
l^{(F),d}_{n}(0,\eta_{F})&=&a^{(F),d+2(n-1)}_{2n}(0,\eta_{F})
\\\nonumber
l^{(B),d}_{n}(\omega,\eta)&=&a^{(FB),d-2}_{0,n}(0,0,\omega,\eta)
\\\nonumber
l^{(FB),d}_{n,m}(0,\eta_{F},\omega,\eta)&=&a^{(FB),d+2(n-1)}_{n,m}(0,\eta_{F},\omega,\eta)
\\\nonumber
m^{(F),d}_{4}(\omega_{F},\eta_{F})&=&\gamma^{(F),d}_{2}(\omega_{F},\eta_{F}).
\end{eqnarray}

The threshold functions defined above are not mutually independent. By partial integration
one can obtain relations linking some of the threshold functions above. Let us content ourselves
with two examples which are useful for a comparison with results in the literature,
\begin{equation}
\beta^{(FBB),d}_{n,m_{1},m_{2}}(F,1,2)=\frac{d}{2n}a^{(FBB),d-2}_{n,m_{1},m_{2}}(F,1,2)
-\frac{m_{1}}{n}b^{(FBB),d}_{n,m_{1},m_{2}}(F,1,2)-\frac{m_{2}}{n}b^{(FBB),d}_{n,m_{2},m_{1}}(F,2,1).
\end{equation}
In the special case $m_{1}=m_{2}=0$, $d=2n=4$ this reduces to
\begin{equation}
\label{equ::universal1}
\beta^{(F),4}_{2}(F)=a^{(F),2}_{2}(F).
\end{equation}
Moreover,
\begin{eqnarray}
\!\!\!\!\!\!\epsilon^{(FBB),d}_{n,m_{1},m_{2}}(F,1,2)=&-&\frac{d+2}{2}\beta^{(FBB),d}_{n,m_{1},m_{2}}(F,1,2)
+(n+1)\gamma^{(FBB),d}_{n,m_{1},m_{2}}(F,1,2)
\\\nonumber
&+&m_{1}\delta^{(FBB),d}_{n,m_{1},m_{2}}(F,1,2)+m_{2}\delta^{(FBB),d}_{n,m_{1},m_{2}}(F,2,1),
\end{eqnarray}
considering the same case as above and using relation \eqref{equ::universal1} we find
\begin{equation}
\label{equ::universal2}
\epsilon^{(F),d}_{2}(F)+xa^{(F),2}_{2}(F)+y\beta^{(F),4}_{2}(F)=3\gamma^{(F),4}_{2}(F)=3m^{(F),4}_{4}(F),
\quad \textrm{for} \,\,(x+y)=3.
\end{equation}

Finally, let us come to a property we need when studying the pointlike limit for
the bosons.
For large boson masses the inverse propagator acts
like $\omega^{2}$. This gives us the asymptotic properties\footnote{This argument and the corresponding
properties are only valid at fixed anomalous dimension.},
\begin{eqnarray}
a^{(FBB),d}_{n,m_{1},m_{2}}\!\!
&=&\!\!-\frac{1}{2}k^{2(n+m_{1}+m_{2}-1)-d}\,\tilde{\partial}_{t}\int dx x^{\frac{d}{2}}F^{-n}(x)\omega^{-2m_{1}}_{1}(x)\omega^{-2m_{2}}_{2}(x)
\\\nonumber
&=&\frac{1}{\omega^{2m_{1}}_{1}\omega^{2m_{2}}_{2}}a^{(F),d}_{n}
\\\nonumber
b^{(FBB),d}_{n,m_{1},m_{2}}\!\!&=&\!\!-\frac{1}{2}k^{2(n+m_{1}+m_{2}-1)-d}\,\tilde{\partial}_{t}\int dx x^{\frac{d}{2}}
F^{-n}(x)\dot{P}_{1}(x)\omega^{-2(m_{1}+1)}_{1}(x)\omega^{-2m_{2}}_{2}(x)
\\\nonumber
&=&\frac{1}{\omega^{2(m_{1}+1)}_{1}\omega^{2m_{2}}_{2}}b^{(F),d}_{n},
\end{eqnarray}
and similar for all other threshold functions.
\section{UV Regularization -- ERGE Scheme}\label{app::regularization}
\subsection{Effect of UV Regularization}
\begin{figure}[!t]
\begin{center}
\begin{picture}(220,150)(10,10)
\Text(210,35)[]{\scalebox{1.3}[1.3]{$\frac{q^2}{\Lambda^{2}}$}}
\includegraphics[width=8cm]{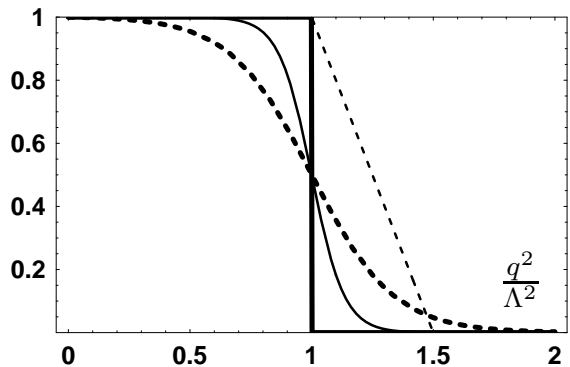}
\end{picture}
\end{center}
\caption{In a UV regularized theory not all modes contribute completely.
This plot schematically depicts ``how much'' each mode contributes.
The thick line is for the sharp momentum cutoff. All modes with $q^{2}\leq\Lambda^{2}$ are included completely.
Other UV regularizations (dashed, thin dashed and thin solid line) typically not only include a small fraction
of the high momentum modes, $q^2>\Lambda^{2}$, but in addition leave out a small fraction
of the low momentum modes.}
\label{fig::regu}
\end{figure}

UV regularization can be implemented by a sharp cutoff in all integrals
over momentum space\footnote{When choosing a UV regularization one has to
be careful to take one compatible with the symmetries of the theory in question. E.g.
for gauge theories the sharp momentum cutoff violates gauge symmetry and is therefore
not a suitable choice.}. Yet, this
is not the only possibility. Indeed, it is often not the most practical regularization.
E.g. in perturbation theory dimensional regularization is often much more convenient.
But, even at fixed spacetime dimension  there are other regularization methods. Prominent
examples are the Pauli-Villars and Schwinger proper time
regularization (for details s. e.g. \cite{Zinn-Justin:2002vj}).
More or
less any modification (compatible with the symmetries) of the short
distance behavior of the propagators which renders all
Feynman diagrams finite, can be called a UV regularization.
In general, the modes with $q^2>\Lambda^2$ are not completely left out, only suppressed, as
depicted in Fig. \ref{fig::regu}.

From another point of view, a modification of the propagator can be implemented in the action.
This gives then an ``UV regularized classical action''. The function describing the suppression
of the UV modes (cf. Fig. \ref{fig::regu}) in some way appears in this UV regularized action, e.g.
the sharp cutoff limit can be implemented by multiplying all terms in the action by appropriate
$\Theta(1-\frac{q^2}{\Lambda^{2}})$-functions. Hence, different UV regularizations usually correspond
to different ``classical actions''. Therefore, it is no surprise, that different
UV regularizations give different results. In particular, this
is true for the critical coupling in our NJL model Eq. \eqref{equ::faction}, as one
can see by comparing Tabs. \ref{tab::crit}, \ref{tab::crit2}
with Tabs. \ref{tab::crit3}, \ref{tab::crit4}, calculated using a sharp UV cutoff at $\Lambda$
and a UV regularization by the ERGE scheme (s. below) with the linear cutoff Eq. \eqref{equ::cutoff}, respectively.
Only in renormalizable theories\footnote{This includes critical behavior at second-order phase
transitions (cf. Chap. \ref{chap::quest}), in particular critical exponents and other universal quantities.}
we can remove the cutoff and obtain results independent of the
specifics of the UV regularization.

\subsection{ERGE Scheme}
In Sect. \ref{app::finite} we have evaluated the threshold functions for a theory
which is UV regularized by a sharp cutoff in momentum space. It turned out that the threshold
functions depended on the ratio $s=\frac{k^2}{\Lambda^2}$ in a rather complicated way.
Constant threshold functions would be desirable, among other things to simplify
numerical calculations.

We already noted, that for $\Lambda=\infty$ the cutoff functions are indeed constant for all $k$,
because $s=0$ for all finite $k$. Moreover, for an IR cutoff decreasing sufficiently fast in the UV
all threshold functions are finite, even for $\Lambda=\infty$. The flow
equation \eqref{equ::flow1}, \eqref{equ::flow2} is UV finite.
Now, let us remember that we have chosen the IR cutoff such that
only the modes around $q^2=k^2$ effectively contribute to the flow (Eq. \eqref{equ::contribcond}).
Starting the flow at $k_{0}=\Lambda$ and integrating to $k=0$ we have included only modes
with $q^2\lesssim\Lambda^2$. This is exactly what we expect for an UV regularization.
More precisely, the UV regularization is now implemented in the finite choice
for the initial conditions at $k=k_{0}=\Lambda$. This is the so called
ERGE scheme for the UV regularization.

As flow equations are much simpler with constant threshold functions,
this is the method of choice to implement UV regularization in ERGE. Nevertheless, this is bought
at the prize that we cannot any longer compare non-universal quantities between different
IR regularizations, as they automatically lead to different UV regularizations.

Although it is usually not the simplest method, we can invoke UV regularization by the
ERGE scheme also in the context of perturbation theory or SDE. This follows
along the lines indicated in Sect. \ref{sec::rg} for perturbation theory.
Typically any expression can be written in terms of inverse propagators $P$, internal
momenta $q$ we integrate over, and external momenta $p$ we do not integrate over,
\begin{equation}
\label{equ::unreg}
\int_{q} F(P,q,p).
\end{equation}
A specific ERGE scheme is specified by the choice of the IR regulator $R_{k}$.
Replacing the inverse propagator $P$ by the IR regularized inverse propagator $P+R_{k}$
we can calculate the contribution from each scale $k$, $k^{-1}\tilde{\partial}_{t}F(P+R_{k},q,p)$.
Integrating over all scales from $k_{0}=\Lambda$ to $k=0$ we obtain the UV regularized
expression,
\begin{equation}
\int^{0}_{k_{0}=\Lambda} dk\,k^{-1}\tilde{\partial_{t}}\left[\int_{q} F(P+R_{k},q,p)\right].
\end{equation}

\chapter{Flow Equations for Sect. \ref{sec::beyondlpa}}\label{app::flowbeyond}
In this appendix we list the flow equations for the effective action
\eqref{equ::baction} generalized to include kinetic terms \eqref{equ::kinetic}.
Moreover, we calculate the (momentum-dependent) field redefinitions
necessary to keep a simple form of the action with Yukawa couplings constant in
momentum space and no four-fermion interactions.
\section{Flow Equations at Fixed Fields}
The first diagram that we
have to evaluate is Fig. \ref{fig::mass}, giving the self-energy
$\mu^{2}(p)$. Momentum conservation implies that it depends only
on one momentum variable $p$. Expanding for small values of $p^{2}$ and
abbreviating the arguments of the threshold functions (cf. App.
\ref{app::newthresh}) as $F=(\omega_{F},\eta_{F})$ we find,
\begin{eqnarray}
\pt\mu^{2}_{\sigma}&=&8\vv k^{2}\hsz\af{F},
\\\nonumber
\pt\mu^{2}_{V}&=&8\vv k^{2}\hvz\af{F},\quad\pt\mu^{2}_{A}=8\vv
k^{2}\haz\af{F}
\end{eqnarray}
and
\begin{eqnarray}
\frac{\pt
Z_{\sigma}}{Z_{\sigma}}&=&-4\vv\hsz\gamma^{(F),4}_{2}(F),
\\\nonumber
\frac{\pt
Z_{V}}{Z_{V}}&=&-\frac{16}{3}\vv\hvz\gamma^{(F),4}_{2}(F),\quad
\frac{\pt Z_{A}}{Z_{A}}=-\frac{16}{3}\vv\haz\gamma^{(F),4}_{2}(F).
\end{eqnarray}

In general, the vertex correction depicted in Fig.
\ref{fig::vertex} depends on two momentum variables. As discussed in Sect. \ref{sec::momconf}
we perform the evaluation for
the configuration
$(p_{1},p_{2},p_{3})=(p,\frac{1}{2}p,\frac{1}{2}p)$. With
$\sigma=(\omega_{\sigma},\eta_{\sigma})$,
$V=(\omega_{V},\eta_{V})$, $A=(\omega_{A},\eta_{A})$ this yields,
\begin{eqnarray}
\beta^{(0)}_{\hs}&=&-16\vv\hs\bigg[\hvz\aefb{F}{V}-\haz\aefb{F}{A}\bigg],
\\\nonumber
\beta^{(0)}_{\hv}&=&-2\vv\hv\bigg[\hsz\aefb{F}{\sigma}+2\hvz\aefb{F}{V}+2\haz\aefb{F}{A}\bigg],
\\\nonumber
\beta^{(0)}_{\ha}&=&2\vv\ha\bigg[\hsz\aefb{F}{\sigma}-2\hvz\aefb{F}{V}-2\haz\aefb{F}{A}\bigg],
\end{eqnarray}
and
\begin{eqnarray}
\beta^{(2)}_{\hs}&=&4\vv\hs\bigg[\hvz(\aefb[2]{F}{V}+2\betaefb{F}{V}-\gammaefb{F}{V}+\epsilonefb{F}{V})
\\\nonumber
&&\hspace{1.2cm}-(V\rightarrow A)\bigg],
\\\nonumber
\beta^{(2)}_{\hv}&=&\frac{1}{3}\vv\hv\bigg[\hsz(3\aefb[2]{F}{\sigma}+3\betaefb{F}{\sigma}
-2\gammaefb{F}{\sigma}+2\epsilonefb{F}{\sigma})
\\\nonumber
&&\hspace{1.2cm}+2\hvz(3\aefb[2]{F}{V}+3\betaefb{F}{V}-2\gammaefb{F}{V}+2\epsilonefb{F}{V})
\\\nonumber
&&\hspace{1.2cm}+(V\rightarrow A)\bigg],
\\\nonumber
\beta^{(2)}_{\ha}&=&\frac{1}{3}\vv\ha\bigg[-\hsz(3\aefb[2]{F}{\sigma}+3\betaefb{F}{\sigma}
-2\gammaefb{F}{\sigma}+2\epsilonefb{F}{\sigma})
\\\nonumber
&&\hspace{1.2cm}+2\hvz(3\aefb[2]{F}{V}+3\betaefb{F}{V}-2\gammaefb{F}{V}+2\epsilonefb{F}{V}))
\\\nonumber
&&\hspace{1.2cm}+(V\rightarrow A)\bigg].
\end{eqnarray}

In order to compare the momentum dependence on $s$ and $t$ in the vicinity of $(s,t)=(0,0)$ we have
evaluated both momentum configurations discussed in Sect. \ref{sec::momconf}.

For the configuration $(p_{1},p_{2},p_{3},p_{4})=\frac{1}{2}(p,p,p,p)$ with $t=0$ and $s=p^{2}$ we find
\begin{eqnarray}
\label{equ::schannel1} \bsn&=&-8\vv\hsz\haz\afbb{F}{\sigma}{V},
\\\nonumber
\bvn&=&24\vv\hv\ha\afbb{F}{V}{A},
\\\nonumber
\ban&=&-\vv[\hsv\afb{F}{\sigma}-12\hvv\afb{F}{V}-12\hav\afb{F}{A}],
\end{eqnarray}
\begin{eqnarray}
\label{equ::schannel2}
\bsz&=&\vv\hsz\haz\bigg[2\bfbb{F}{\sigma}{A}-2\cfbb{F}{\sigma}{A}+\dfbb{F}{\sigma}{a}
\\\nonumber
&&\hspace{3cm}+\efbb{F}{\sigma}{A}+(\sigma\leftrightarrow A)\bigg],
\\\nonumber
\bvz&=&-\frac{2}{3}\vv\hvz\haz\bigg[9\bfbb{F}{V}{A}-10\cfbb{F}{V}{A}+5\dfbb{F}{V}{A}
\\\nonumber
&&\hspace{3cm}+\efbb{F}{V}{A}+(V\leftrightarrow A)\bigg],
\\\nonumber
\baz&=&-\frac{1}{6}\vv\bigg[4\hvv(9\bfb{F}{V}-5\cfb{F}{V}+5\efb{F}{V})+(V\rightarrow
A)
\\\nonumber
&&\hspace{1cm}+\hsv(-3\bfb{F}{\sigma}+\cfb{F}{\sigma}-\efb{F}{\sigma}\bigg].
\end{eqnarray}

After the appropriate Fierz transformation the configuration $(p_{1},p_{2},p_{3},p_{4})=\frac{1}{2}(p,-p,-p,p)$
with $s=0$ and $t=p^{2}$ yields,
\begin{eqnarray}
\label{equ::tchannel1}
\bsn&=&\vv\bigg[\hsv\afb{F}{\sigma}+24\hvz\haz\afbb{F}{V}{A}-12\hvv\afb{F}{V}
\\\nonumber
&&\hspace{1cm}-12\hav\afb{F}{A}\bigg],
\\\nonumber
\bvn&=&-\vv\bigg[\frac{1}{2}\hsv\afb{F}{\sigma}+4\hsz\haz\afbb{F}{\sigma}{A}
-12\hvz\haz\afbb{F}{V}{A}
\\\nonumber
&&\hspace{1cm}-6\hvv\afb{F}{V}-6\hav\afb{F}{A}\bigg],
\\\nonumber
\ban&=&-\vv\bigg[\frac{1}{2}\hsv\afb{F}{\sigma}-4\hsz\haz\afbb{F}{\sigma}{A}
-12\hvz\haz\afbb{F}{V}{A}
\\\nonumber
&&\hspace{1cm}-6\hvv\afb{F}{V}-6\hav\afb{F}{A}\bigg],
\end{eqnarray}
\begin{eqnarray}
\label{equ::tchannel2} \bsz&=&\frac{1}{4}\vv\bigg[
\hsv(\afb[2]{F}{\sigma}-4\betafb{F}{\sigma}+3\gammafb{F}{\sigma}-\epsilonfb{F}{\sigma})
\\\nonumber
&&\hspace{1cm}-8\hvz\haz(5\afbb[2]{F}{V}{A}+4\betafbb{F}{V}{A}-\gammafbb{F}{V}{A}
\\\nonumber
&&\hspace{3cm}+3\epsilonfbb{F}{V}{A})-\frac{1}{2}(A\rightarrow
V)-\frac{1}{2}(V\rightarrow A)\bigg],
\\\nonumber
\bvz&=&\frac{1}{12}\vv\bigg[
\hsv(3\afb[2]{F}{\sigma}+3\betafb{F}{\sigma}-2\gammafb{F}{\sigma}+2\epsilonfb{F}{\sigma})
\\\nonumber
&&\hspace{1cm}+4\hsz\hvz(3\afbb[2]{F}{\sigma}{V}+\gammafbb{F}{\sigma}{V}+\epsilonfbb{F}{\sigma}{V})
\\\nonumber
&&\hspace{1cm}+4\hsz\haz(3\afbb[2]{F}{\sigma}{A}+6\betafbb{F}{\sigma}{A}-5\gammafbb{F}{\sigma}{A}
\\\nonumber
&&\hspace{3cm}+3\epsilonfbb{F}{\sigma}{A})
\\\nonumber
&&\hspace{1cm}+8\hvz\haz(9\afbb[2]{F}{V}{A}-21\betafbb{F}{V}{A}+16\gammafbb{F}{V}{A}
\\\nonumber
&&\hspace{3cm}-4\epsilonfbb{F}{V}{A})+\frac{1}{2}(A\rightarrow
V)+\frac{1}{2}(V\rightarrow A)\bigg],
\\\nonumber
\baz&=&\frac{1}{12}\vv\bigg[\hsv(3\afb[2]{F}{\sigma}+3\betafb{F}{\sigma}-2\gammafb{F}{\sigma}
+2\epsilonfb{F}{\sigma})
\\\nonumber
&&\hspace{1cm}-4\hsz\hvz(3\afbb[2]{F}{\sigma}{V}+\gammafbb{F}{\sigma}{V}+\epsilonfbb{F}{\sigma}{V})
\\\nonumber
&&\hspace{1cm}-4\hsz\haz(3\afbb[2]{F}{\sigma}{A}+6\betafbb{F}{\sigma}{A}-5\gammafbb{F}{\sigma}{A}
\\\nonumber
&&\hspace{3cm}+3\epsilonfbb{F}{\sigma}{A})
\\\nonumber
&&\hspace{1cm}+8\hvz\haz(9\afbb[2]{F}{V}{A}-21\betafbb{F}{V}{A}+16\gammafbb{F}{V}{A}
\\\nonumber
&&\hspace{3cm}-4\epsilonfbb{F}{V}{A})+\frac{1}{2}(A\rightarrow V)+\frac{1}{2}(V\rightarrow A)\bigg].
\end{eqnarray}

\section{Field Redefinitions}
Before we start, let us remark that in this section we write down explicit factors of the
wave function renormalization $Z$. To obtain the expressions in the renormalized couplings we simply have to set
$Z=1$.
To keep the form of the effective action simple (more precisely to retain Yukawa couplings constant in
momentum space and $\lambda(p)=0$) we allow for momentum-dependent field redefinitions,
\begin{eqnarray}
\pt\phi(q)&=&-\left(\bar{\psi}\left(\frac{1-\gamma^{5}}{2}\right)\psi\right)(q)\pt\omega_{\sigma}(q)
+\phi(q)\pt\alpha_{\sigma}(q),
\\\nonumber
\pt\phi^{\star}(q)&=&\left(\bar{\psi}\left(\frac{1+\gamma^{5}}{2}\right)\psi\right)(q)\pt\omega_{\sigma}(q)
+\phi^{\star}(q)\pt\alpha_{\sigma}(q),
\\\nonumber
\pt
V^{\mu}(q)&=&(\bar{\psi}\gamma^{\mu}\psi)(q)\pt\omega_{V}(q)+V^{\mu}(q)\pt\alpha_{V}(q),
\\\nonumber
\pt
A^{\mu}(q)&=&(\bar{\psi}\gamma^{\mu}\gamma^{5}\psi)(q)\pt\omega_{A}(q)
+A^{\mu}(q)\pt\alpha_{A}(q).
\end{eqnarray}
Evaluating
\begin{equation}
\pt\Gamma=\pt\gamma\mid+\int_{q}\frac{\delta\Gamma}{\delta\phi(q)}\pt\phi(q)
+\frac{\delta\Gamma}{\delta\phi^{\star}(q)}\pt \phi^{\star}(q)
+\frac{\delta\Gamma}{\delta V^{\mu}(q)}\pt V^{\mu}(q)
+\frac{\delta\Gamma}{\delta A^{\mu}(q)}\pt A^{\mu}(q)
\end{equation}
we find
\begin{eqnarray}
\label{equ::changesigma}
\pt\ls(q)&=&\pt\ls(q)\mid-\hs\pt\omega_{\sigma}(q),
\\\nonumber
\pt\hs(q)&=&\pt\hs(q)\mid+(\mu^{2}_{\sigma}+Z_{\sigma}q^{2})\pt\omega_{\sigma}(q)+\hs\pt\alpha_{\sigma}(q),
\\\nonumber
\pt\mu^{2}_{\sigma}&=&\pt\mu^{2}_{\sigma}\mid+2\mu^{2}_{\sigma}\pt\alpha_{\sigma}(0),
\\\nonumber
\pt Z_{\sigma}(q)&=&\pt Z_{\sigma}(q)\mid+2\mu^{2}_{\sigma}
\frac{\pt\alpha_{\sigma}(q)-\pt\alpha_{\sigma}(0)}{q^{2}}
+2Z_{\sigma}\pt\alpha_{\sigma}(q),
\end{eqnarray}
\begin{eqnarray}
\label{equ::changev}
\pt\lv(q)&=&\pt\lv(q)\mid+2\hv\pt\omega_{V}(q),
\\\nonumber
\pt\hv(q)&=&\pt\hv(q)\mid-(\mu^{2}_{V}+Z_{V}q^{2})\pt\omega_{V}(q)+\hv\pt\alpha_{V}(q),
\\\nonumber
\pt\mu^{2}_{V}&=&\pt\mu^{2}_{V}\mid+2\mu^{2}_{V}\pt\alpha_{V}(q),
\\\nonumber
\pt Z_{V}(q)&=&\pt
Z_{V}(q)\mid+2\mu^{2}_{V}\frac{\pt\alpha_{V}(q)-\pt\alpha_{V}(0)}{q^{2}}+2Z_{V}\pt\alpha_{V}(q),
\end{eqnarray}
and an analogous equation with $(V\rightarrow A)$. As the expressions
for the axial vector boson can always be obtained by this replacement, we write
in the following only the expression for the vector boson.

Imposing the condition
\begin{equation}
\pt\lambda(q)=0,
\end{equation}
determines the functions $\pt\omega$,
\begin{eqnarray}
\pt\omega_{\sigma}(q)&=&\frac{\pt\lambda_{\sigma}(q)}{\hs},\quad\pt\omega_{V}(q)=-\frac{\pt\lv(q)}{2\hv}.
\end{eqnarray}
The requirement for a constant Yukawa coupling reads,
\begin{eqnarray}
\pt\left(\frac{\hs(q)-\hs(0)}{q^{2}}\right)&=&0,\quad\pt\left(\frac{\hv(q)-\hv(0)}{q^{2}}\right)=0.
\end{eqnarray}
This fixes the functions $\pt\alpha(q)$ up to a constant in $q$,
\begin{eqnarray}
\frac{\pt\alpha_{\sigma}(q)-\pt\alpha_{\sigma}(0)}{q^{2}}&=&-\frac{1}{\hs}
\bigg[\pt\left(\frac{\hs(q)-\hs(0)}{q^{2}}\right)\mid
\\\nonumber
&&\hspace{1cm}+\frac{\mu^{2}_{\sigma}}{\hs}
\pt\left(\frac{\ls(q)-\ls(0)}{q^{2}}\right)\mid+\frac{Z_{\sigma}}{\hs}\pt\lambda_{\sigma}(q)\mid\bigg],
\\\nonumber
\frac{\pt\alpha_{V}(q)-\pt\alpha_{V}(0)}{q^{2}}&=&-\frac{1}{\hv}
\bigg[\pt\left(\frac{\hv(q)-\hv(0)}{q^{2}}\right)\mid
\\\nonumber
&&\hspace{1cm}+\frac{\mu^{2}_{V}}{2\hv}\pt\left(\frac{\lv(q)-\lv(0)}{q^{2}}\right)\mid
+\frac{Z_{V}}{2\hv}\pt\lv(q)\mid\bigg].
\end{eqnarray}
The remaining constant can be fixed by requiring that our fields are always renormalized
if they are so at the beginning, i.e.
\begin{eqnarray}
\pt Z(0)=0.
\end{eqnarray}
Using our definition of the anomalous dimension, $\eta=2\pt\alpha^{(0)}$,
we find,
\begin{eqnarray}
\label{equ::zero}
\eta_{\sigma}=2\pt\alpha^{(0)}_{\sigma}&=&-\frac{\pt
Z_{\sigma}(0)\mid}{Z_{\sigma}} +\frac{2\mu^{2}_{\sigma}}{Z_{\sigma}\hs}\bigg[\pt
h^{(2)}_{\sigma}\mid
+\frac{\mu^{2}_{\sigma}}{\hs}\pt\lambda^{(2)}_{\sigma}\mid+\frac{Z_{\sigma}}{\hs}\pt\lambda^{(0)}_{\sigma}\mid\bigg],
\\\nonumber
\eta_{V}=2\pt\alpha^{(0)}_{V}&=&-\frac{\pt Z_{V}(0)\mid}{Z_{V}}
+\frac{2\mu^{2}_{V}}{Z_{V}\hv}\bigg[\pt
h^{(2)}_{V}\mid+\frac{\mu^{2}_{V}}{2\hv}\pt\lambda^{(2)}_{V}\mid
+\frac{Z_{V}}{2\hv}\pt\lambda^{(0)}_{V}\mid\bigg].
\end{eqnarray}
Inserting this into Eqs. \eqref{equ::changesigma}, \eqref{equ::changev}
we obtain our final form for the flow equations,
\begin{eqnarray}
\pt h^{2}_{\sigma}&=&\eta_{\sigma}h^{2}_{\sigma}+2h_{\sigma}\pt h_{\sigma}|
+2\mu^{2}_{\sigma}\pt \lambda^{(0)}_{\sigma}|,
\\\nonumber
\pt h^{2}_{V}&=&\eta_{V}h^{2}_{V}+2h_{V}\pt h_{V}|
+\mu^{2}_{V}\pt \lambda^{(0)}_{V}|,
\end{eqnarray}
\begin{eqnarray}
\pt \mu^{2}_{\sigma}=\eta_{\sigma}\mu^{2}_{\sigma}+\pt\mu^{2}_{\sigma}|,
\\\nonumber
\pt \mu^{2}_{V}=\eta_{V}\mu^{2}_{V}+\pt\mu^{2}_{V}|.
\end{eqnarray}

\end{appendix}

\pagestyle{empty} \cleardoublepage \thispagestyle{empty}

\chapter*{Thanks...}

First of all, to Prof. Dr. Christof Wetterich, for the kind and encouraging supervision of this thesis and
his ongoing optimism. For answering a nearly infinite amount of questions, while leaving room
for my own ideas.

\noindent
To my second supervisor, Prof. Dr. Michael G. Schmidt, for the friendly interest in my work and his willingness
to referee this thesis.

\noindent
I am deeply indebted to Dr. Holger Gies for many fruitful discussions, collaboration on wild ideas
(``Renormalizable Standard Model''), for proofreading and many suggestions.

\noindent
Special thanks go to Dr. Michael Doran for collaboration on quantum
corrections to Quintessence potentials resulting in my first
paper \cite{Doran:2002bc,Doran:2002qd} (yeah, that's the fine art
of highly unnecessary self-citation), a nice trip to Les
Arcs (``Jacques, you know I don't want to bother you, but...'') and many interesting discussions.

\noindent
I am very grateful to Dr. Tobias Baier for many discussions on the ``Fierz ambiguity'' and for reading
this manuscript.

\noindent
To Dr. J\"{u}rgen Berges and Prof. Dr. Berthold Stech for illuminating explanations.

\noindent
Of course, thanks to the rest of the Les Arcs gang, Lala Adueva, Jan Schwindt, \mbox{Dr. Frank Steffen}
(thanks for the nice workshop!), and all the others, Juliane Behrend, Dr. Eike Bick, Sebastian Diehl,
Dietrich Foethke, Felix H\"{o}fling, Bj\"{o}rn O. Lange,
Christian M. M\"{u}ller, Markus M. M\"{u}ller, Christian Nowak, Gregor Sch\"{a}fer
and Kai Schwenzer, who were always ready to discuss matters in and out of physics -- in short, it was fun.

\noindent
Finally, to Dr. Christiane J\"{a}ckel, Bengt J\"{a}ckel and Kurt Fippinger for being family.

\end{document}